\documentclass[suppldata]{interact}
\usepackage{epstopdf}
\usepackage{placeins}
\usepackage{color}
\usepackage{csquotes}
\usepackage{natbib}

\setcitestyle{comma}
\usepackage{dsfont}   
\usepackage{changepage}
\usepackage{amsmath}

\theoremstyle{plain}

\theoremstyle{definition}

\theoremstyle{remark}

\usepackage{xcolor}
\definecolor{ao(english)}{rgb}{0.0, 0.5, 0.0}
\definecolor{deeppink}{rgb}{1.0, 0.08, 0.58}
 \newcommand{\Hquad}{\hspace{0.25em}} 

\usepackage{amssymb}
\usepackage{amsthm}
\usepackage{amsmath}

\usepackage{amsfonts}
\usepackage{subfig}
\usepackage{enumerate}
\newcommand{\R}{\mbox{$\mathbb{R}$}}

\newcommand{\bqan}{\begin{eqnarray}}
	\newcommand{\eqan}{\end{eqnarray}}
\newcommand{\vmu}{\boldsymbol{\mu}}
\newcommand{\vSigma}{\boldsymbol{\Sigma}}

\newcommand{\vX}{\boldsymbol{X}}
\newcommand{\vC}{\boldsymbol{C}}
\newcommand{\vF}{\boldsymbol{F}}
\newcommand{\vp}{\boldsymbol{p}}
\newcommand{\vph}{\boldsymbol{\hat{p}}}

\newcommand{\vY}{\boldsymbol{Y}}
\newcommand{\vZ}{\boldsymbol{Z}}
\newcommand{\vR}{\boldsymbol{R}}
\newcommand{\vM}{\boldsymbol{M}}
\newcommand{\vI}{\boldsymbol{I}}
\newcommand{\vJ}{\boldsymbol{J}}

\newcommand{\vT}{\boldsymbol{T}}
\newcommand{\bqa}{\begin{eqnarray*}}
	\newcommand{\vVh}{\boldsymbol{\hat{V}}}
		\newcommand{\vV}{\boldsymbol{V}}
	\newcommand{\eqa}{\end{eqnarray*}}
\numberwithin{equation}{section}
\theoremstyle{plain}
\newtheorem{thm}{Theorem}[section]

\newtheorem*{assumption*}{\assumptionnumber}
\providecommand{\assumptionnumber}{}
\makeatletter
\newenvironment{assumption}[2]
{%
	\renewcommand{\assumptionnumber}{Assumption #1$\mathcal{#2}$}%
	\begin{assumption*}%
		\protected@edef\@currentlabel{#1$\mathcal{#2}$}%
	}
	{%
	\end{assumption*}
}

\begin{document}

\title{Incompletely observed nonparametric factorial designs with repeated measurements: A wild bootstrap approach}

\author{%
	Lubna Amro$^{1*}$, \footnote{\\$^1$Department of Statistics, TU Dortmund University, Dortmund, Germany
		\\$^2$Institute of Biometry and Clinical Epidemiology, Charit\`e—Universit\"atsmedizin Berlin, Berlin, Germany\\
		$^3$Berlin Institute of Health (BIH), Berlin, Germany\\
		$^4$UA Ruhr, Research Center Trustworthy Data Science and Security,  Dortmund, Germany
	}%
	Frank Konietschke$^{2,3}$ \footnote{\\{\bf Corresponding author}:\\
	Lubna Amro,	Department of Statistics, TU Dortmund University, Dortmund, Germany.\\
		Email: lubna.amro@tu-dortmund.de}%
	\and Markus Pauly$^{1,4}$ \footnote{}%
}

\maketitle

\begin{abstract}
In many life science experiments or medical studies, subjects are repeatedly observed  and measurements are collected in factorial designs with multivariate data.  
 The analysis of such multivariate data is typically based on multivariate analysis of variance (MANOVA) or mixed models, requiring complete data, and certain 
assumption on the underlying parametric distribution such as continuity or a specific covariance structure, e.g., compound symmetry. However, these methods are {\color{black} usually} not applicable when discrete data or even ordered categorical data are present. In such cases, nonparametric rank-based methods that do not require stringent distributional assumptions are the preferred choice. However, in the multivariate case, most rank-based approaches have only been developed for complete observations. It is the aim of this work is to develop asymptotic correct procedures that are capable of handling missing values, allowing for singular covariance matrices and {\color{black}are} applicable for ordinal or ordered categorical data. This is achieved by applying a wild bootstrap procedure {\color{black}in combination with quadratic form-type test statistics}. 
{\color{black}Beyond proving their asymptotic correctness, extensive simulation studies  
validate their applicability for small samples}. Finally, two real data examples are analyzed.

\end{abstract}


\section{Introduction}\label{int}

Factorial designs have a long history in several scientific fields {\color{black}such as biology, ecology or medicine} \citep{traux1965personality, nesnow1998lung, wildsmith2001maximization, biederman2008nematode, lekberg2018relative}. {\color{black}The reasons are simple: They are an efficient way to study main and interaction effects between different factors. Typically, such designs are inferred by parametric mean-based procedures} such as linear mixed models, or multivariate analysis of variance (MANOVA). These procedures, {\color{black} however}, rely on restrictive distributional assumptions, such as  multivariate normality, or special dependencies \citep{lawley1938generalization, bartlett1939note, davis2002statistical,johnson2007applied,fitzmaurice2012applied}.\\
  However, assessing multivariate normality or {\color{black}a specific} type of covariance matrix is difficult {\color{black} in practice} \citep{micceri1989unicorn,qian2005comparison,bourlier2008remodeling, xu2008robustified}, especially when the sample sizes are small. In particular, \cite{erceg2008modern} pointed out that \enquote{\it Researchers relying on statistical tests (e.g., Levene’s test) to identify assumption violations may frequently fail to detect deviations from normality and homoscedasticity that are large enough to seriously affect the type-I error rate and power of classic 
parametric tests}(p. 600). {\color{black}In fact}, classical MANOVA procedures are usually non robust to deviations and may result in inaccurate decisions caused by possibly conservative or inflated type-I error rates \citep{arnau2012using,pesarin2012review, brombin2013robust,konietschke2015parametric,pauly2015analysis,friedrich2017permuting,friedrich2018mats}. Moreover, if ordinal, or ordered categorical data are present, mean-based approaches are not applicable. A tempting alternative are nonparametric rank-based methods, which are  applicable for non-normal data, in particular discrete data or even ordered categorical data. Their key features are their robustness and their invariance under monotone transformations of the data. Consequently, \cite{akritas1994fully,akritas1997unified,munzel2000nonparametric,brunner2001nonparametric,Akritas2011,brunner2017rank}, and \cite{friedrich2017wild} proposed nonparametric ranking methods {\color{black}for all kinds of factorial designs. Thereby, hypotheses are no longer formulated in terms of means (which do not exist for ordinal data), distribution functions are used instead.} However, all these methods are only applicable for completely observed factorial designs data and cannot be used to analyze multivariate data with missing values.\\

In contrast, there are only a few methods which are applicable in case of missing values, require no parametric assumptions and also lead to valid inferences in case of arbitrary covariance structures, skewed distributions or unbalanced experimental designs. {\color{black} Examples in the nonparametric framework for the matched pairs designs include the ranking methods proposed by \cite{akritas2002nonparametric,konietschke2012ranking,fong2018rank}. A promising approach for factorial designs with repeated measurements is} given by the proposals of \cite{brunner1999rank} and  \cite{domhof2002rank}  who recommended two types of quadratic forms
 for testing nonparametric hypotheses in terms of distribution functions: rank based Wald and ANOVA-type {\color{black}statistics}. The Wald-type test is an asymptotically valid test {\color{black}which usually needs large sample sizes to obtain accurate test decisions, see  \cite{brunner2001asymptotic, friedrich2017wild} for the case of complete observations and our simulation study below.}
  Apart from that, the ANOVA-type test is 
 based upon an approximation of its distribution with scaled
  $\chi^2$-distributions. As the latter does not coincide with the ANOVA-type test limiting distribution under $H_0$, the test is in general not asymptotically correct and also exhibits a more or less conservative behaviour under small sample sizes, see  \cite{brunner2001asymptotic} for the case of complete observations. Another applicable  technique is the nonparametric imputation method proposed by  \cite{gao2007nonparametric}, which possesses a good type-I error control but a lower power behavior than the methods by  \cite{brunner1999rank} and  \cite{domhof2002rank}, see the simulation
  	study of \cite{gao2007nonparametric}.\\ 
  	
{\color{black}Thus, {\bf the aims of the present paper} are (i)} to provide statistical tests {\color{black} for hypotheses formulated in distribution functions} that are capable of treating missing values in factorial designs{\color{black}, (ii) work without} parametric assumptions such as continuity of the distribution functions, or nonsingular covariance matrices{\color{black}, (iii) are asymptotically correct while (iv) showing a satisfactorily}  type-I error control and good power properties. To accomplish this, we {\color{black} propose three different quadratic-form type test statistics and equip them  with a nonparametric Wild bootstrap procedure for calculating critical values to enhance their small sample performance. Here, we throughout assume missing completely at random (MCAR) mechanism, when constructing the test statistics
	and developing their related theories. However, in the simulation studies, we investigate the effects of some missing at random (MAR) scenarios on the test statistics performance under small sample sizes, see the supplement for the explicit definition of the missing mechanisms. \\

The paper is organized as follows: First we introduce the statistical model and the hypotheses of interest {\color{black}in the next section}. In Section \ref{SecStat}, we {\color{black}introduce the test statistics and analyze their asymptotic behaviour.} In Section \ref{SecBootstrap},  {\color{black}the proposed wild bootstrap technique is explained.} Section \ref{Secsim} {\color{black}displays the results from our extensive simulation study and two real} data examples from a fluvoxamine study and a skin disorder clinical trial are analyzed in Section \ref{Secexample}. {\color{black}All proofs as well as additional simulation results can be found in the supplement.}\\

{\color{black} To facilitate the presentation, we introduce the following notations: Let $\vI_d$ denote the d-dimensional unit matrix, $\vJ_d$ denote the $d\times d$ matrix of $\boldsymbol{1}$'s i.e. $\vJ_d=\boldsymbol{1}_d \boldsymbol{1}_d^T$, where $ \boldsymbol{1}_d=(1, ..., 1)^T_{d\times1}$ denotes the 	d-dimensional column vector and $\boldsymbol{P}_d=\vI_d-\frac{1}{d}\vJ_d$ is the so-called d-dimensional centering matrix. Finally, by $\boldsymbol{A}\otimes\boldsymbol{B}$ we denote the Kronecker product of the  matrices $\boldsymbol{A}$ and $\boldsymbol{B}$.}
 
  \section{Statistical model and nonparametric hypotheses} \label{Secmod}
We consider a nonparametric repeated measures model with $a$ independent and potentially unbalanced treatment groups and $d$ different time points {\color{black}given by independent} random vectors
\bqan
 \vX_{ik}=(X_{i1k}, ...,X_{idk})^T,  \quad    i=1,..., a, \quad k=1, ...,n_i,
 \eqan
 where $X_{ijk}\sim F_{ij}(x)=\frac{1}{2}[F^+_{ij}(x)+F^-_{ij}(x)], i=1, ...,a, j=1, ...,d, k=1, ...,n_i$. Here, $F^+_{ij}(x)=P(X_{ijk}\leq x)$ is the right continuous version and $F^-_{ij}(x)=P(X_{ijk}<x)$ is the left continuous version of the distribution function.  Using the normalized version $F_{ij}(x)$ is useful for handling ties and including continuous as well as discontinuous distribution functions \citep{brunner2018rank}. To include the case of missing values, {\color{black}we follow the notation of \cite{brunner1999rank} and let}
\bqan \label{missing}
 \lambda_{ijk}=  \begin{cases}
	 1, \quad \text{if $X_{ijk}$ is observed}	\\
	 	 0, \quad \text{if $X_{ijk}$ is non-observed}
 \end{cases}
 i=1, ...,a, j=1, ...,d, k=1, ...,n_i.
\eqan

Moreover, let $n=\sum_{i=1}^{a}n_i$ denote the total number of subjects and let\\$N=\sum_{i=1}^{a}\sum_{j=1}^{d}\sum_{k=1}^{n_i}\lambda_{ijk}$ denote the total number of observations.\\
 
To formulate the null hypothesis in this nonparametric setup,  let $\boldsymbol{F}=(F_{11}, ...,F_{ad})^T$ denote the vector of the distribution functions $F_{ij}, i=1, ..., a, j=1, ..., d.$ and $\boldsymbol{C}$ denote a contrast matrix , i.e., $\boldsymbol{C1}=\boldsymbol{0}$ where $ \boldsymbol{1}=(1, ..., 1)^T$ and  $\boldsymbol{0}=(0, ..., 0)^T$. 
Then, the null hypotheses are formulated by  $H_0:\{\vC\boldsymbol{F}=\boldsymbol{0}\}$. {\color{black}This framework covers different factorial repeated measures designs. 
	For example, the hypothesis of no treatment group effect, i.e. $H_0^G:\{\bar{F}_{1.}=...=\bar{F}_{a.}\}$ , is equivalently written in matrix notation as $H_0^G:\{(\boldsymbol{P}_a\otimes\frac{1}{d}\boldsymbol{1}_d^T)\vF=\boldsymbol{0}\}$. Similarly, the hypothesis of no time effect, i.e. $H_0^T:\{\bar{F}_{.1}=...=\bar{F}_{.d}\}$ is equivalently written as $H_0^T:\{(\frac{1}{a}\boldsymbol{1}_a^T \otimes \boldsymbol{P}_d)\vF=\boldsymbol{0}\}$, and the hypothesis of no interaction effect {\color{black} between treatment and time is written as}  $H_0^{GT}:\{(\boldsymbol{P}_a\otimes \boldsymbol{P}_d)\vF=\boldsymbol{0}\}$.}\\

 To entail a parameter for describing differences between distributions, \cite{brunner1999rank} and \cite{domhof2002rank} considered the relative marginal effects
\bqan 
 p_{ij}=\int{H(x)dF_{ij}(x)},
 \eqan
  where $H(x)=N^{-1}\sum_{i=1}^{a}\sum_{j=1}^{d} \sum_{k=1}^{n_i}\lambda_{ijk}F_{ij}(x)$ is the weighted average of all distribution functions in the experiment. Estimators thereof are given by plugging-in the empirical versions of $F_{ij}(x)$ and $H(x)$ 
\begin{align}
\hat{F}_{ij}(x)&=\frac{1}{\lambda_{ij.}}\sum_{k=1}^{n_i}\lambda_{ijk}\hat{F}_{ijk}(x)\\
               &=\frac{1}{\lambda_{ij.}}\sum_{k=1}^{n_i}\lambda_{ijk}c(x-X_{ijk}), \quad \lambda_{ij.}=\sum_{k=1}^{n_i}\lambda_{ijk},\\
               \hat{H}(x)&=\frac{1}{N}\sum_{i=1}^{a}\sum_{j=1}^{d}\sum_{k=1}^{n_i}c(x-X_{ijk}),
\end{align}

where, $c(u)$ is the normalized version of the counting function, i.e. $c(u)=0, 1/2$ or $1$ according as $u<0, u=0$ or $u> 0$ and  $c(x-X_{ijk})$ is assumed to equal $0$ if the observation $X_{ijk}$ is missing. Note that $\hat{F}_{ij}(x)=0$ in case of $\lambda_{ij.}=0$. Thus, the relative marginal effects $p_{ij}$ are estimated by 
\begin{align}
\hat{p}_{ij}&=\int\hat{H}d\hat{F}_{ij}=\frac{1}{\lambda_{ij.}}\sum_{k=1}^{n_i}\lambda_{ijk}\hat{H}(X_{ijk})\\
&=\frac{1}{\lambda_{ij.}}\sum_{k=1}^{n_i}\lambda_{ijk}\frac{1}{N}(R_{ijk}-\frac{1}{2}),
\end{align}

where $R_{ijk}$ is the mid-rank of $X_{ijk}$ among all $N$ dependent and independent observations.  For {\color{black}convenience}, the relative marginal effect estimators are collected in the vector $\hat{\vp}=(\hat{p}_{11}, ...,\hat{p}_{ad})^T$.\\

In the sequel, we derive  asymptotic theory under the following sample size assumption and missing values:

\begin{assumption}{1}{}\label{Assump1} $\frac{\lambda_{ij.}}{n} \rightarrow \kappa_{i} \in (0,1) \quad i=1,..., a, \quad  j=1,..., d$, as $n_0\equiv min \{\lambda_{ij.}\} \rightarrow \infty$.

\end{assumption}


This ensures that {\color{black}there are 'enough' subjects without missing values in each group.}\\

\textbf{Asymptotic distribution.} \cite{brunner1999rank} derived the asymptotic distribution of $\sqrt{n}\vC\hat{\vp}$ under the null hypothesis $H_0:\vC \vF= \boldsymbol{0}$. Setting $\hat{\vY}_{ijk}=\hat{H}(X_{ijk})$ and $\vY_{ijk}=H(X_{ijk})$, they proved that under $H_0$, $\sqrt{n}\vC\hat{\vp}$ and $\sqrt{n}\vC\bar{\vY}_.$ are asymptotically equivalent. Accordingly, they showed that $\sqrt{n}\vC\hat{\vp}$ follows under $H_0$, asymptotically, a multivariate normal distribution with expectation $\boldsymbol{0}$ and covariance matrix $\boldsymbol{C\vV_n C^T}$. Here 
\bqan
\vV_n= \bigoplus\limits_{i=1}^d \frac{n}{n_i} \vV_{i} = \bigoplus\limits_{i=1}^d \frac{n}{n_i^2} \sum_{k=1}^{n_i} \boldsymbol{\Lambda}_{ik} \vV_{ik}\boldsymbol{\Lambda}_{ik},
\eqan
where $\boldsymbol{\Lambda}_{ik}=n_idiag\{\frac{\lambda_{i1k}}{\lambda_{i1.}},...,\frac{\lambda_{idk}}{\lambda_{id.}}\}$, and $\vV_{ik}=Cov(\vY_{ik})$ {\color{black}denotes the covariance matrix} of  the random vectors $\vY_{ik}=(H(X_{i1k}),...,H(X_{idk}))^T$.\\


{\color{black}\cite{brunner1999rank} propose to estimate }the covariance matrix $\vV_n$  by
\bqan
\vVh_n=\bigoplus\limits_{i=1}^r\frac{n}{n_i}\vVh_{i},
\eqan
where $\vVh_{i}=[\hat{v}_i(j,j')]$ with diagonal and off-diagonal elements $\hat{v}_i(j,j)$ and $\hat{v}_i(j,j')$, respectively, defined as

\bqan\label{Vhat1}
\hat{v}_i(j,j)=\frac{n_i\sum_{k=1}^{n_i}\lambda_{ijk}[R_{ijk}-\bar{R}_{ij.}]^2}{(N^2)\lambda_{ij.}(\lambda_{ij.}-1)},
\eqan

\bqan\label{Vhat2}
\hat{v}_i(j,j')=\frac{n_i \sum_{k=1}^{n_i}\lambda_{ijk}\lambda_{ij'k}[(R_{ijk}-\bar{R}_{ij.})(R_{ij'k}-\bar{R}_{ij'.})]}{(N^2)((\lambda_{ij.}-1)(\lambda_{ij'.}-1)+\Delta_{i,jj'}-1)}	.
\eqan

Here, $\bar{R}_{ij.}=\frac{1}{\lambda_{ij.}}\sum_{k=1}^{n_i}\lambda_{ijk}(R_{ijk})$, and $\Delta_{i,jj'}=\sum_{k=1}^{n_i}\lambda_{ijk}\lambda_{ij'k}$. {\color{black}Under Assumption 1,  \cite{brunner1999rank} have shown that $\vVh_{i}$ is a consistent estimator of $\vV_i$.}

\section{Statistics and Asymptotics}
\label{SecStat}
In this section, we propose three different quadratic forms for testing the null hypothesis: a Wald-type statistic (WTS), an ANOVA-type statistic (ATS) as suggested in \cite{brunner1999rank} and \cite{domhof2002rank} as well as a modified ANOVA-type statistic (MATS). {\color{black}For mean-based analyses, the latter was proposed by \cite{friedrich2018mats}.}\\

The rank version of the WTS  is defined as
\bqan
T_W=n\vph^T\vC^T[\vC\vVh_n\vC^T]^+\vC\vph,
\eqan
where $[\boldsymbol{B} ]^+$ denotes the Moore-penrose inverse of a matrix $\boldsymbol{B}$. Its asymptotic null distribution is summarized below.

\begin{thm}\label{theorem 1}
Under Assumption $(1)$ and  $\vV_n \rightarrow \vV >0$ as $n_0 \rightarrow \infty$, the statistic $T_W$  has under the null hypothesis $H_0:\vC\vF=\boldsymbol{0}$, asymptotically, as $n_0 \rightarrow \infty$, a central $\chi^2_f$-distribution with $f=rank(\vC)$ degrees of freedom.

\end{thm}
{\color{black}Wald-type statistics of similar form are used in many diferent situations,  e.g., in heteroscedatic mean-based analyses} \citep{krishnamoorthy2010parametric, xu2013parametric, konietschke2015parametric, friedrich2018mats,amro2019asymptotic} {\color{black} and even more complex regression models in survival analyses \citep{martinussen2007dynamic, dobler2019confidence}}. However, the convergence of the WTS to its limiting $\chi^2$-distribution is usually slow and large sample sizes are required to obtain adequate results \citep{vallejo2010analysis, konietschke2015parametric, pauly2015asymptotic, smaga2017bootstrap}.  Thus,  \cite{brunner1999rank} and \cite{domhof2002rank} proposed an alternative quadratic form by deleting the variance $\vVh_n$  involved in the computation of the WTS, resulting in the ANOVA-type statistic (ATS) defined as
\bqan
T_A= \frac{1}{tr(\vT\vVh_n)}n\vph^T\vT\vph,
\eqan

where $\vT=\vC^T[\vC\vC^T]^+\vC$ is a projection matrix. Note that $H_0:\vT\vF=\boldsymbol{0}\Leftrightarrow\vC\vF=\boldsymbol{0}$ because $\vC^T[\vC\vC^T]^+$ is a generalized inverse of $\vC$.\\
It is also worth to note, that different to $T_W$, the asymptotic distribution of $T_A$ can also be derived if $\vV$ is singular.

\begin{thm}\label{thmanova}

Under Assumption $(1)$ and under the null hypothesis $H_0:\vC\vF=\boldsymbol{0}$, the test statistic $T_A$  has asymptotically, as $n_0 \rightarrow \infty$, the same distribution as the random variable
	\bqan
	\boldsymbol{A}= \sum_{i=1}^{a}\sum_{j=1}^{d} \zeta_{ij}B_{ij}/ tr(\vT\vV_n),
	\eqan
\end{thm}
where $B_{ij} \overset{\text{i.i.d}}{\sim} \chi^2_1 $ and the weights $\zeta_{ij}$ are the eigenvalues of $\vT\vV_n$.\\

{\color{black} The limiting distribution is approximated by a scaled $g\chi^2_f$ distribution. \cite{brunner1999rank} proposed a \cite{box1954some}-type approximation such that the first two moments of $T_A$ and $g\chi^2_f$ approximately coincide where $f$ can be estimated by $\hat{f}=\frac{(tr(\vT\vVh_n))^2}{tr(\vT\vVh_n\vT\vVh_n)} $. Thus,  the distribution of $T_A$ is approximated by a central $F(\hat{f}, \infty)$-distribution.} The corresponding ANOVA-type test $\phi_A=\mathds{1}\{T_A> F_\alpha(\hat{f}, \infty) \}$, where $ F_\alpha(\hat{f}, \infty)$ denotes the $(1-\alpha)$-quantile of the $F(\hat{f}, \infty)$-distribution.\\

 Despite its advantage of being applicable in case of singular covariance matrices, the ATS has the drawback of being an approximative test and thus, even its asymptotic exactness cannot be guaranteed.\\
 
 Another possible test statistic is the modified version of the ANOVA type-statistic (MATS) that was developed by \cite{friedrich2018mats} for {\color{black}mean-based} MANOVA models. Here, we transfer it to the ranked-based set-up, where it is given by 
 \bqan
T_M=n\vph^T\vC^T[\vC\hat{\boldsymbol{D}}_n\vC^T]^+\vC\vph,
\eqan
 
 where $\hat{\boldsymbol{D}}_n=diag(\frac{n}{n_i}\hat{v}_i(j,j)), i=1,..., a, j=1,..., d$. In this way it is a compromise between the WTS and the ATS as it only uses the diagonal entries of $\hat{\vV}_n$ for the multivariate studentization. In fact, {\color{black} the  nonsingularity assumption of $\vV$} is not needed to derive its asymptotics.  {\color{black}It is replaced by the weaker requirement that $\mathbf{D} = {\color{black} diag(\frac{n}{n_i}v_i(j,j))}>0, {\color{black}i=1,..., a, j=1,..., d}$, which is fulfilled  when all {\color{black}the} diagonal elements {\color{black}$v_i(j,j)$ of $\vV_{i}$} are positive. This is the same as $Var(H(X_{ij1}))>0$ for all $ i \in \{1, ..., a\}, j\in\{1, ..., d\}$. It is supposed to be met in a  wide range of  applicable settings, except only cases where, for example, at least one time point of any of the treatment groups is a discrete variable with very few distinct values.}

\begin{thm}
	\label{themmats}
	Under Assumption $(1)$ and {\color{black} assuming that $v_i(j,j)>0$  for all $ i \in \{1, ..., a\}, j\in\{1, ..., d\}$, the test statistic $T_M$, has under the null} hypothesis $H_0:\vC\vF=\boldsymbol{0}$, asymptotically, as $n_0\to \infty$, the same distribution as the random variable
	\begin{equation*}
	\vM= \sum_{i=1}^{a}\sum_{j=1}^{d} \tilde{\zeta}_{ij}\tilde{B}_{ij} ,
	\end{equation*}
	where $\tilde{B}_{ij} \overset{\text{i.i.d}}{\sim} \chi_1^2 $ and the weights $\tilde{\zeta}_{ij}$ are the eigenvalues of $\vV_n^{1/2}\vC^T[\vC\boldsymbol{D}\vC^T]^+ \mathbf{C} \vV_n^{1/2}$ and $\mathbf{D} = diag({\color{black}\frac{n}{n_i}v_i(j,j))}$.\\
\end{thm} 
 
 Since, the limit distributions of both, the ATS and the MATS depend on unknown weights $\zeta_i$ and $\tilde{\zeta}_i$, we cannot directly calculate critical values. In addition, the $\chi^2_f-$approximation to $T_W$ is rather slow. To this end, we  develop  asymptotically correct testing procedures based on bootstrap versions of $T_W, T_A,$ and  $T_M$ in the subsequent section.

\section{Wild bootstrap approach} 
\label{SecBootstrap}
We consider a wild bootstrap approach to derive new {\color{black}asymptotically valid} testing procedures with good finite sample properties. 
 To this end, let $\vZ_{ik}=(\vR_{ik}-\bar{\vR}_{i.})$
 denote the centered rank vectors,  where $\vR_{ik}=(R_{i1k},..., R_{idk})^T$, $\bar{\vR}_{i.}=(\bar{R}_{i1.},..., \bar{R}_{id.})^T$ and $\bar{R}_{ij.}=\frac{1}{\lambda_{ij.}}\sum_{k=1}^{n_i}\lambda_{ijk}(R_{ijk})$. Moreover, let $W_{ik}$ denote independent and identically distributed random weights with $E(W_{ik})=0$ and $Var(W_{ik})=1$. Although, there are different possible choices for these random weights \citep{mammen1992bootstrap,davidson2008wild}, some particular choices have become popular. {\color{black}Following the investigations in \cite{friedrich2017wild} for the corresponding complete case scenario, we use Rademacher random variables, which are defined by $ P(W_{ik}=-1) = P(W_{ik}=1)=1/2$}. Then, a wild bootstrap sample is defined as
   \bqan \vZ_{ik}^*=W_{ik}.\vZ_{ik}, \quad  i=1,..., a, \quad k=1,..., n_i.
   \eqan  
  
  {\color{black}In other words,} $\vZ_{ik}^*$ is a symmetrization of the rank vector $\vZ_{ik}$. 
Now, we can define the bootstrap version of the relative effect estimator $\hat{p}_{ij}$ 
 \bqan
 \hat{p}_{ij}^{*}=\frac{1}{\lambda_{ij.}}\sum_{k=1}^{n_i}\lambda_{ijk}\frac{1}{N}(Z_{ijk}^*)=\frac{1}{\lambda_{ij.}}\sum_{k=1}^{n_i}\lambda_{ijk}\frac{1}{N}(W_{ik} (R_{ijk}-\bar{R}_{ij.})).
 \eqan
 
{\color{black}We pool them into the vector  $\hat{\vp}^*=(\hat{p}_{11}^*, ...,\hat{p}_{ad}^*)^T$ to obtain the bootstrap counterpart of $\hat{\vp}$.}\\
In the same way, the bootstrap covariance matrix estimator $\vVh_{i}^*=[\hat{v}_i^*(j,j')]$ with diagonal and off-diagonal elements $\hat{v}_i^*(j,j)$ and $\hat{v}_i^*(j,j')$, respectively, is given by

\bqan
\hat{v}_i^*(j,j)=\frac{n_i\sum_{k=1}^{n_i}\lambda_{ijk}[Z^*_{ijk}-\bar{Z}^*_{ij.}]^2}{(N^2)\lambda_{ij.}(\lambda_{ij.}-1)},
\eqan

\bqan
\hat{v}_i^*(j,j')=\frac{n_i\sum_{k=1}^{n_i}\lambda_{ijk}\lambda_{ij'k}[(Z^*_{ijk}-\bar{Z}^*_{ij.})(Z^*_{ij'k}-\bar{Z}^*_{ij'.})]}{(N^2)((\lambda_{ij.}-1)(\lambda_{ij'.}-1)+\Delta_{i,jj'}-1)}.	
\eqan

From this, the bootstrapped versions of the quadratic forms, i.e. the Wald-type statistic $T_W^*$, the ANOVA-type statistic $T_A^*$  and the modified ANOVA-type statistic $T_M^*$ are computed as

\begin{align}
T_W^*=n\vph^{*T}\vC^T[\vC\vVh_n^*\vC^T]^+\vC\vph^*,
\label{PBWald}\\
T_A^*=\frac{1}{tr(\vT\vVh^*_n)}n\vph^{*T}\vT\vph^*,
\label{ANOVABoot} \\
T_M^*=n\vph^{*T}\vC^T[\vC\hat{\boldsymbol{D}}^*_n\vC^T]^+\vC\vph^*, \label{MATSBoot}
\end{align} 

where $\vVh_n^*=\bigoplus\limits_{i=1}^a\frac{n}{n_i}\vVh_{i}^*$ and $\hat{\boldsymbol{D}}^*_n=diag({\color{black}\frac{n}{n_i}\hat{v}_i^*(j,j)), i=1,..., a, j=1,..., d}$. To get an asymptotically valid bootstrap test, we have to assure that the conditional distribution of the Wald,  ANOVA-type, and MATS-type bootstrap statistics $T_W^*$, $T_A^*$, and $T_M^*$ approximate the null distribution of  $T_W$, $T_A$, and $T_M$, respectively. 
 
\begin{thm}\label{theorem 3}
Under Assumption $(1)$, the following results hold:
\begin{enumerate}
	\item For $i=1,..., a$, the conditional distribution of $\sqrt{n}\vph^*_i$, given the data  $\vX$, converges weakly to the multivariate $N(0,\kappa_i^{-1}\vV_i)$-distribution in probability.
	\item The conditional distribution of $\sqrt{n}\vph^*$, given the data $\vX$, converges weakly to the  multivariate $N(0,\bigoplus\limits_{i=1}^a\kappa_i^{-1}\vV_i)$-distribution in probability. 
\end{enumerate}

\end{thm}
Thus, the distributions of $\sqrt{n}\vC\vph^*$ and $\sqrt{n}\vC(\vph-\vp)$ asymptotically coincide under the null hypothesis $H_0$. 

\begin{thm}\label{ThmPBWald}
Under  Assumption $(1)$, for any choice $[-] \in \{A,M,W\}$, the conditional distribution of $T_{[-]}^*$ converges weakly to the null distribution of $T_{[-]}$ in probability for any choice of $\boldsymbol{p}\in\R^{ad}$. In particular we have that
	\begin{align*}
 	\sup_{x\in\R}|P_{\boldsymbol{p}}(T_{[-]}^*\leq x|\vX)-P_{H_0}(T_{[-]}\leq x)| \xrightarrow{\text{p}} 0
 \end{align*}
 {\color{black} holds, provided that  $\vV>0$ or $v_i(j,j)>0$  for all $ i \in \{1, ..., a\}, j\in\{1, ..., d\}$ is fulfilled in case of $T_W$ or $T_M$, respectively}.  
\end{thm}

Therefore, the corresponding wild bootstrap tests are given by $\phi^*_W=\mathds{1}\{T_{W} > c^*_w\}$, $\phi^*_A=\mathds{1}\{T_{A} > c^*_A\}$, and $\phi^*_M=\mathds{1}\{T_{M} > c^*_M\}$, where $c^*_w$, $c^*_A$, and $c^*_M$ denoting the $(1-\alpha)$ quantiles of the conditional bootstrap distributions of $T_W^*$, $T_A^*$, and $T_M^*$  given the data, respectively.\\

Theorem \ref{ThmPBWald} implies that the wild bootstrap tests are of asymptotic level $\alpha$ under the null hypothesis and consistent for any fixed alternative $H_1:\vC\vF\neq\boldsymbol{0}$, i.e., they have asymptotically power $1$. In addition, it follows from \cite{jansen2003local} that they have the same local power under contiguous alternatives as their original tests. \\

\begin{figure}[h!]
	\begin{center}
		\includegraphics[scale=0.6]{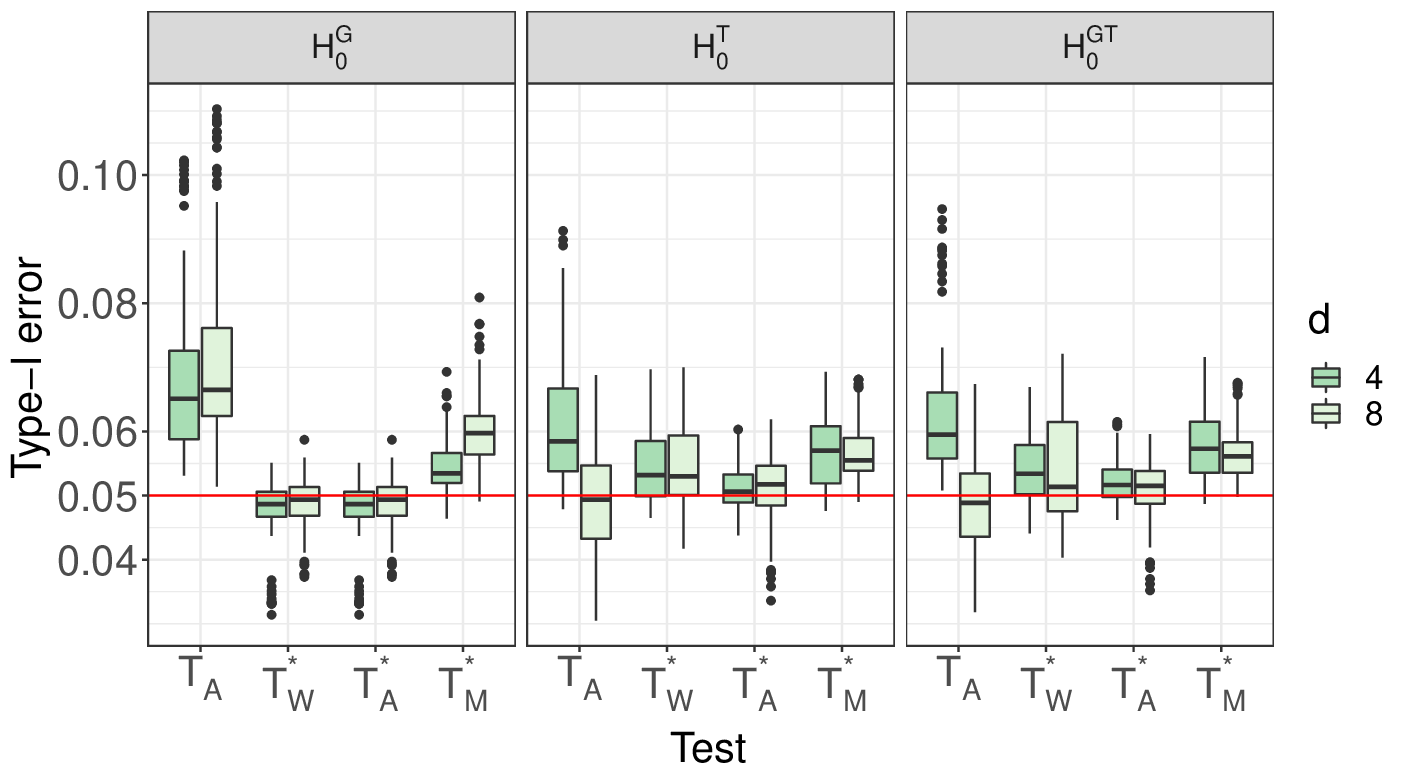}	
	\end{center}
	\caption{Type-I error of the asymptotic ANOVA-type test $T_A$ and the bootstrapped tests $T_W^*$, $T_A^*$, and $T_M^*$ based on several hypotheses of interest for varying time points $d\in\{4,8\}$. Each boxplot {\color{black}summarizes the type-I error results from 96 different simulation scenarios for this hypothesis}. For all individual simulations, see the tables in the supplement.}
	\label{fig:BoxplotHyp}	
\end{figure}

\begin{figure}[h!]
	\begin{center}
		\includegraphics[scale=0.6]{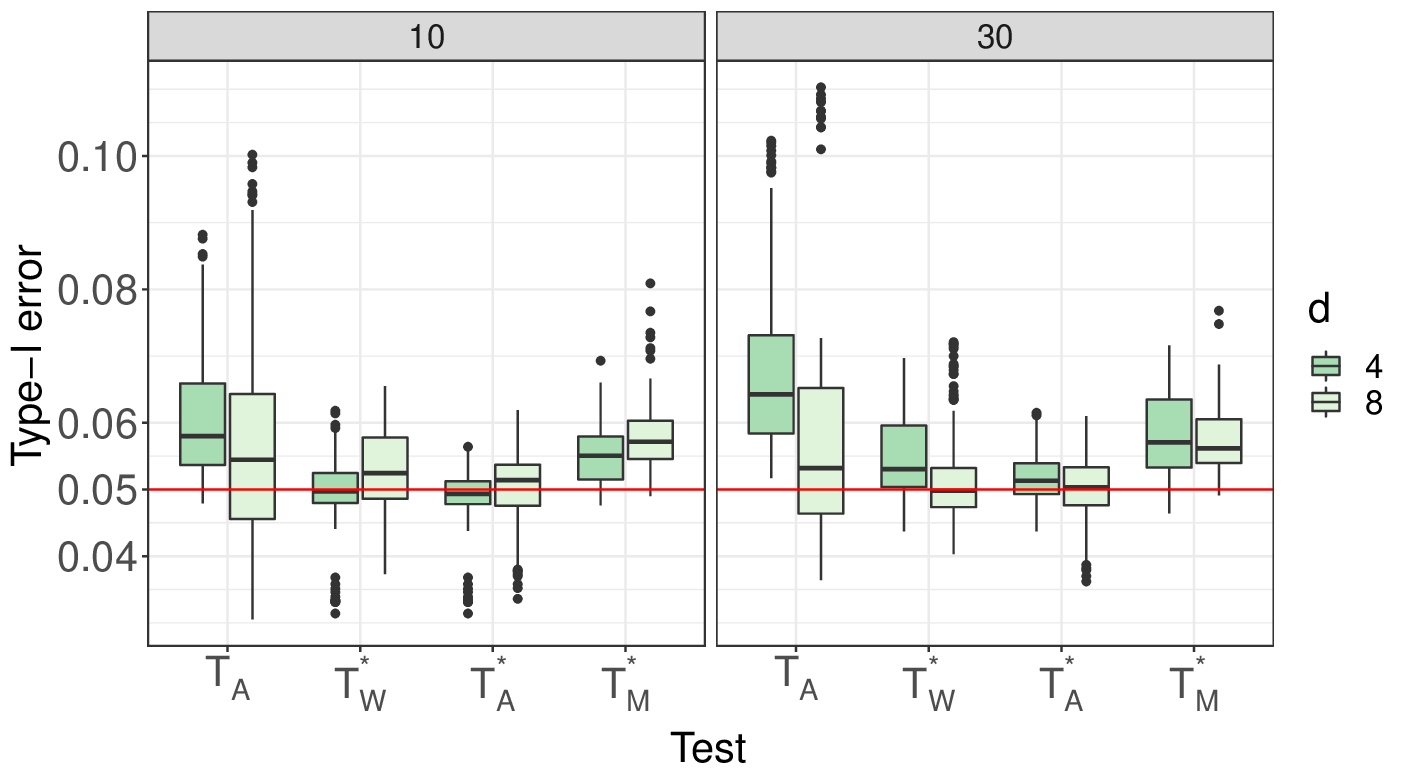}	
	\end{center}
	\caption{Type-I error of the asymptotic ANOVA-type test $T_A$ and the bootstrapped tests $T_W^*$, $T_A^*$, and $T_M^*$ based on missing rates $r\in\{10,30\}$ for varying time points $d\in\{4,8\}$. Each boxplot summarizes the type-I error results from 144 different simulation scenarios for this missing rate. For all individual simulations, see the tables in the supplement.}
	\label{fig:BoxplotMiss}	
\end{figure}

\begin{figure}[h!]
	\centering
	\hspace*{-1.2in}
	\subfloat[Normal]{{\includegraphics[scale=0.45, clip=true,]{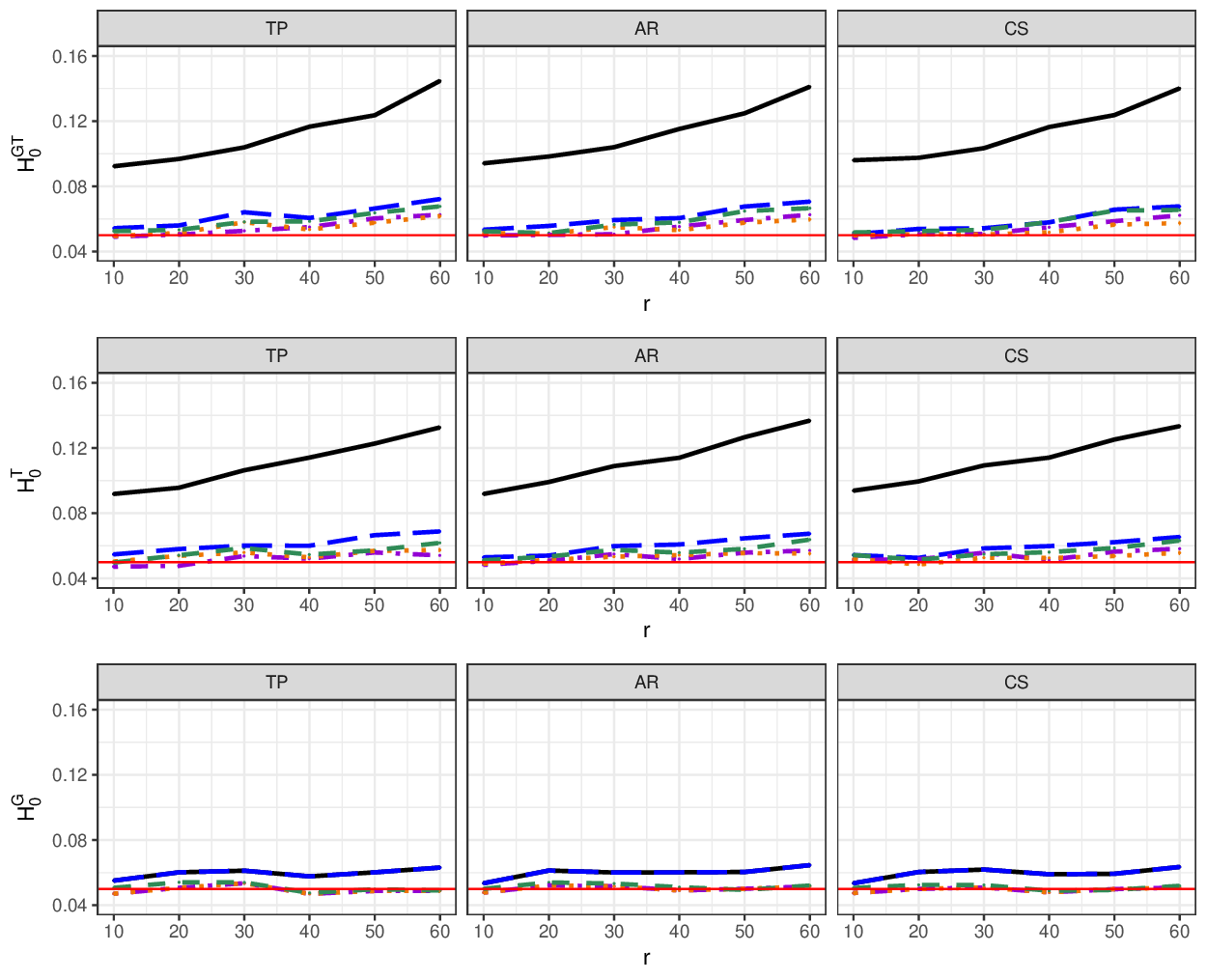} }}%
\Hquad
	\hspace*{-0.1in}
	\subfloat[Chisquare]{{\includegraphics[scale=0.45, clip=true,]{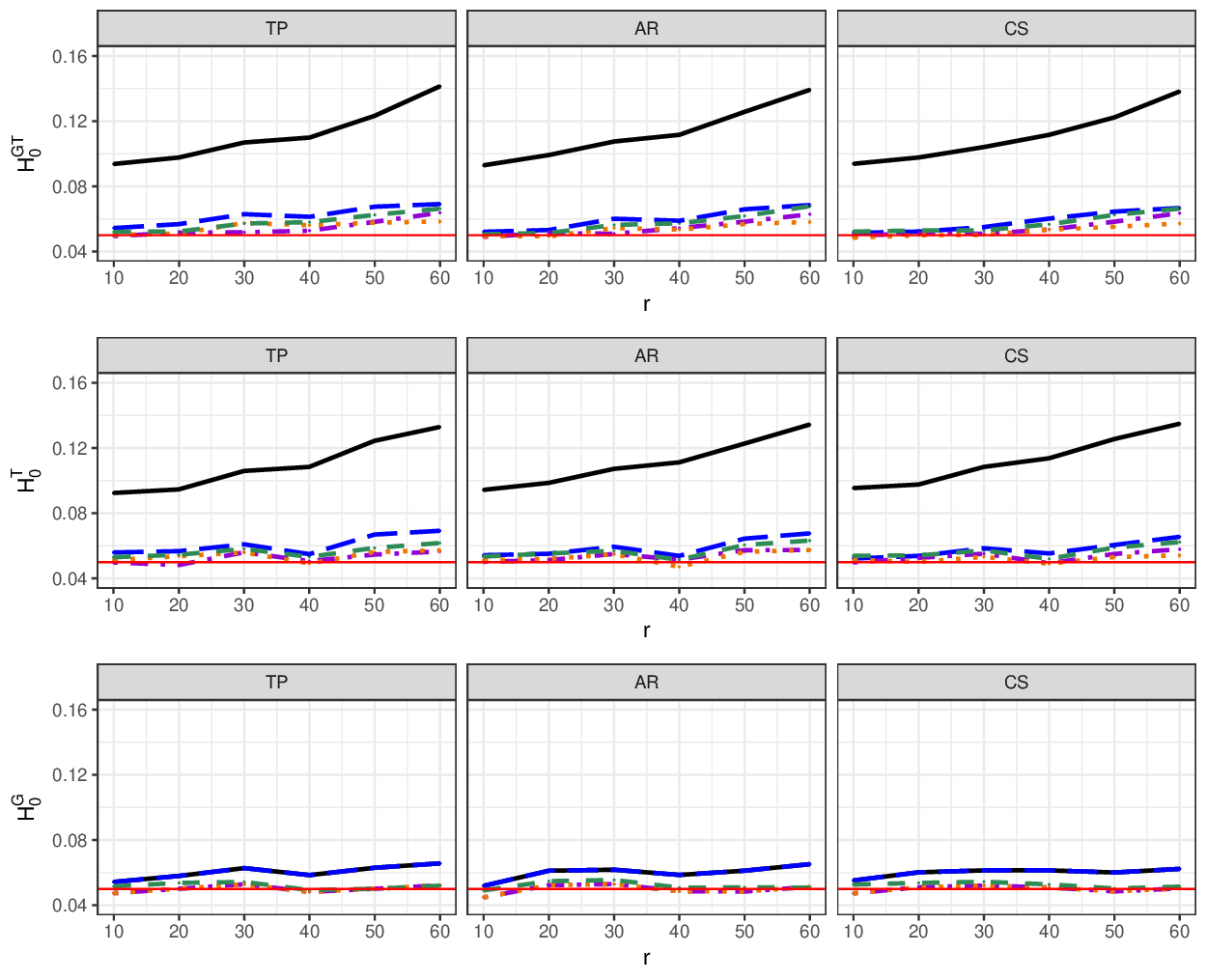} }}%
		\hspace*{-1.1in}
	\caption{	Type-I error simulation results ($\alpha=0.05$) of the tests $T_W$ $({\color{black}\textbf{\textendash\textendash\textendash}})$, $T_A$ $({\color{blue} \textbf{ \textendash\textendash \quad \textendash\textendash}})$, 	$T_W^*$ $({\color{violet}\textbf{--}\boldsymbol{\cdot}\textbf{--}})$, $T_{A}^*$ $({\color{brown}\boldsymbol{\cdots}})$, and $T_{M}^*$ $({\color{ao(english)} \textbf{-- - --} })$ under different covariance structures with sample sizes $(n_1,n_2)=(15,15)$  and $d=4$ for varying percentages of MCAR data  $r\in\{10\%,20\%,30\%,40\%,50\%,60\%\}$ with observations generated from (a) a Normal  and (b) a $\chi^2_{15}$  distribution, respectively.}%
	\label{fig:NHypIncmiss}%
\end{figure}

\section{Monte-Carlo simulations} \label{Secsim}
The above results are valid for large sample sizes. To analyze the finite sample behavior of the  asymptotic quadratic tests and their wild bootstrap counterparts described in Sections \ref{SecStat}  and \ref{SecBootstrap}, we  conduct extensive simulations. As an assessment criteria, all procedures were studied with respect to their 
\begin{enumerate}[(i)]
	\item type-I error rate control at level $\alpha = 5\%$ and their
	\item  power to detect deviations from the null hypothesis. 
\end{enumerate}

All simulations were operated by means of the R computing environment, version 3.2.3  \cite{Rlanguage} and each setting was based on $10,000$ simulation runs and $B=999$ bootstrap runs. The algorithm for the computation of the $p$-value of the wild bootstrap tests {\color{black}in any of the test statistics $T\in\{T_A,T_W, T_M \}$ is} as follows:
\begin{enumerate}
	\item For the given incomplete multi-factor data, calculate the observed test statistic, say $T$. 
	\item Compute the rank values $\vZ_{ik}$.
	\item Using i.i.d Rademacher weights ${W}_{i1}, ..., {W}_{in_i}$, generate  bootstrapped rank values $\vZ_{ik}^*=W_{ik}.\vZ_{ik}$.
	\item Calculate the value of the test statistic for the bootstrapped sample $T^*$. 
	
	\item Repeat the Steps $3$ and $4$ independently $B=999$ times and collect the observed test statistic values in ${T^*_{b}, b=1,.....,B}$.
	\item  Finally, estimate the wild bootstrap $p$-value as   $P-value=\frac{\sum_{b=1}^{B}I(T^*_{b}>= T)}{B}$.
\end{enumerate}

We considered a two-way layout design with $a=2$ independent groups and two different time points $d \in \{4,8\}$ underlying discrete and continuous distributions. 
{\color{black} As in \cite{konietschke2015parametric} and \cite{bathke2018testing} for the case of complete observations,} we investigated balanced situations with sample size vectors $(n_1,n_2)\in\{(5,5),(10,10), (20,20)\}$, and an unbalanced situation with sample size vector $(n_1,n_2)= (10,20)$. In particular, we studied three different kinds of hypotheses: the hypothesis of no group effect $H_0^G:(\boldsymbol{P}_a\otimes\frac{1}{d}\boldsymbol{1}_d^T)$, no time effect $H_0^T:(\frac{1}{a}\boldsymbol{1}_a^T \otimes \boldsymbol{P}_d)$, as well as no interaction effect $H_0^{GT}:(\boldsymbol{P}_a\otimes \boldsymbol{P}_d)$.\\

For each scenario, we generated missingness  within MCAR as well as MAR frameworks as described below: For the MCAR mechanism,  we simulated $\vX_{ik}=({\color{black}\delta_{i1k}}X_{i1k},..., \delta_{idk}X_{idk}), \quad k=1,...,n_i$,  for  independent Bernoulli distributed $\delta_{i1k}\sim B(r)$ and a zero entry was interpreted as a missing observation. The missing probability was chosen from $r\in\{10\%, 30\%\}$.\\

{\color{black}
For the {\it MAR mechanism}, we considered several pairs of features $\{X_{obs},X_{miss}\}$. For each pair, there is a determining feature $X_{obs}$ that determines the missing pattern of its corresponding $X_{miss}$ \citep{santos2019generating}. Thus, for $d=4$, we define the pairs $\{X_{i1k}, X_{i2k}\}$ and $\{X_{i3k}, X_{i4k}\}$. Whereas, for $d=8$, we define the following pairs $\{X_{i1k}, X_{i2k}\}, \{X_{i1k}, X_{i3k}\}, \{X_{i6k}, X_{i7k}\},$ and $\{X_{i6k}, X_{i8k}\} $. Two different MAR scenarios are investigated - MAR1 and MAR2 -. In MAR1, for each pair, we divided $\mathbf{X}$ into three groups based on their $X_{obs}$ values: 
the first group is given by $\{\vX_{ik}: X_{obs}\in (-\infty, -2\hat{\sigma}_1), k=1,.., n_i\}$, the second by $\{\vX_{ik}: X_{obs}\in (-2\hat{\sigma}_1,2\hat{\sigma}_1), k=1,.., n_i\}$ and the last group by $\{\vX_{ik}: X_{obs}\in  (2\hat{\sigma}_1,\infty), k=1,.., n_i\}$, where $\hat{\sigma}_1$ is the estimated standard deviation of $X_{obs}$. Then, we randomly inserted missing values on $X_{miss}$   based on the following missing percentages: $15\%$ for group one and three and $30\%$ for the second group. \\

For generating MAR2 scenario, we considered the median of each $X_{obs}$ to
define the missing pattern of $X_{miss}$ \citep{zhu2012robust, pan2015missing}. For each pair from above, two groups
were defined: the first one is $\{\vX_{ik}: X_{obs}\in (-\infty, median(X_{obs})], k=1,.., n_i\}$, and the second is $\{\vX_{ik}: X_{obs}\in (median(X_{obs}),\infty), k=1,.., n_i\}$. Then, missing values were inserted on $X_{miss}$ as follows: $10\%$ for group one and $30\%$ for the second group. }\\

\subsection{Continuous data}
To investigate the type-I error control of the suggested methods, data samples were generated from the model

\begin{center}
	$\vX_{ik}\sim\vF_i(\vmu_0,\vSigma_i),\quad i=1,2, \quad k=1,... ,n_i ,$
	
\end{center}

where, $\vF_i(\vmu_0,\vSigma_i)$ represents a multivariate distribution with expectation vector $\vmu_0$ and covariance matrix $\vSigma_i$. Marginal data distributions were generated from different symmetric distributions (normal, double exponential) as well as skewed distributions (lognormal, $\chi^2_{15}$) {\color{black}(similar to \cite{pauly2015asymptotic})}. We used normal copulas to generate the dependency structures of the repeated measurements using the R package {\bf copula} \citep{yan2007enjoy}. For the covariance matrix $\boldsymbol{\Sigma}_i$, we investigated the three following covariance structures 
\begin{description}
	
	\item [(AR) Setting 1:] 	 $\vSigma_i= (\rho^{|l-j|})_{l,j\leq d},  \rho=0.6$ for $i=1,2$,
	
	\item	[(CS) Setting 2:]  	$\vSigma_i = \vI_t$, for $i=1,2,$
	
	\item [(TP) Setting 3:] $\vSigma_i = (d-|l-j|)_{l,j\leq d},$ for $i=1,2$,
\end{description}

representing an autoregressive structure (Setting $1$), compound symmetry pattern (Setting $2$) and a linear Toeplitz covariance structure (Setting $3$). {\color{black}These covariance settings  were inspired by the simulations studies in \cite{konietschke2015parametric,friedrich2017wild} and \cite{umlauft2017wild}.} \\

{\bf Type-I Error Results.}
The type-I error simulation results of the studied procedures for testing the hypotheses of no time effect $H_0^T$, no group $\times$ time interaction $H_0^{GT}$, and no group effect $H_0^{G}$ {\color{black} under the MCAR and MAR frameworks} are shown in Tables S.1 - S.12 (MCAR framework) and Tables S.13 - S.24 (MAR framework) in the supplement. 
It can be readily seen that the suggested bootstrap approaches based upon $T_W^*, T_A^*$ and $T_M^*$ tend to result in quite accurate type-I error rate control {\color{black}for most hypotheses} under symmetric as well as skewed distributions and under MCAR and MAR mechanisms. The type-I error control is surprisingly not affected by less stringent missing mechanisms and the bootstrap tests are robust under fairly large amounts of missing observations. In addition, the data dependency structures do not affect the quality of the approximations. Moreover, there was no significant impact of the sample sizes balanced or unbalancedness in the control of type-I error. On the other hand, the asymptotic ANOVA-type test $T_A$ also shows {\color{black}a quite accurate type-I error control for large sample sizes.} However, under small sample sizes, $T_A$  tends to be sensitive to the missing rates. In particular, it exhibits a liberal behavior for larger missing rates. In contrast, the asymptotic Wald test $T_W$ shows an extremely liberal behavior in all considered situations and under all investigated missing mechanisms. A closer look at the type-I error simulation results for all considered settings under the MCAR framework is provided. Compact boxplots factorized in terms of hypotheses of interest or missing rates are given in Figure \ref{fig:BoxplotHyp} and \ref{fig:BoxplotMiss}, respectively. {\color{black} They are based upon different sample size settings (4) x different distributions (4) x different missing rates (2) x different covariance structures (3) x different time points (2) x different hypotheses (3). }Due to the extreme liberal behavior of the asymptotic Wald test, it has been removed from those plots in order to get a  better view of the remaining tests behavior. It can be clearly seen that the asymptotic ANOVA-type test  has the less accurate control among all other considered tests. The hypothesis of interest and number of time points obviously impact the quality of the approximations of the asymptotic ANOVA-type test. Furthermore, the type-I error control behavior of the bootstrapped MATS depends on the hypothesis of interest. Consequently, the bootstrapped Wald and ANOVA-type tests are recommended over all considered methods.\\  

In order to cover the {\it effect of increasing missing rates}, we additionally studied type-I error control for $a=2$ groups, $d=4$ time points, $(n_1=15,n_2=15)$ sample sizes with  $r\in\{10\%,20\%,30\%,40\%,50\%,60\%\}$ covering missingness in observations ranging from $10\%$ to $60\%$. Figure \ref{fig:NHypIncmiss} and Figures S.1 - S.2 in the supplement summarize type-I error rate control for these settings under symmetric and asymmetric distributions.  The results indicate that the asymptotic Wald test $T_{W}$ tends to be liberal in all considered situations. In particular, It is extremely liberal when testing the hypotheses of no time effect ($H_0^T$) and no interaction effect ($H_0^{GT}$). In contrast, the asymptotic ANOVA-type test $T_A$ tends to be sensitive to missing rates and hypothesis type. In particular, it exhibits  an accurate or liberal behavior for small or large missing rates, respectively. Moreover, it shows a quite constant liberal behavior when testing the hypothesis of no group effect ($H_0^G$). Its behaviour appears to be independent of the covariance pattern. In contrast, the suggested bootstrap approaches tend to control type-I error rate more accurate over the range of missing rates $r$ for almost all settings.\\ 

\begin{figure}[h!]
	\begin{center}
		\includegraphics[scale=0.7]{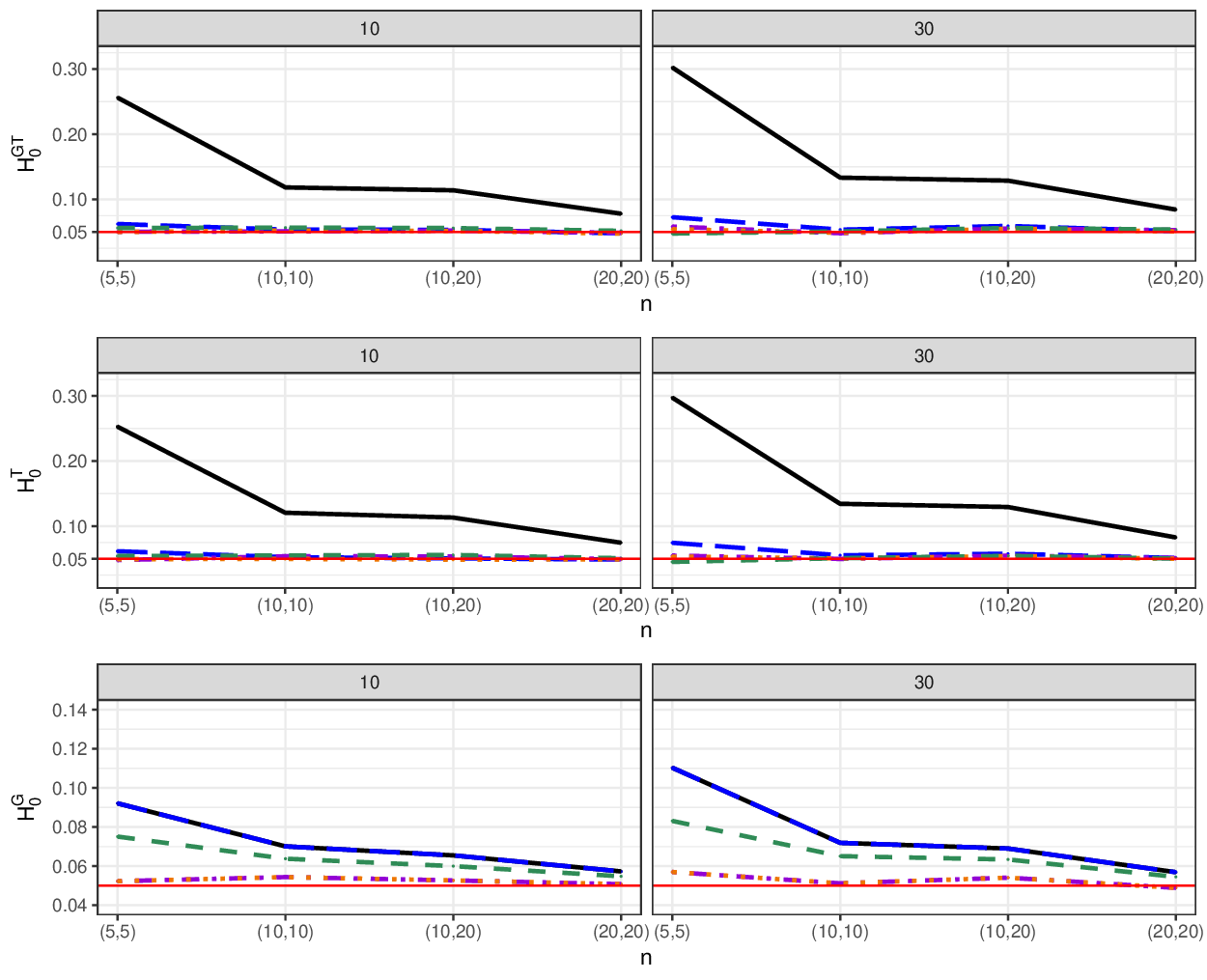}	
	\end{center}

	\caption{	Type-I error simulation results ($\alpha=0.05$) of the tests $T_W$ $({\color{black}\textbf{\textendash\textendash\textendash}})$, $T_A$ $({\color{blue} \textbf{ \textendash\textendash \quad \textendash\textendash}})$, 	$T_W^*$ $({\color{violet}\textbf{--}\boldsymbol{\cdot}\textbf{--}})$, $T_{A}^*$ $({\color{brown}\boldsymbol{\cdots}})$, and $T_{M}^*$ $({\color{ao(english)} \textbf{-- - --} })$ for ordinal data  under MCAR framework with sample sizes $(n_1,n_2)=\{(5,5),(10,10),(10,20),(20,20)\}$  and $d=4$.}
	\label{fig:Ordinalt4}	
\end{figure}

\subsection{Ordinal data}

In order to address all the goals outlined above, we simulated ordinal data. 
The observations were simulated {\color{black}similar to \cite{brunner2000nonparametric}} as follows:
\bqa
\vX_{ijk}=int \big( 4 \frac{cZ_{ik} + Y_{ijk}}{c+1} \big) +1, \quad i=1,..., a, \quad k=1,..., n_i, \quad j=1,..., d,
\eqa

where $Z_{ik}$ and $Y_{ijk}$ are independently uniformly distributed in the interval $[0,1]$, $c>0$ is a constant and $int(x)$ indicates the integer part of $x$.  The elements of $\vX_{ijk}$ take values between $1$ and $4$. The correlation between $\vX_{ij^`k}$ and $\vX_{ijk}$ is determined by the choice of the constant $c$. In our simulation study, we considered $c=1$ assuring a compound symmetric covariance structure. We considered $a=2$ groups and $d\in\{4,8\}$ dimensions, and the same sample sizes as in the continuous data settings.\\

 The type-I error results of the considered methods under MCAR and MAR frameworks are summarized in Figure \ref{fig:Ordinalt4} and Figures S.3 - S.5 in the supplement. It can be seen that the asymptotic Wald test is too liberal in most situations and under all considered hypotheses. The  ANOVA-type test of \cite{brunner1999rank} shows much better behavior. In contrast, the type-I error control for the bootstrap-based procedures is the best, particularly for the bootstrapped Wald and ANOVA-type tests {\color{black}which are also less effected by an increased missing rate or strict MAR assumptions.}
\begin{figure}[h!]
	\begin{center}
		\includegraphics[scale=0.7]{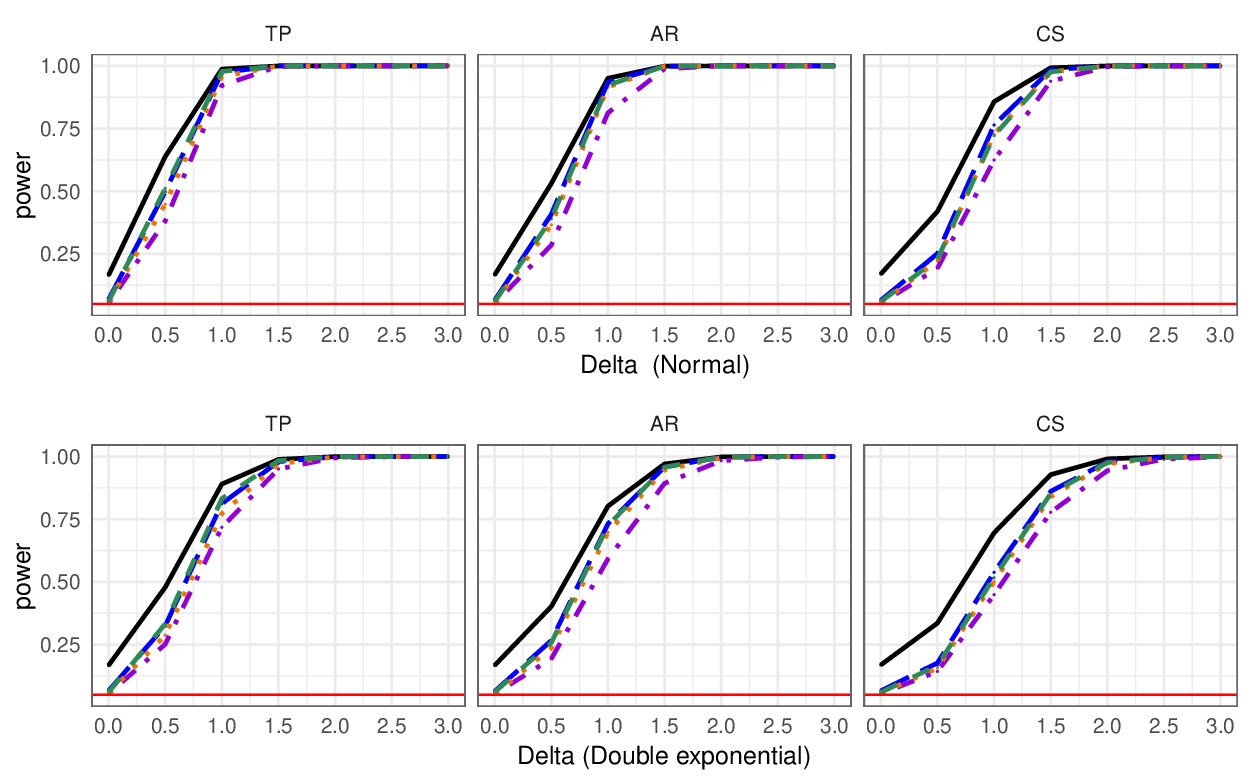}	
	\end{center}
	\caption{Power simulation results of the tests $T_W$ $({\color{black}\textbf{\textendash\textendash\textendash}})$, $T_A$ $({\color{blue} \textbf{ \textendash\textendash \quad \textendash\textendash}})$ , 	$T_W^*$ $({\color{violet}\textbf{--}\boldsymbol{\cdot}\textbf{--}})$, $T_{A}^*$ $({\color{brown}\boldsymbol{\cdots}})$, and $T_{M}^*$ $({\color{ao(english)} \textbf{-- - --} })$ under different covariance structures with sample size $n=15$  and $d=4$ under alternative $1$, under the MCAR framework with missing rate $r=30\%$ with observations generated from a Normal (upper row) and a Double exponential (bottom row) distribution, respectively.}
	\label{fig:PowerNDMiss30HT}	
\end{figure}

\begin{figure}[h!]
	\begin{center}
		\includegraphics[scale=0.7]{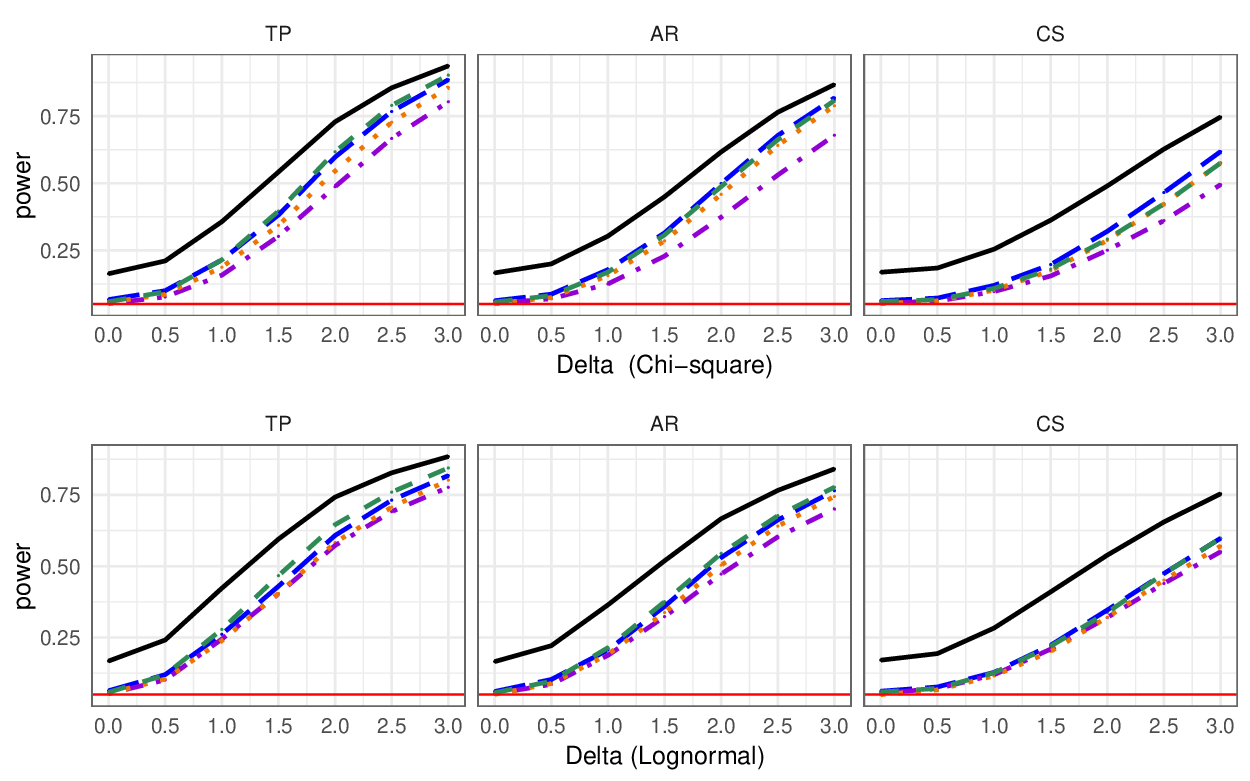}	
	\end{center}
	\caption{Power simulation results of the tests $T_W$ $({\color{black}\textbf{\textendash\textendash\textendash}})$, $T_A$ $({\color{blue} \textbf{ \textendash\textendash \quad \textendash\textendash}})$ , 	$T_W^*$ $({\color{violet}\textbf{--}\boldsymbol{\cdot}\textbf{--}})$, $T_{A}^*$ $({\color{brown}\boldsymbol{\cdots}})$, and $T_{M}^*$ $({\color{ao(english)} \textbf{-- - --} })$ under different covariance structures with sample size $n=15$  and $d=4$ under alternative $1$, under the MCAR framework with missing rate $r=30\%$ with observations generated from a $\chi^2_{15}$ (upper row) and a Lognormal (bottom row) distribution, respectively.}
	\label{fig:PowerCLMiss30HT}	
\end{figure}

\subsection{Power}

In order to assess the empirical power of all studied methods, we considered a one-sample repeated measures design with $d=4$ repeated measures, sample size  $n=15$, and covariance structures as given in Settings $1-3$ for various distributions. Data was generated by
\bqa
{\color{black}
[\vX_{1k}\sim\vF_1(\vmu_0,\vSigma_1)] + \boldsymbol{\mu}_1,}\quad k=1,... ,15 ,
\eqa

where, we were interested in detecting two specific alternatives 	
\begin{itemize}
	
	\item { Alternative 1:  $\boldsymbol{\mu}_1=(0,0,\zeta,\zeta)$,}
	\item Alternative 2: $\boldsymbol{\mu}_1=(0,0,0,\zeta)$,
\end{itemize}
  for varying shift parameter $\zeta\in\{0,0.5,1,1.5,2,2.5,3\}$.\\

The power analysis results of the considered methods under the MCAR framework for several distributions, involving various covariance settings for detecting Alternative $1$ are summarized in Figures \ref{fig:PowerNDMiss30HT} - \ref{fig:PowerCLMiss30HT}  (missing rate $r=30\%$) and Figures S.6 - S.7 in the supplement  ($r=10\%$). The simulation results for investigating  Alternative $2$ are displayed in Figures S.8 - S.11 in the supplement. The power analysis results of the considered methods under the respective MAR framework are summarized
in Figures S.12 - S.15 (MAR1 scenario)  and  Figures S.16 - S.19 (MAR2 scenario). \\

As the asymptotic Wald test is too liberal compared to the other studied methods, its power function is larger. Moreover, the bootstrapped Wald test exhibits the lowest power behavior, while the bootstrapped MATS and ANOVA-type have a quite similar power behavior as the \cite{brunner1999rank} ANOVA-type test. {\color{black}The differences between the procedures is less pronounced} under the MAR framework.\\

To sum up, we recommend the bootstrap ANOVA-type statistic. It exhibits the overall best type-I-error control combined with a good power behaviour and needs the less stringent assumptions for application.

\begin{figure}[h!]
	\centering
	\hspace*{-1.2in}
	\subfloat[{\bf Side effect}]{{\includegraphics[scale=0.35, clip=true,trim=0cm 4.5cm 0cm 4cm]{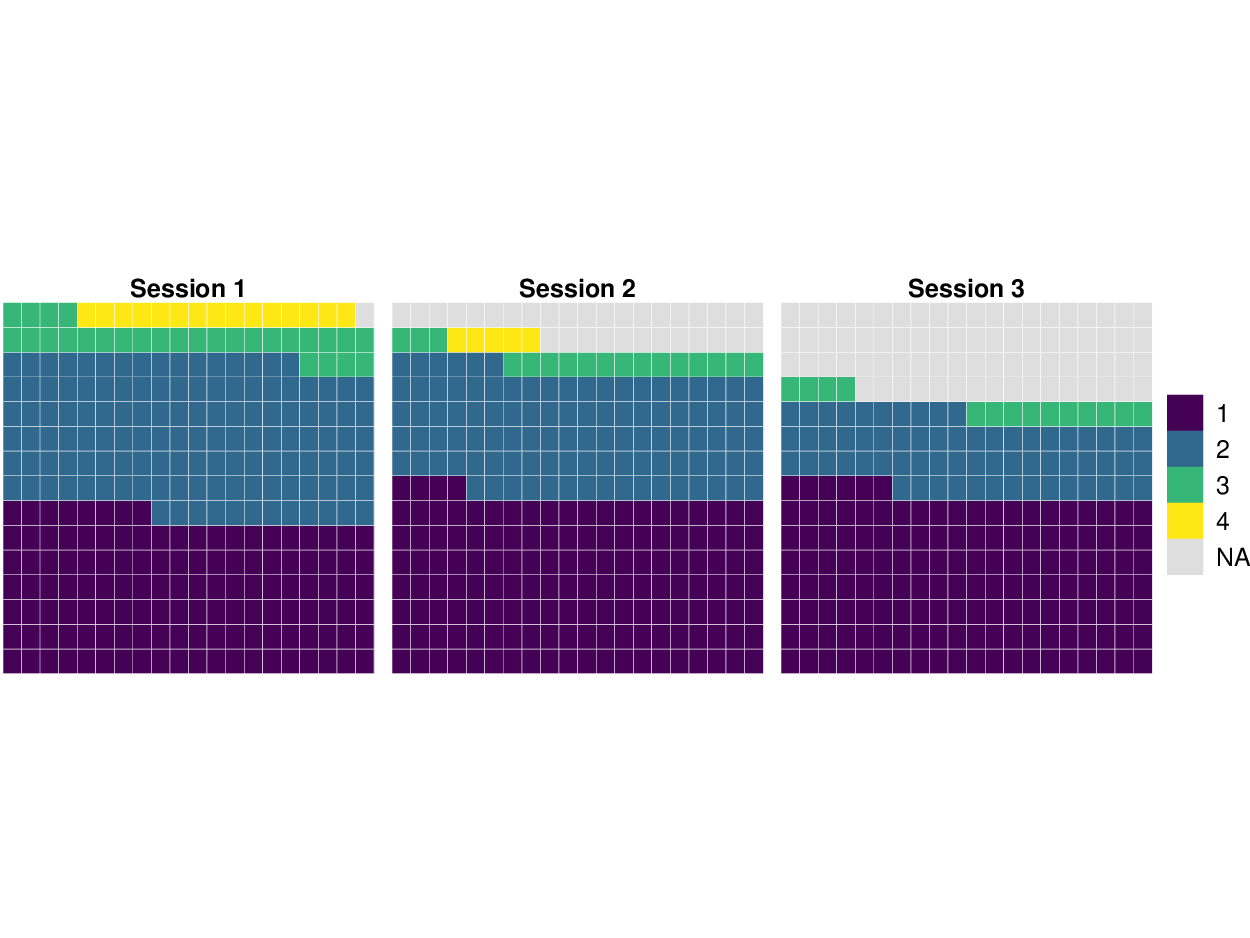} }}%
	\Hquad
	\hspace*{-0.1in}
	\subfloat[{\bf Therapeutic effect}]{{\includegraphics[scale=0.35, clip=true,trim=0cm 4.5cm 0cm 4cm]{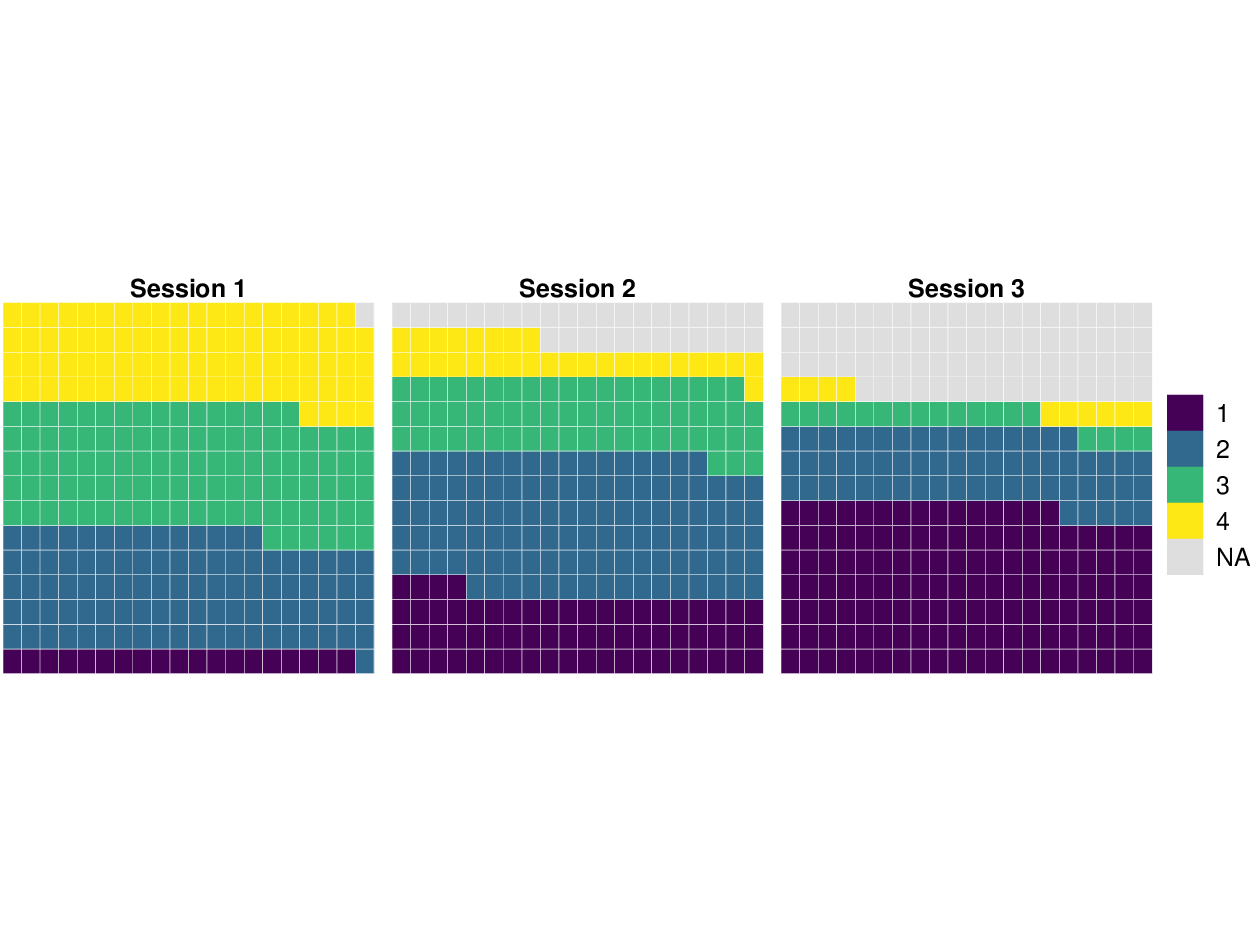} }}%
	\hspace*{-1.1in}
	\caption{Frequencies of the {\bf side effect} and {\bf therapeutic effect}  scores observed in the fluvoxamine trial.}%
	\label{fig:fluvoxamine}%
\end{figure}

\section{Application to empirical data}\label{Secexample}
\subsection{The fluvoxamine trial}
In this section, we re-examine a clinical fluvoxamine study. It  has already been analyzed by \cite{molenberghs1994marginal, molenberghs1997analysis, van2001local, jansen2003local, molenberghs2004meaningful}, and \cite{ molenberghs2007missing}.
The study has been performed to 
establish the profile of fluvoxamine in ambulatory clinical psychiatric operations. Hereby, 
 315 patients who  suffer from depression, panic disorder, and/or obsessive-compulsive disorder were scored every two weeks over six weeks  of treatment (d=3). At each clinical visit, scores for both {\it side effect} and {\it therapeutic effect} scales which are based upon about 20 psychiatric symptoms were recorded. The side effect scale ranges from $1$ to $4$. The lower the score,  the better the clinical record. For example, score $1$ stands for \enquote{no side effect} while score $4$ indicates that \enquote{the side effect surpasses the therapeutic effect}. Similarly, the therapy effect is a $4$ category ordinal scale: (1) \enquote{no improvement or worsening}; (2) \enquote{minimal improvement, not changing functionality}; (3) \enquote{moderate improvement, partial disappearance of symptoms}; and (4) \enquote{important improvement, almost disappearance of symptoms}.  The higher the score, the better the patients health. \\

 Several patients missed the recording of their measurements in some sessions which leaded to a large amount of missing values.  A closer look to our data shows that, from the total of 315 initially recruited patients, 14 patients dropped off, 31 patients were scored on the first session only, and 44 patients on Session 1 and Session 3 only. Thus, only 224 patients have complete observations. Two patients were excluded from the analysis due to their non-monotone missing pattern.  This leaves us with 299 patients. Waffle plots representing the distributions of the side effect and therapeutic effect among the three sessions are shown in Figure \ref{fig:fluvoxamine}. We aim to test the hypotheses whether side effect or therapeutic effect scores  are significantly different between the three sessions for patients with psychiatric disorder. To this end, we applied all  considered testing methods; asymptotic Wald and ANOVA-type tests ($T_W, T_A$), and the bootstrap procedures ($T_W^*$, $T_A^*$, $T_{M}^*$) to detect the null hypothesis $H_0^T:\{\vC\vF = \boldsymbol{0}\}$. The results are summarized in Table \ref{pvalFluxovamine}. It can be seen that all tests indicate a significant difference between the therapeutic effect scores as well as the side effect scores of the three sessions (two-sided p-value $< 0.00001$). Moreover, we conclude that the clinical outcome of the patients significantly improves after three sessions. {\color{black}These findings coincide with that in \cite{molenberghs2007missing}}.
\begin{table*}[ht]%
	\centering
	\caption{P-values of  the fluvoxamine study.\label{Pvalfluv}}%
	\begin{tabular*}{0.65\linewidth}{lccccc}
		\cmidrule{1-6}
		\textbf{Effect}&$T_W$& 	$T_A$ &	$T_W^*$ & 	$T_A^*$ & 	$T_M^*$ \\
		\cmidrule{1-6}
		\textbf{Therapeutic effect}&$2.3e^{-79}$&$2.2e^{-86}$&0&0&0\\
		\textbf{Side effect}&$2.3e^{-10}$& $4.1e^{-12}$&0&0&0\\

		\cmidrule{1-6}
	\end{tabular*}
	\label{pvalFluxovamine}
\end{table*}


\subsection{The skin disorder trial}

Here, we study data from a  randomized, multi-center, parallel group study for treating a skin condition. Treating skin conditions can be tough, thus the goal of the study was to assess the severe rate of the skin condition over time and to compare the efficiency and safety of two continuous therapy treatments drug and placebo.  Patients were randomly assigned to drug or placebo therapy treatment.  Prior to treatment, patients were assessed to determine the initial severity of the skin condition (moderate or severe). At three  follow-up visits, the treatment outcome was measured according to a five-point
ordinal response scale that assess the extent of improvement (1 = rapidly improving,
2 = slowly improving, 3 = stable, 4 = slowly worsening, 5 = rapidly
worsening). The study consists of 88 and 84 subjects allocated to the active treatment group and the placebo group, respectively. And, the proportion of missing observations is around $30\%$. The distribution of patients improvement across the treatment groups and the follow-up visits is displayed in Figure \ref{fig:placebo}. The study is described in  full detail in \cite{landis1988some} and is published in \cite{davis2002statistical}.\\
\begin{figure}[h!]
	\begin{center}
		\includegraphics[scale=0.6]{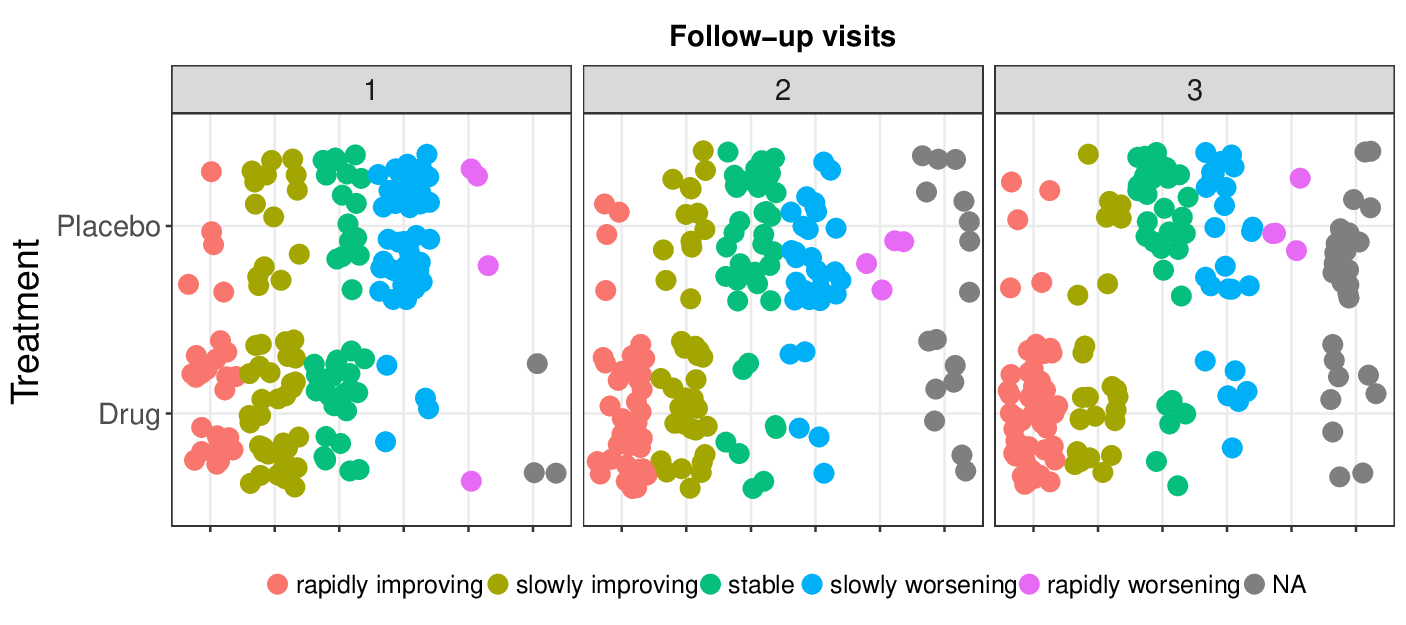}	
	\end{center}
	\caption{Frequencies of observed patients treatment outcome in the skin disorder trial.}
	\label{fig:placebo}	
\end{figure}
Similar to the fluvoxamine study above, we applied all asymptotic and bootstrap
procedures to infer the following null hypotheses: \enquote{no group effect}, \enquote{no
time effect} and \enquote{no group $\times$ time effect}. The results are summarized in Table \ref{pvalplacebo}: All approaches 
reject the null hypothesis of no group effect,  the null hypothesis of no time
effect and the null hypothesis of no treatment
group $\times$  time interaction. This implies that the clinical outcome of the patients significantly improves with time and  this progression is significantly different between the two treatment groups drug and placebo.  Therefore, the data is further analyzed and split by the factor initial severity and the analysis is replicated separately for each baseline severity level (moderate or severe). The results are provided in Table \ref{pvalplacebobaseline}. \\

It can be seen from Table \ref{pvalplacebobaseline} that  all approaches detect a significant group effect as well as a significant time effect in both moderate and severe groups. In contrast, a significant  group $\times$  time interaction effect arises only in the moderate severity group.This indicates that the change in clinical outcomes of patients of moderate severity varies over time depending on treatment group membership. All five approaches share the previous findings.

\begin{table*}[ht]%
	\centering
	\caption{P-values of  the skin disorder trial.}%
	\begin{tabular*}{0.6\linewidth}{lccccc}
		\cmidrule{1-6}
		\textbf{Hypothesis}&$T_W$& 	$T_A$ &	$T_W^*$ & 	$T_A^*$ & 	$T_M^*$ \\
		\cmidrule{1-6}
		\textbf{Group}&0&0&0&0&0\\
		\textbf{Time}&0&0&0&0&0\\
		\textbf{Group $\times$ Time}&$0.032$& $0.016$& $0.026$& $0.01$& $0.009$\\
		
		\cmidrule{1-6}
	\end{tabular*}
	\label{pvalplacebo}
\end{table*}

\begin{table*}[ht]%
	\caption{P-values of  the split skin disorder trial based on initial severity.}%
	
	\begin{tabular*}{0.65\linewidth}{lccccccccccc}
		\cmidrule{1-12}
		\textbf{Hypothesis}& \multicolumn{5}{c}{Initial severity = Moderate} &&  \multicolumn{5}{c}{Initial severity  = Severe}  \\
		\cmidrule{2-6}
		\cmidrule{8-12}
		
		&$T_W$&$T_A$ &	$T_W^*$ & 	$T_A^*$ & 	$T_M^*$&&$T_W$&$T_A$ &	$T_W^*$ & 	$T_A^*$ & 	$T_M^*$  \\
		\cmidrule{1-12}
		\textbf{Group}&0&0&0&0&0&&0&0&0&0&0\\
		\textbf{Time}&$0.003$ &$0.004$& $0.003$& $0.002$ & $0.002$&&0&0&0&0&0\\
		\textbf{Group $\times$ Time}&$0.041$& $0.032$ &$0.043$& $0.027$& $0.029$&&$0.423$&$0.318$& $0.447$& $0.338$& $0.335$\\
		\cmidrule{1-12}
	\end{tabular*}
	\label{pvalplacebobaseline}
\end{table*}

\subsection{Simulations for data with similar attributes to the trials data}
 It was also interesting to discover the type-I error rate control of the tests under similar attributes to the data sets of the fluvoxamine and skin disorder clinical trials. The data sets reflect large sample sizes and a small or moderate amount of missing values from either a one sample repeated measurements design or a two way layout design. Simulation results for the type-I error rate of the studied procedures for  $(n = 299, d = 3)$ and ($n_1=88,  n_2=84, d=3$) sample sizes are presented in Table S.25 and Table S.26 in the supplement, respectively. It can be seen that most tests are robust under both settings and control type-I error rate accurately. Only the Wald type tests exhibit a liberal behaviour for some of the lognormal settings.\\


\section{Summary} \label{Secdis}
Multi-group repeated measures design with ordinal or skewed observations 
are quite common. If the observations are additionally subject to missing values existing methods for testing null hypotheses in terms of distribution function may be either liberal (Wald-type statistics) or run (asymptotically) on a wrong type-I-error level (ANOVA-type statistic, \citep{domhof2002rank}). To this end we investigated three alternatives based upon resampling. We proved their asymptotic validity and analyzed their small sample behaviour regarding type-I-error control and power in extensive simulations. Under all of the $5$ considered methods, an ANOVA-type statistic with critical values calculated by means of a Wild bootstrap approach exhibits the best behaviour and is recommended. 

In the future, we will include the present methodology into the R package nparLD \citep{noguchi2012nparld}. Moreover, we plan to extend our investigations to general MANOVA settings, e.g., extending the results of \cite{dobler2019nonparametric} to the situation with missing values.

{\color{black}
\section*{Acknowledgement} 
	Lubna Amro and Markus Pauly acknowledge the support of the German Research Foundation (DFG) (No. PA 2409/3-2). Frank Konietschke’s work was also supported by the German Research Foundation (DFG) (No. KO 4680/3-2). Lubna Amro also acknowledges the support of the project "From Prediction to Agile Interventions in the Social Sciences (FAIR)", which receives funding from the programme "Profilbildung 2020", an initiative of the Ministry of Culture and Science of the State of Northrhine Westphalia. The sole responsibility for the content of this publication lies with the authors.
	}


\FloatBarrier
\bibliographystyle{apa}
\bibliography{References}

\end{document}


\title{Supplementary material to: Incompletely observed nonparametric factorial designs with repeated measurements: A wild bootstrap approach}

\maketitle

 In this supplementary material, we present the proofs of all theorems from the paper. {\color{black} Furthermore, we recall the definition of
 	the different missing mechanisms and} present additional type-I error and power simulation results of the asymptotic quadratic form tests and their wild bootstrap counterparts as  described in Sections 3  and 4 of the paper. 

\section{Proofs}\label{Proofs}

\textit{Proof of Theorem 3.1:}\\ 
The proof is stated in \cite{brunner1999rank}, we will explain it here shortly: first, we know that we have convergence in distribution $\sqrt{n}\vC\hat{\vp} \xrightarrow{\text{d}}N(\boldsymbol{0,C\vV C^T})$ as $n_0\rightarrow\infty$ under $H_0$. Hence, using the CMT, the quadratic form $\tilde{T}_W=n\vph^T\vC^T[\vC\vV\vC^T]^+\vC\vph$ has asymptotically a central $\chi^2_f$ distribution with $f=rank(\vC)$ degrees of freedom {\color{black}since $\vV>0$}. Moreover, as $\vVh_n$ is a consistent estimator for $\vV$, the result follows from Slutzky's theorem noting that the probability of the set $\{\vVh_n>0\}$ converges to 1.\\

\textit{Proof of Theorem 3.2:}\\ 
Applying again the CMT, it follows that $tr(\vT\vV) \cdot T_A=n\vph^T\vT\vph$ has asymptotically the same distribution as  $\sum_{i=1}^{a}\sum_{j=1}^{d} \zeta_{ij}B_{ij}$ \citep{graybill1976theory, brunner2001nonparametric}. Then, the result follows from the invariance of the multivariate standard normal distribution under orthogonal transformations {\color{black}(as $\vC$)} and the consistency of $\hat{\vV}_n$ and thus of $tr(\vT\hat{\vV}_n)$ by using Slutzky theorem.\\

\textit{Proof of Theorem 3.3:}\\
Following similar arguments as prescribed in \cite{friedrich2018mats}, we can obtain that $( \vC \hat{\mathbf{D}}_n \vC^\top )^+\xrightarrow{\text{p}} ( \vC \mathbf{D} \vC^\top )^+$ as $\hat{\boldsymbol{D}}_n=diag(\vVh_n)_{ii} \overset{p}{\to} diag(\vV)_{ii}=\vD$ where $D$ is of full rank by assumption. Thus, the result follows  from the representation theorem of quadratic forms \citep{rao1972generalized}. \\

\textit{Proof of Theorem 4.1:}\\
First, we prove (i). We show that given the data, the expression $\sqrt{n}\vph^*_i$ 
is asymptotically multivariate $N(0,\kappa_i^{-1}\vVh_i)$ distributed. 

The results follows by applying the conditional CLT for the wild bootstrap given in Theorem A.1. in  \cite{beyersmann2013weak}. In order to show the conditional asymptotic normality of $\sqrt{n}\vph^*_i$, we need to recall that $\hat{Y}_{ijk}=\hat{H}(X_{ijk})=\frac{1}{N}(R_{ijk}-\frac{1}{2}),  \bar{\hat{Y}}_{ij.}=\frac{1}{\lambda_{ij.}}\sum_{k=1}^{n_i}\lambda_{ijk}\hat{Y}_{ijk} $, $\hat{\vY}_{ik}=(Y_{i1k},...,Y_{idk})^T$, $\bar{\hat{\vY}}_{i.}=(\bar{\hat{Y}}_{i1.}, ..., \bar{\hat{Y}}_{id.})$, and $\boldsymbol{\Lambda}_{ik}=n_idiag\{\frac{\lambda_{i1k}}{\lambda_{i1.}},...,\frac{\lambda_{idk}}{\lambda_{id.}}\}$.\\

 Now, it remains to show that all conditions of their Theorem A.1 are fulfilled:\\
 
\begin{itemize}
\item [A)] As $1 \leq n_i/\lambda_{ij.}\leq n_i$ is bounded, $|\hat{Y}_{ijk}|\leq1$, and $ n/n_i^2 \to 0$ it follows that 
\bqa
\max_{{1\leq i\leq a}}\frac{\sqrt{n}||\Lambda_{ik}(\hat{\vY}_{ik}-\bar{\hat{\vY}}_{i.})||}{n_i} \xrightarrow{n \to \infty} 0 \quad \text{in probability.}
\eqa

\item [B)] We need to study the convergence of
\bqa
\begin{split}
&\frac{n}{n_i^2}\sum_{k=1}^{n_i}[\Lambda_{ik}(\hat{\vY}_{ik}-\bar{\hat{\vY}}_{i.})][\Lambda_{ik}(\hat{\vY}_{ik}-\bar{\hat{\vY}}_{i.})]^T\\
&=\frac{n}{n_i}\frac{1}{n_i}\sum_{k=1}^{n_i}\Lambda_{ik}(\hat{\vY}_{ik}-\bar{\hat{\vY}}_{i.})(\hat{\vY}_{ik}-\bar{\hat{\vY}}_{i.})^T\Lambda_{ik}.
\end{split}
\eqa

First, we consider the diagonal elements which are given by

\begin{equation*}
\begin{split}
&\frac{n}{n_i}\frac{1}{n_i}\sum_{k=1}^{n_i}n_i^2\frac{\lambda_{ijk}^2}{\lambda_{ij.}^2}(\hat{Y}_{ijk}-\bar{\hat{Y}}_{ij.})^2\\
&=\frac{n}{n_i}\frac{n_i}{\lambda_{ij.}^2}\sum_{k=1}^{n_i}\lambda_{ijk}(\hat{Y}_{ijk}-\bar{\hat{Y}}_{ij.})^2 \quad \text{(since} \hspace{0.5em} \lambda_{ijk}^2=\lambda_{ijk} \hspace{0.5em} \text{by definition)}\\
&=\frac{n(\lambda_{ij.}-1)}{n_i\lambda_{ij.}}\bigg[\frac{n_i}{\lambda_{ij.}(\lambda_{ij.}-1)}\sum_{k=1}^{n_i}\lambda_{ijk}(\hat{Y}_{ijk}-\bar{\hat{Y}}_{ij.})^2\bigg]\\
&=\frac{n(\lambda_{ij.}-1)}{n_i\lambda_{ij.}}\bigg[\frac{n_i}{\lambda_{ij.}(\lambda_{ij.}-1)}\sum_{k=1}^{n_i}\lambda_{ijk}\big(\frac{1}{N}(R_{ijk}-\frac{1}{2})-\frac{1}{N}(\bar{R}_{ij.}-\frac{1}{2})\big)^2\bigg]\\
&=\frac{n(\lambda_{ij.}-1)}{n_i\lambda_{ij.}}\bigg[\frac{n_i}{(N^2)\lambda_{ij.}(\lambda_{ij.}-1)}\sum_{k=1}^{n_i}\lambda_{ijk}(R_{ijk}-\bar{R}_{ij.})^2\bigg]\\
&=\frac{n(\lambda_{ij.}-1)}{n_i\lambda_{ij.}}\hat{v}_{i}(j,j).\\
\end{split}
\end{equation*}

Therefore, we get (recalling that $\lambda_{ij.}\leq n_i,\hspace{0.5em} \frac{ \lambda_{ij.}}{n} \rightarrow \kappa_{i},$ and $\hat{v}_{i}(j,j)$ is  consistent for $v_{i}(j,j)$)\\
\bqa
\frac{n}{n_i}\frac{1}{n_i}\sum_{k=1}^{n_i}n_i^2\frac{\lambda_{ijk}^2}{\lambda_{ij.}^2}(\hat{Y}_{ijk}-\bar{\hat{Y}}_{ij.})^2\xrightarrow{n_0 \to \infty} \frac{1}{\kappa_i} v_i(j,j) \hspace{0.5em} \text{in probability.} \quad \\
\eqa

Similarly, we show it for the off-diagonal elements.

\bqa
\begin{split}
  &\quad\frac{n}{n_i}\frac{1}{n_i}\sum_{k=1}^{n_i}n_i^2\frac{\lambda_{ijk}\lambda_{ij'k}}{\lambda_{ij.}\lambda_{ij'.}}(\hat{Y}_{ijk}-\bar{\hat{Y}}_{ij.})(\hat{Y}_{ij'k}-\bar{\hat{Y}}_{ij'.})\\
  &\quad=\frac{n}{n_i}\frac{n_i}{\lambda_{ij.}\lambda_{ij'.}}\sum_{k=1}^{n_i}\lambda_{ijk}\lambda_{ij'k}(\hat{Y}_{ijk}-\bar{\hat{Y}}_{ij.})(\hat{Y}_{ij'k}-\bar{\hat{Y}}_{ij'.})\\
  &\quad=\frac{n[(\lambda_{ij.}-1)(\lambda_{ij'.}-1)+\Delta_{i,jj'}-1]}{n_i\lambda_{ij.}\lambda_{ij'.}}\bigg[\frac{n_i\sum_{k=1}^{n_i}\lambda_{ijk}\lambda_{ij'k}(\hat{Y}_{ijk}-\bar{\hat{Y}}_{ij.})(\hat{Y}_{ij'k}-\bar{\hat{Y}}_{ij'.})}{[(\lambda_{ij.}-1)(\lambda_{ij'.}-1)+\Delta_{i,jj'}-1]}\bigg]\\
&\quad=\frac{n[(\lambda_{ij.}-1)(\lambda_{ij'.}-1)+\Delta_{i,jj'}-1]}{n_i\lambda_{ij.}\lambda_{ij'.}}\\
&\quad\times\bigg[\frac{n_i\sum_{k=1}^{n_i}\lambda_{ijk}\lambda_{ij'k}(\frac{1}{N}(R_{ijk}-\frac{1}{2})-\frac{1}{N}(\bar{R}_{ij.}-\frac{1}{2}))(\frac{1}{N}(R_{ij'k}-\frac{1}{2})-\frac{1}{N}(\bar{R}_{ij'.}-\frac{1}{2}))}{[(\lambda_{ij.}-1)(\lambda_{ij'.}-1)+\Delta_{i,jj'}-1]}\bigg]\\  
\vspace{2em}
&\quad=\frac{n[(\lambda_{ij.}-1)(\lambda_{ij'.}-1)+\Delta_{i,jj'}-1]}{n_i\lambda_{ij.}\lambda_{ij'.}}\bigg[\frac{n_i\sum_{k=1}^{n_i}\lambda_{ijk}\lambda_{ij'k}(R_{ijk}-\bar{R}_{ij.})(R_{ij'k}-\bar{R}_{ij'.})}{(N^2)[(\lambda_{ij.}-1)(\lambda_{ij'.}-1)+\Delta_{i,jj'}-1]}\bigg]\\  
&\quad=\frac{n[(\lambda_{ij.}-1)(\lambda_{ij'.}-1)+\Delta_{i,jj'}-1]}{n_i\lambda_{ij.}\lambda_{ij'.}}\hat{v}_i(j,j').\\  
\end{split}
\eqa

As $\Delta_{i,jj'}\leq n_i, \lambda_{ij.}\leq n_i$, $\frac{ \lambda_{ij.}}{n} \rightarrow \kappa_{i}$, and $\hat{v}_i(j,j')$ is consistent for $v_i(j,j')$ it follows that
\bqa
\centering
\frac{n}{n_i}\frac{1}{n_i}\sum_{k=1}^{n_i}n_i^2\frac{\lambda_{ijk}\lambda_{ij'k}}{\lambda_{ij.}\lambda_{ij'.}}(\hat{Y}_{ijk}-\bar{\hat{Y}}_{ij.})(\hat{Y}_{ij'k}-\bar{\hat{Y}}_{ij'.}) \xrightarrow{n_0 \to \infty} \frac{1}{\kappa_i} v_i(j,j') \hspace{0.5em} \text{in probability.}
\eqa

So, summing up, we get that
\bqa
\centering
\frac{n}{n_i}\frac{1}{n_i}\sum_{k=1}^{n_i}\Lambda_{ik}(\hat{\vY}_{ik}-\bar{\hat{\vY}}_{i.})(\hat{\vY}_{ik}-\bar{\hat{\vY}}_{i.})^T\Lambda_{ik} \xrightarrow{n_0 \to \infty} \frac{1}{\kappa_i} \vV_i \hspace{0.5em} \text{in probability.}
\eqa
Thus, we have conditional weak convergence given the data $\vX$
\bqa
\centering
\frac{\sqrt{n}}{n_i}\sum_{k=1}^{n_i}W_{ik}\Lambda_{ik}(\hat{\vY}_{ik}-\bar{\hat{\vY}}_{i.})\xrightarrow{d} N(0,\frac{1}{\kappa_i} \vV_i),
\eqa

in probability and therefore $\sqrt{n}\vph^*_i$  is asymptotically multivariate
normally distributed with expectation $0$ and asymptotic covariance matrix $\frac{1}{\kappa_i} \vV_i$. Finally, a point-wise application of Slutzky shows that $\sqrt{n}\vph^*$ converges in distribution to $N(0,\bigoplus\limits_{i=1}^a\kappa_i^{-1}\vV_i)$ {\color{black}in probability given the data}.\\

\end{itemize}

\textit{Proof of Theorem 4.2:}

It has already been shown that $\vVh_n-\vV \xrightarrow{p} 0$ \citep[Theorem 3.5]{brunner1999rank}. So, it suffices to show that $\vVh_n^*-\vVh \xrightarrow{p} 0$. Thus, consider 
 \bqa
\hat{p}_{ij}^{*}=\frac{1}{N}\bar{Z}^{*}_{ij.}=\frac{1}{\lambda_{ij.}}\sum_{k=1}^{n_i}\lambda_{ijk}\frac{1}{N}(W_{ik}Z_{ijk})=\frac{1}{\lambda_{ij.}}\sum_{k=1}^{n_i}\lambda_{ijk}\frac{1}{N}(W_{ik} (R_{ijk}-\bar{R}_{ij.})).
\eqa

Then,
\bqa
\mathbb{E}\Big( \hat{p}_{ij}^{*}|\vX\Big)=\mathbb{E}\Big(\frac{1}{\lambda_{ij.}}\sum_{k=1}^{n_i}W_{ik}\lambda_{ijk}\frac{1}{N}Z_{ijk}|\vX\Big)= \frac{1}{N\lambda_{ij.}}\sum_{k=1}^{n_i}\mathbb{E}\Big(W_{ik}\Big).\lambda_{ijk}Z_{ijk}=0.
\eqa
And, as $\lambda_{ij.}/n \rightarrow \kappa_i$
\bqa
\begin{split}
Var\Big( \hat{p}_{ij}^{*}|\vX\Big)&=\frac{1}{(N\lambda_{ij.})^2}\sum_{k=1}^{n_i} Var\Big( W_{ik}\Big) \Big(\lambda_{ijk}Z_{ijk}\Big)^2 \quad \text{(} W_{ik}Z_{ijk} \hspace{0.5em} \text{are conditionally independent given} \hspace{0.25em} \vX \text{.)}\\ &= \frac{1}{(N\lambda_{ij.})^2}\sum_{k=1}^{n_i} \lambda_{ijk}\Big(Z_{ijk}\Big)^2\\&\leq 
\frac{1}{N^2\lambda_{ij.}^2}n_i (N-1)^2\leq \frac{n}{\lambda_{ij.}^2}(\frac{N-1}{N})^2 \xrightarrow{p} 0. 
\end{split}
\eqa
Consequently, by using Chebyshev’s inequality, we get  $ \hat{p}_{ij}^{*} \rightarrow 0$ in probability for all $i=1,..., a$. Now, we show by tedious calculations that  the diagonal elements  

\bqa
\begin{split}
\hat{v}_i(j,j)-\hat{v}^*_i(j,j)&=\frac{n_i \sum_{k=1}^{n_i}\lambda_{ijk}[R_{ijk}-\bar{R}_{ij.}]^2}{(N^2)\lambda_{ij.}(\lambda_{ij.}-1)}- \frac{n_i \sum_{k=1}^{n_i}\lambda_{ijk}[Z^*_{ijk}-\bar{Z}^*_{ij.}]^2}{(N^2)\lambda_{ij.}(\lambda_{ij.}-1)}\\
&=\frac{n_i }{(N^2)\lambda_{ij.}(\lambda_{ij.}-1)} \sum_{k=1}^{n_i}\lambda_{ijk} \bigg[Z^2_{ijk} -[Z^{*2}_{ijk}-2Z^*_{ijk}\bar{Z}^*_{ij.}+\bar{Z}^{*2}_{ij.}]\bigg]\\
&=\frac{n_i }{(N^2)\lambda_{ij.}(\lambda_{ij.}-1)} \sum_{k=1}^{n_i}\lambda_{ijk} \bigg[2Z^*_{ijk}\bar{Z}^*_{ij.}-\bar{Z}^{*2}_{ij.}\bigg] \quad (\text{since} \hspace{0.25em} W_{ik}^2=1)\\
&=\frac{n_i }{(N^2)\lambda_{ij.}(\lambda_{ij.}-1)} \bigg[2\bar{Z}^*_{ij.}\sum_{k=1}^{n_i}\lambda_{ijk} Z^*_{ijk}-\lambda_{ij.}\bar{Z}^{*2}_{ij.}\bigg]\\
&=\frac{n_i }{(N^2)\lambda_{ij.}(\lambda_{ij.}-1)} \lambda_{ij.}\bar{Z}^{*2}_{ij.}\\
&=\frac{n_i}{\lambda_{ij.}-1}\frac{\bar{Z}^{*}_{ij.}}{N}\frac{\bar{Z}^{*}_{ij.}}{N}\\
&=\frac{n_i}{\lambda_{ij.}-1}\hat{p}_{ij}^{*}\hat{p}_{ij}^{*}\\
&\xrightarrow{n_0 \to \infty} 0 \hspace{0.5em} \text{in probability} \quad (\hat{p}_{ij}^{*} \xrightarrow{p} 0, \hspace{0.5em} \frac{ \lambda_{ij.}}{n} \rightarrow \kappa_{i}, \hspace{0.5em} n_i\leq n).
\end{split}
\eqa

Furthermore, using similar arguments as above, we can show that an analogous result holds for the off-diagonal elements.
And, thus $\vVh_n^*-\vVh \xrightarrow{p} 0$. Following the same steps as in the proofs of Theorems 3.1-3.3 , this concludes the proof.

{\color{black}
	
		\section{Missing Data Mechanism}

\cite{rubin1976inference}  identified three missing mechanisms for the data based on the relationship between the missing values and observed values. Let $\vX$ denote a data set which can be decomposed into observed and unobserved portions $\vX=(\vX_{obs},\vX_{mis})$. Let $\vlambda$ be a binary matrix whose components indicate whether $\vX$ is observed or missing. The three missing data mechanisms are:

\begin{enumerate}
	\item \textbf{Missing completely at random (MCAR):}\\
	The probability of an observation being missing does not depend on the values of any observed or unobserved data, i.e., $P(\vlambda|\vX_{obs},\vX_{mis}) = P(\vlambda)$. This implies that the conditional and marginal distributions can always be accurately estimated from the observed data.

	\item \textbf{ Missing at random (MAR):}\\
	The probability of the missingness can depend on the observed data but not on the unobserved data, i.e. $P(\vlambda|\vX_{obs},\vX_{mis}) = P(\vlambda|\vX_{obs})$. Thus, the missing data is due to an external effect, not the variable itself. Note that MCAR is a special case of MAR.
	
	\item  \textbf{Missing not at random (MNAR):}\\
	The probability of the missingness can depend on the unobserved data, i.e. $P(\vlambda|\vX_{obs},\vX_{mis}) \neq P(\vlambda|\vX_{obs})$. MNAR is also known as the non-ignorable case \citep{little2019statistical}  since the missing observation is dependent on the outcome result. 
\end{enumerate}

For additional information on the various missing mechanisms, we refer to  \cite{little2019statistical}.

}

	\section{Type-I Error and Power Results }
In the sequel, we present some additional type-I error and power results of the Monte Carlo simulation study, that is described in detail in Section 5 of the paper, for testing the hypothesis of no group effect $H_0^G$, no time effect $H_0^T$, as well as no interaction effect $H_0^{GT}$ for incomplete nonparametric factorial designs with repeated measurements under the MCAR, and MAR schemes. 
The type-I error results under MCAR scheme, $d=4$ time points for the hypotheses $H_0^T$ and $H_0^{GT}$ are presented in Tables \ref{TableTypErrorNormalt4MCAR}, \ref{TableTypErrorDexpt4MCAR}, \ref{TableTypErrorChisqt4MCAR}, and \ref{TableTypErrorLognormt4MCAR} for the normal, double exponential, lognormal and chi-square distribution, respectively. The results for
$d=8$ time points are in Tables \ref{TableTypErrornormalt8MCAR}, \ref{TableTypErrordexpt8MCAR}, \ref{TableTypErrorchisqt8MCAR}, and \ref{TableTypErrorlognormalt8MCAR} for the normal, double exponential, lognormal and chi-square distribution, respectively.  The results for the hypothesis $H_0^G$ are presented in Tables \ref{TableErrorMCARNormalHG} - \ref{TableErrorMCARLognormHG}. Moreover, the type-I error results under MAR scheme, for the hypotheses $H_0^T$ and $H_0^{GT}$ for symmetric and skewed distributions are presented in Tables \ref{TableErrorMARNormalt4} - \ref{TableErrorMARLognormalt4} ($d=4$ time points) and Tables \ref{TableErrorMARNormalt8} - \ref{TableErrorMARlognormt8} $(d=8)$.  The results for the hypothesis $H_0^G$ are in Tables \ref{TableErrorMARNormalHG}, \ref{TableErrorMARDexpHG}, \ref{TableErrorMARChisqHG}, and \ref{TableErrorMARLognormHG}  for the normal, double exponential, chi-square and lognormal distribution, respectively.\\

Simulation results for the type-I error rate under similar attributes to the fluvoxamine trial data and the skin disorder trial data for sample sizes $(n = 99; d = 3)$ and $(n_1=88, n_2=84; d=3)$ are presented in Tables \ref{Tablea1Flov} and \ref{TableToenail}, respectively.\\

\begin{figure}[h!]
	\begin{center}
		\includegraphics[scale=0.7, clip=true,]{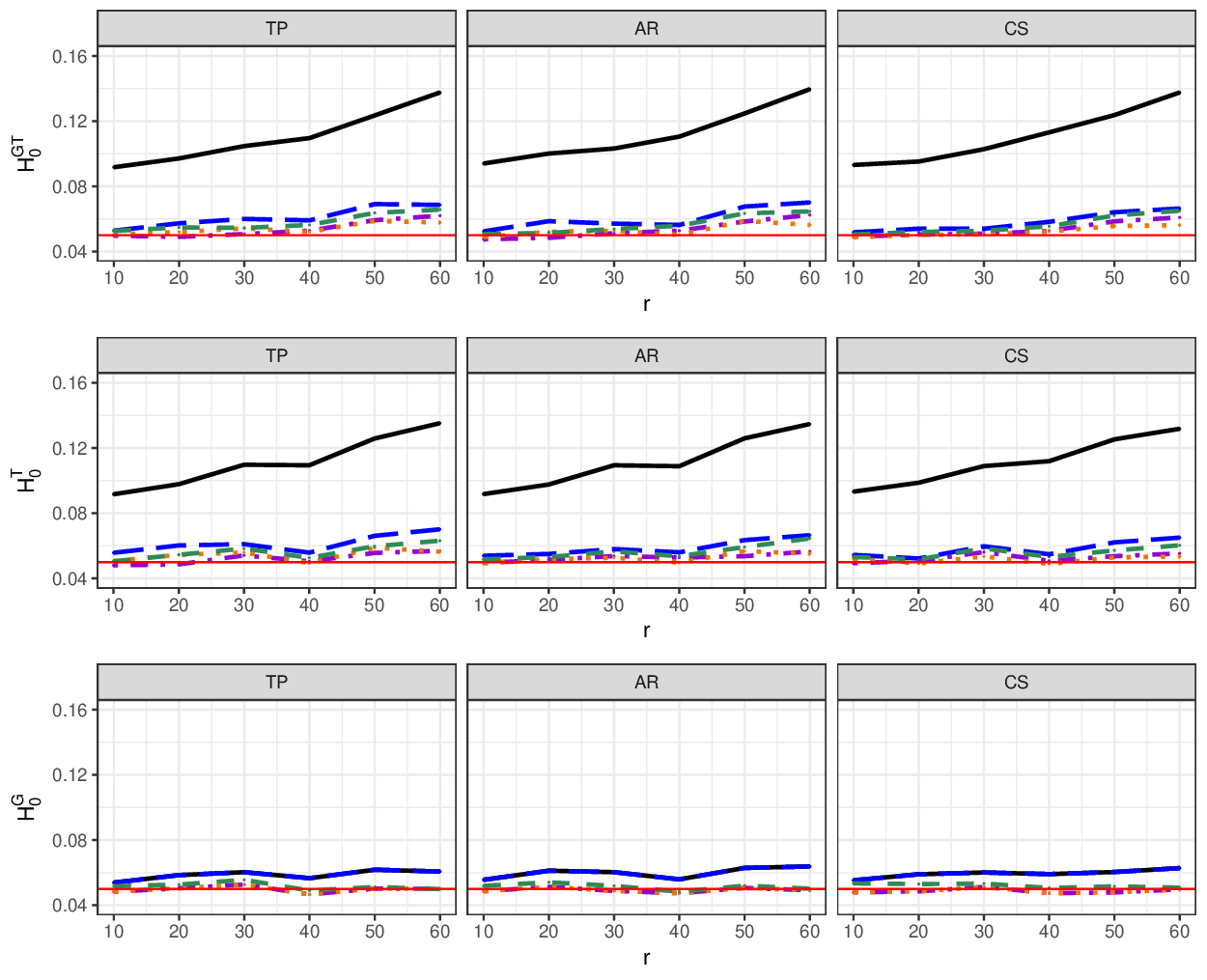}	
	\end{center}
	\caption{	Type-I error simulation results ($\alpha=0.05$) of the tests $T_W$ $({\color{black}\textbf{\textendash\textendash\textendash}})$, $T_A$ $({\color{blue} \textbf{ \textendash\textendash \quad \textendash\textendash}})$, 	$T_W^*$ $({\color{violet}\textbf{--}\boldsymbol{\cdot}\textbf{--}})$, $T_{A}^*$ $({\color{brown}\boldsymbol{\cdots}})$, and $T_{M}^*$ $({\color{ao(english)} \textbf{-- - --} })$ for double exponential distribution under different covariance structures with sample sizes $(n_1,n_2)=(15,15)$  and $d=4$ for varying percentages of MCAR data  $r\in\{10\%,20\%,30\%,40\%,50\%,60\%\}$.}
	\label{figInMissingrateDEXP}	
\end{figure}

\begin{figure}[h!]
	\begin{center}
		\includegraphics[scale=0.7, clip=true,]{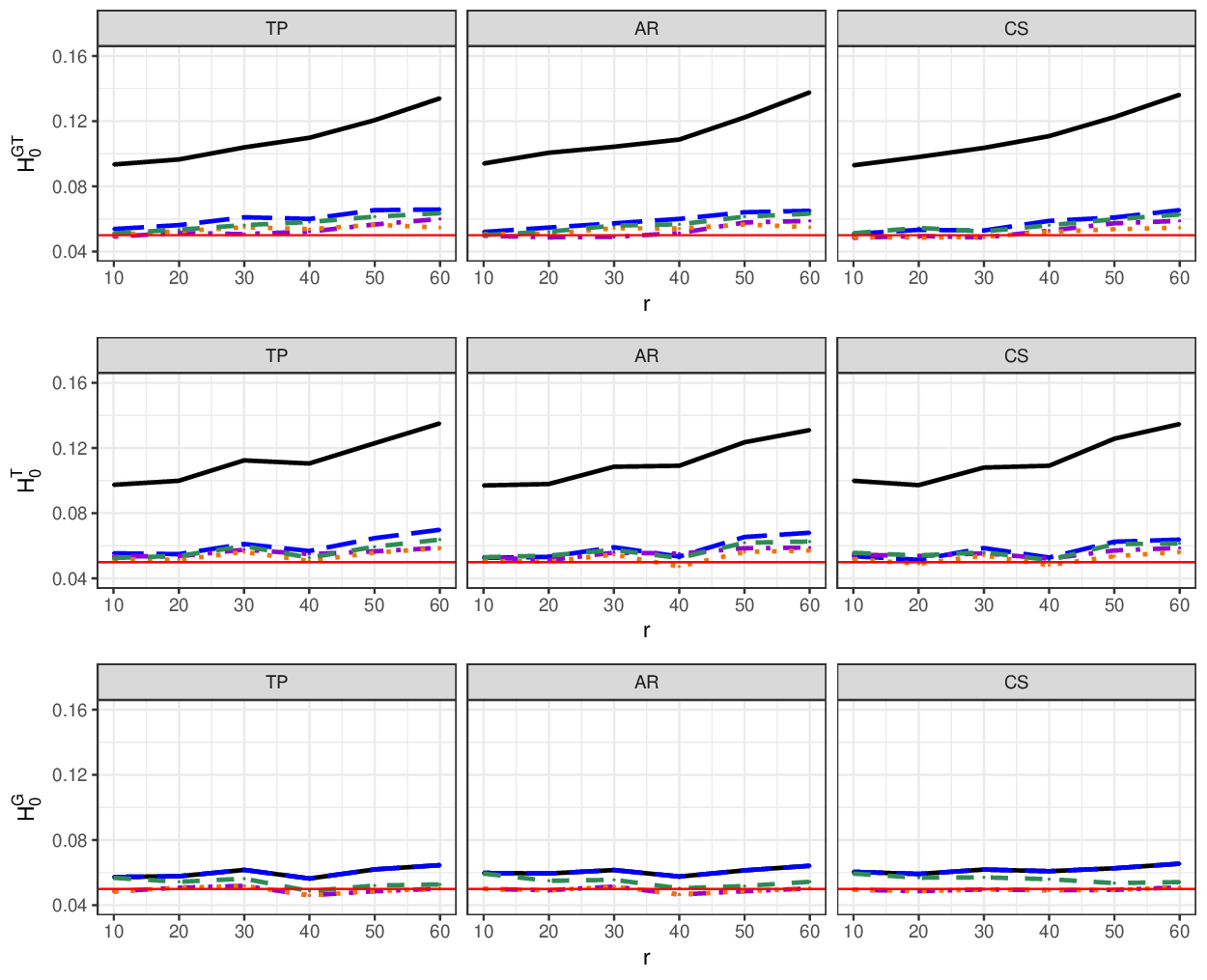}	
	\end{center}
	\caption{Type-I error simulation results ($\alpha=0.05$) of the tests $T_W$ $({\color{black}\textbf{\textendash\textendash\textendash}})$, $T_A$ $({\color{blue} \textbf{ \textendash\textendash \quad \textendash\textendash}})$, 	$T_W^*$ $({\color{violet}\textbf{--}\boldsymbol{\cdot}\textbf{--}})$, $T_{A}^*$ $({\color{brown}\boldsymbol{\cdots}})$, and $T_{M}^*$ $({\color{ao(english)} \textbf{-- - --} })$ for lognormal distribution under different covariance structures with sample sizes $(n_1,n_2)=(15,15)$  and $d=4$ for varying percentages of MCAR data  $r\in\{10\%,20\%,30\%,40\%,50\%,60\%\}$.}
	\label{figInMissingrateLogn}	
\end{figure}

 Further, type-I error control results for detecting the effect of increasing missing rates under MCAR covering missingness in observations ranging from $10\%$ to $60\%$ are summarized in Figures \ref{figInMissingrateDEXP} and \ref{figInMissingrateLogn} for the double exponential and lognormal distribution, respectively. \\
 
\begin{figure}[h!]
	\begin{center}
		\includegraphics[scale=0.7]{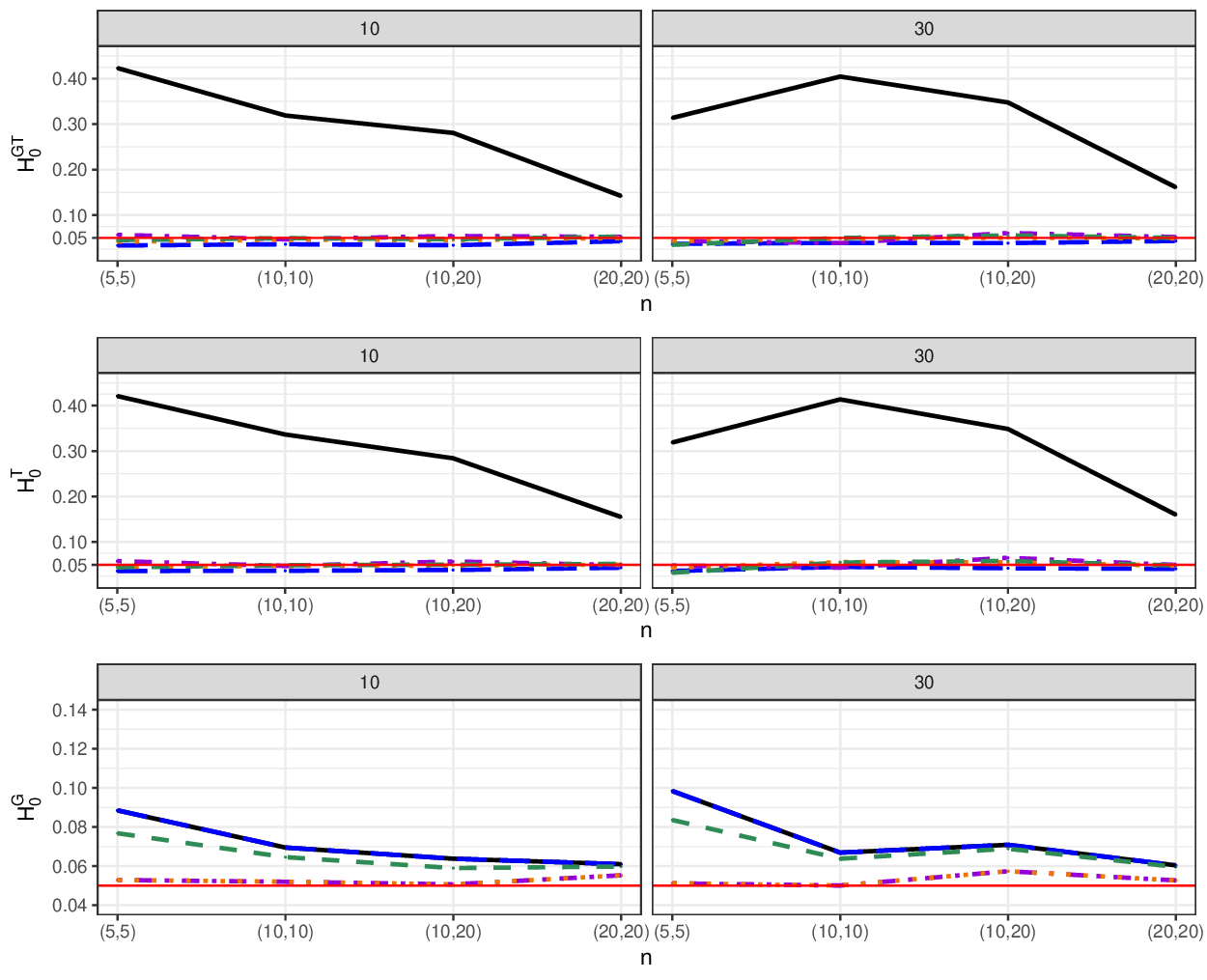}	
	\end{center}

	\caption{	Type-I error simulation results ($\alpha=0.05$) of the tests $T_W$ $({\color{black}\textbf{\textendash\textendash\textendash}})$, $T_A$ $({\color{blue} \textbf{ \textendash\textendash \quad \textendash\textendash}})$, 	$T_W^*$ $({\color{violet}\textbf{--}\boldsymbol{\cdot}\textbf{--}})$, $T_{A}^*$ $({\color{brown}\boldsymbol{\cdots}})$, and $T_{M}^*$ $({\color{ao(english)} \textbf{-- - --} })$ for ordinal data  under MCAR framework with sample sizes $(n_1,n_2)=\{(5,5),(10,10),(10,20),(20,20)\}$  and $d=8$.}
	\label{fig:OrdinalMCARt8}	
\end{figure}

\begin{figure}[h!]
	\begin{center}
		\includegraphics[scale=0.7]{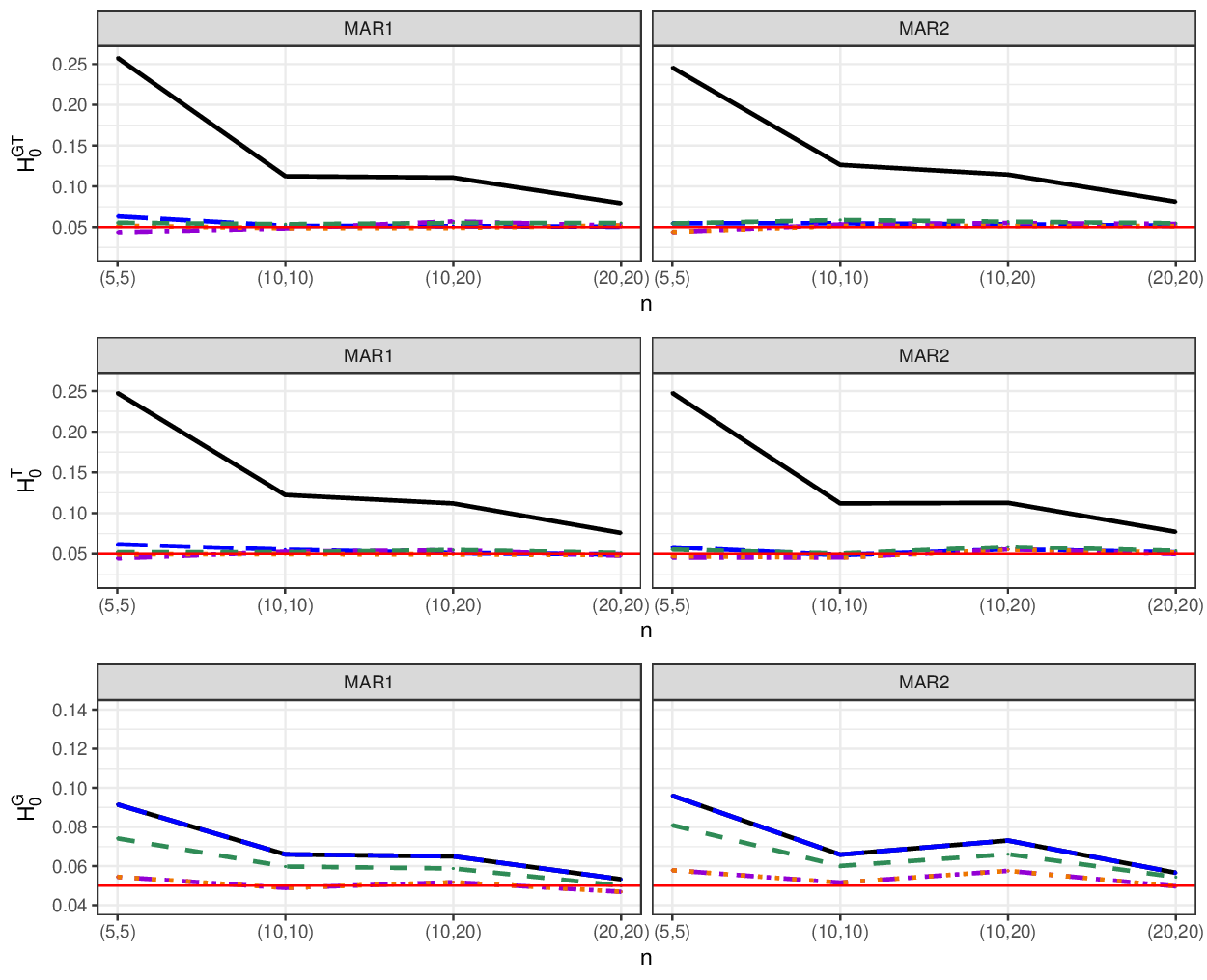}	
	\end{center}

	\caption{	Type-I error simulation results ($\alpha=0.05$) of the tests $T_W$ $({\color{black}\textbf{\textendash\textendash\textendash}})$, $T_A$ $({\color{blue} \textbf{ \textendash\textendash \quad \textendash\textendash}})$, 	$T_W^*$ $({\color{violet}\textbf{--}\boldsymbol{\cdot}\textbf{--}})$, $T_{A}^*$ $({\color{brown}\boldsymbol{\cdots}})$, and $T_{M}^*$ $({\color{ao(english)} \textbf{-- - --} })$ for ordinal data  under MAR framework with sample sizes $(n_1,n_2)=\{(5,5),(10,10),(10,20),(20,20)\}$  and $d=4$.}
	\label{fig:OrdinalMARt4}	
\end{figure}

\begin{figure}[h!]
	\begin{center}
		\includegraphics[scale=0.7]{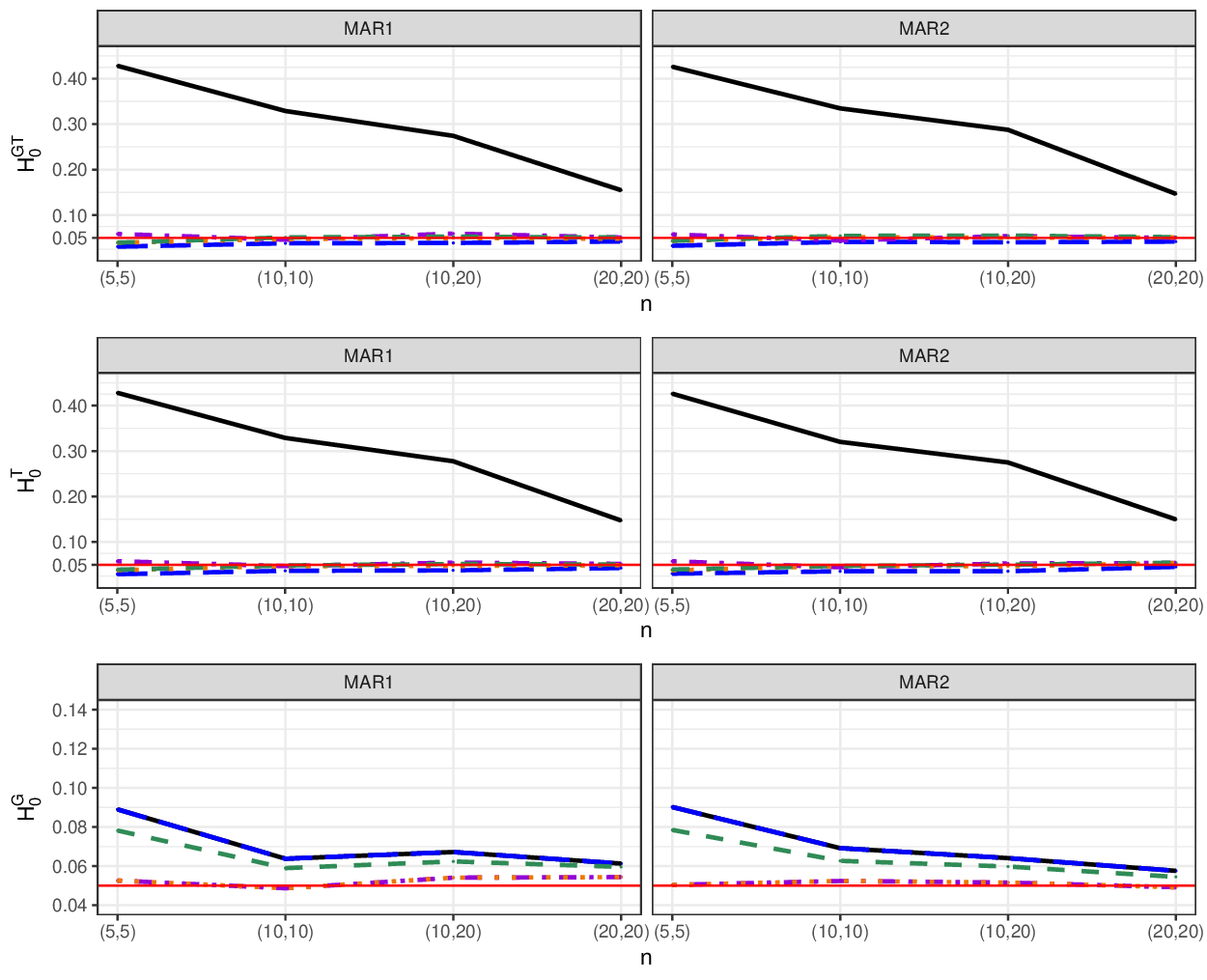}	
	\end{center}

	\caption{	Type-I error simulation results ($\alpha=0.05$) of the tests $T_W$ $({\color{black}\textbf{\textendash\textendash\textendash}})$, $T_A$ $({\color{blue} \textbf{ \textendash\textendash \quad \textendash\textendash}})$, 	$T_W^*$ $({\color{violet}\textbf{--}\boldsymbol{\cdot}\textbf{--}})$, $T_{A}^*$ $({\color{brown}\boldsymbol{\cdots}})$, and $T_{M}^*$ $({\color{ao(english)} \textbf{-- - --} })$ for ordinal data  under MAR framework with sample sizes $(n_1,n_2)=\{(5,5),(10,10),(10,20),(20,20)\}$  and $d=8$.}
	\label{fig:OrdinalMARt8}	
\end{figure}

 Furthermore, the type-I error results for ordinal data under MCAR framework for $d=8$ time points are presented in Figure \ref{fig:OrdinalMCARt8}. And,  the  results for ordinal data under MAR framework are summarized in Figure \ref{fig:OrdinalMARt4} ($d=4$ time points) and Figure \ref{fig:OrdinalMARt8} ($d=8$).\\

\begin{figure}[h!]
	\begin{center}
		\includegraphics[scale=0.7]{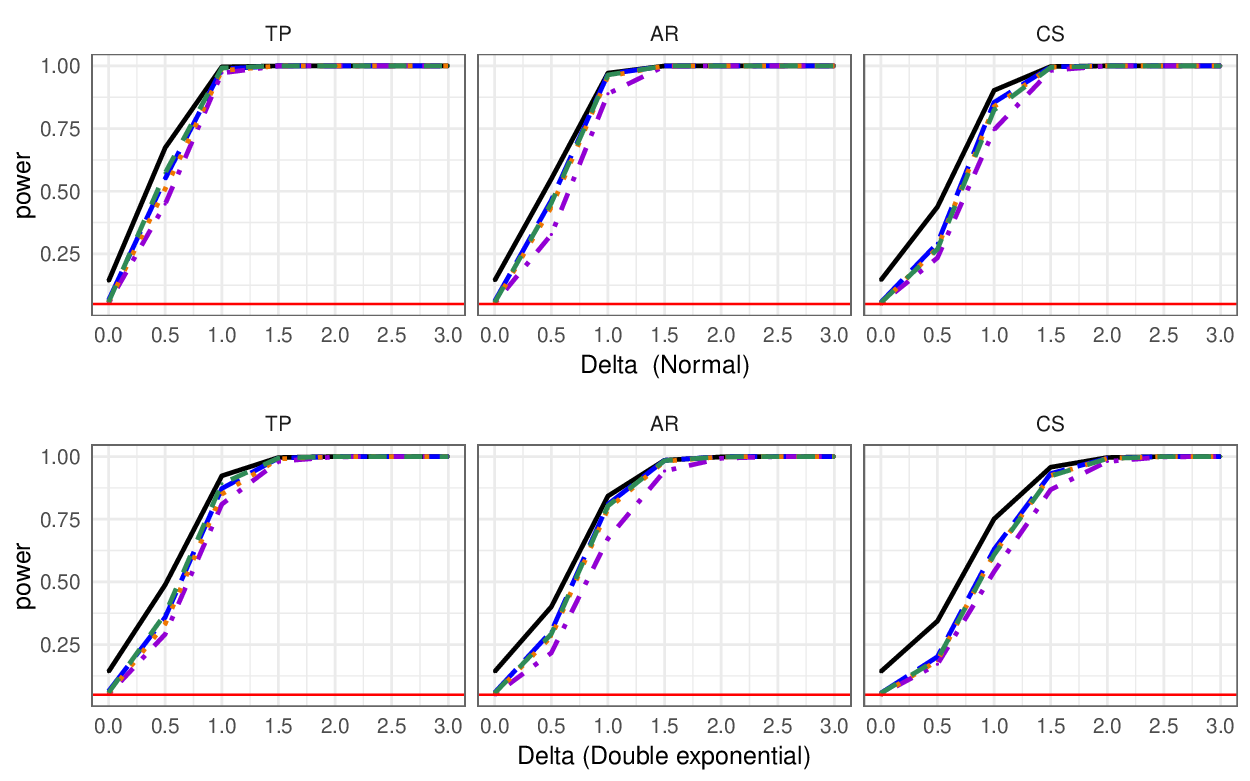}	
	\end{center}
	\caption{Power simulation results of the tests $T_W$ $({\color{black}\textbf{\textendash\textendash\textendash}})$, $T_A$ $({\color{blue} \textbf{ \textendash\textendash \quad \textendash\textendash}})$ , 	$T_W^*$ $({\color{violet}\textbf{--}\boldsymbol{\cdot}\textbf{--}})$, $T_{A}^*$ $({\color{brown}\boldsymbol{\cdots}})$, and $T_{M}^*$ $({\color{ao(english)} \textbf{-- - --} })$ under different covariance structures with sample size $n=15$  and $d=4$ under alternative $1$, for MCAR data with  missing rate $r=10\%$ with observations generated from a Normal (upper row) and a Double exponential (bottom row) distribution, respectively.}
	\label{fig:PowerNDMiss10HT}	
\end{figure}

\begin{figure}[h!]
	\begin{center}
		\includegraphics[scale=0.7]{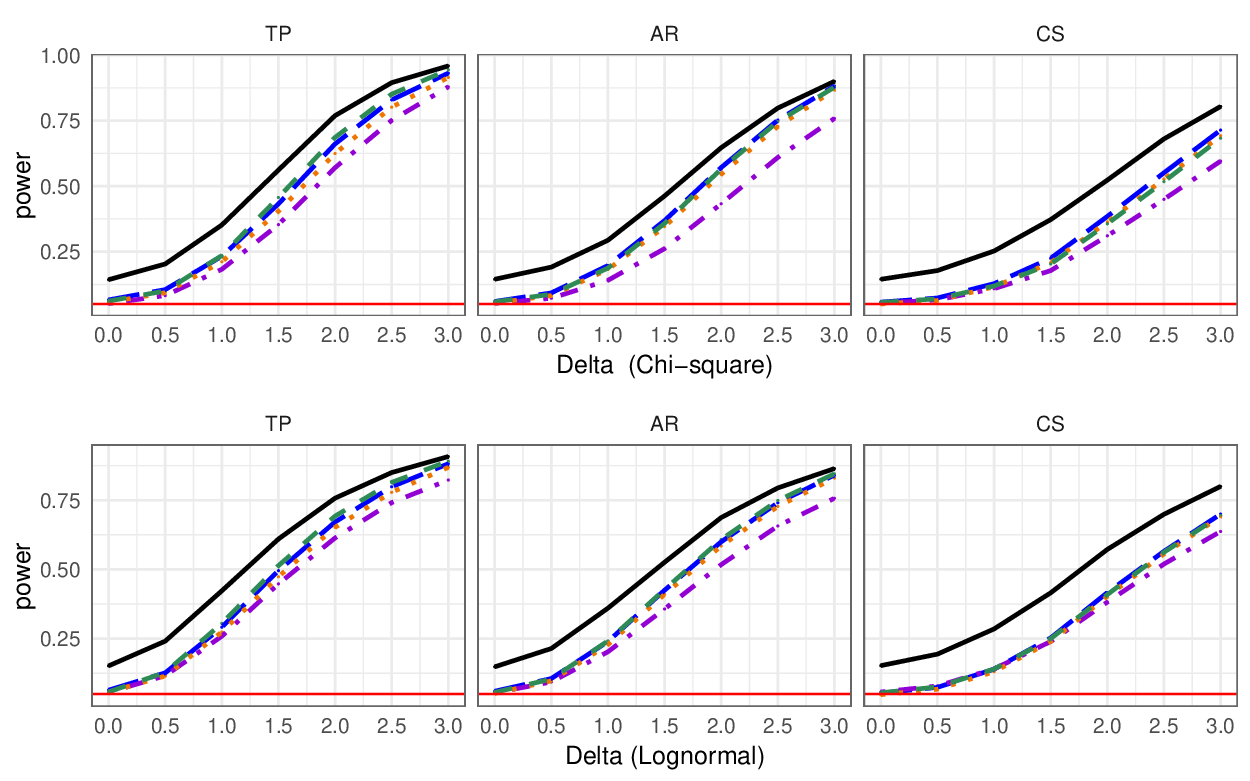}	
	\end{center}
	\caption{Power simulation results of the tests $T_W$ $({\color{black}\textbf{\textendash\textendash\textendash}})$, $T_A$ $({\color{blue} \textbf{ \textendash\textendash \quad \textendash\textendash}})$ , 	$T_W^*$ $({\color{violet}\textbf{--}\boldsymbol{\cdot}\textbf{--}})$, $T_{A}^*$ $({\color{brown}\boldsymbol{\cdots}})$, and $T_{M}^*$ $({\color{ao(english)} \textbf{-- - --} })$ under different covariance structures with sample size $n=15$  and $d=4$ under alternative $1$, for MCAR data with  missing rate $r=10\%$ with observations generated from a $\chi^2_{15}$ (upper row) and a Lognormal (bottom row) distribution, respectively.}
	\label{fig:PowerCLMiss10HT}	
\end{figure}

\begin{figure}[h!]
	\begin{center}
		\includegraphics[scale=0.7]{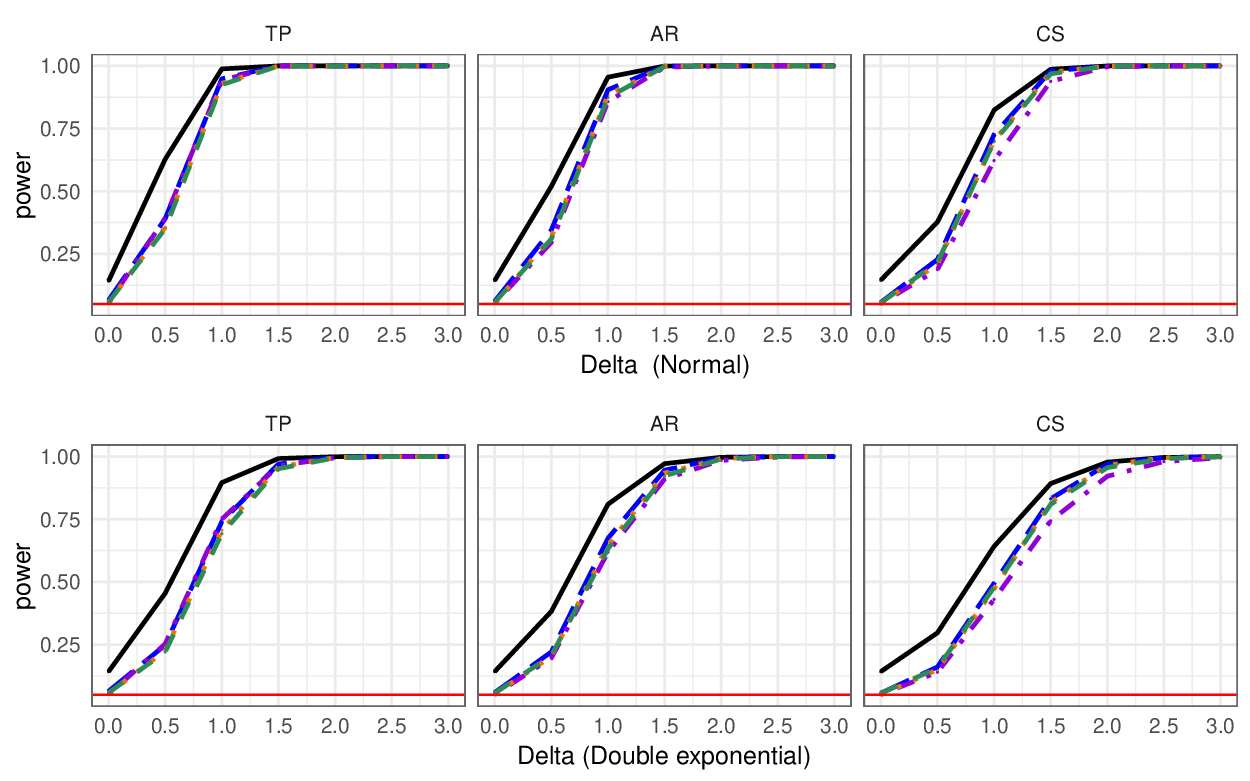}	
	\end{center}
	\caption{Power simulation results of the tests $T_W$ $({\color{black}\textbf{\textendash\textendash\textendash}})$, $T_A$ $({\color{blue} \textbf{ \textendash\textendash \quad \textendash\textendash}})$ , 	$T_W^*$ $({\color{violet}\textbf{--}\boldsymbol{\cdot}\textbf{--}})$, $T_{A}^*$ $({\color{brown}\boldsymbol{\cdots}})$, and $T_{M}^*$ $({\color{ao(english)} \textbf{-- - --} })$ under different covariance structures with sample size $n=15$  and $d=4$ under alternative $2$, for MCAR data with  missing rate $r=10\%$ with observations generated from a Normal (upper row) and a Double exponential (bottom row) distribution, respectively.}
	\label{figPowerNDMiss10HTalt2}	
\end{figure}

\begin{figure}[h!]
	\begin{center}
		\includegraphics[scale=0.7]{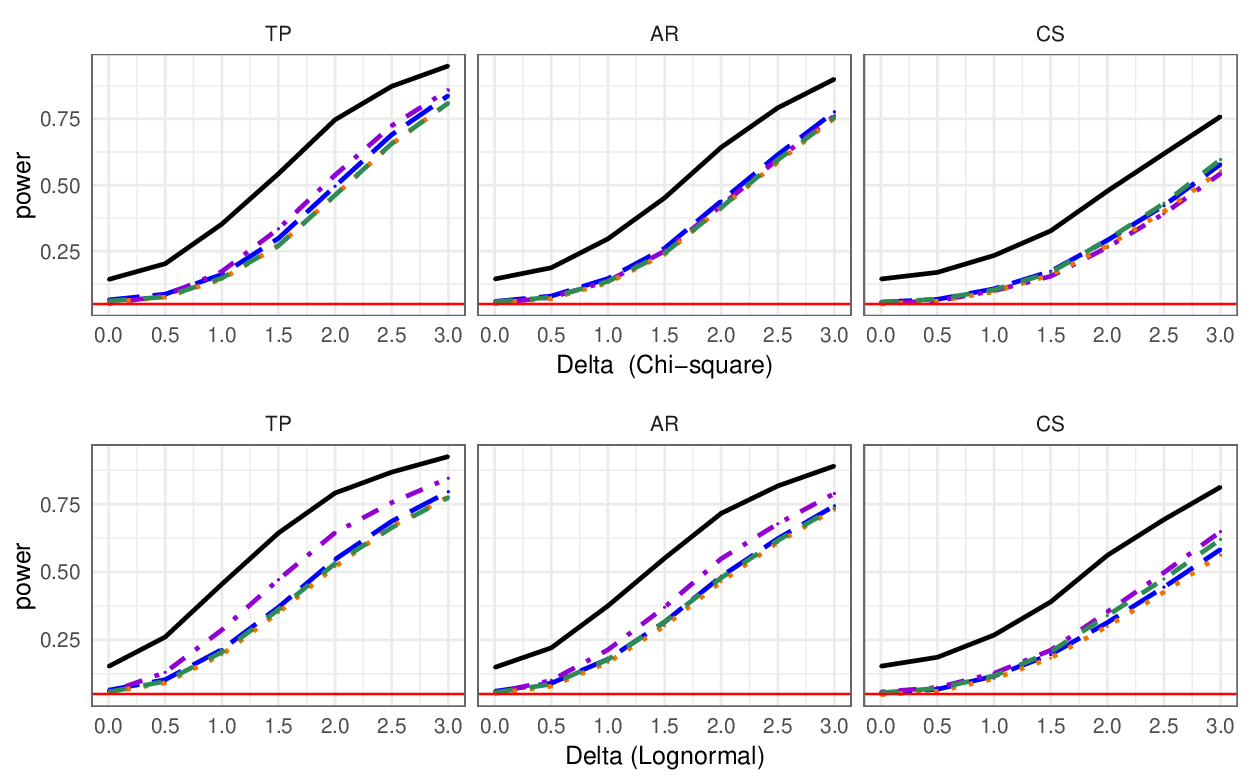}	
	\end{center}
	\caption{Power simulation results of the tests $T_W$ $({\color{black}\textbf{\textendash\textendash\textendash}})$, $T_A$ $({\color{blue} \textbf{ \textendash\textendash \quad \textendash\textendash}})$ , 	$T_W^*$ $({\color{violet}\textbf{--}\boldsymbol{\cdot}\textbf{--}})$, $T_{A}^*$ $({\color{brown}\boldsymbol{\cdots}})$, and $T_{M}^*$ $({\color{ao(english)} \textbf{-- - --} })$ under different covariance structures with sample size $n=15$  and $d=4$ under alternative $2$, for MCAR data with missing rate $r=10\%$ with observations generated from a $\chi^2_{15}$ (upper row) and a Lognormal (bottom row) distribution, respectively.}
	\label{figPowerCLMiss10HTalt2}	
\end{figure}

\begin{figure}[h!]
	\begin{center}
		\includegraphics[scale=0.7]{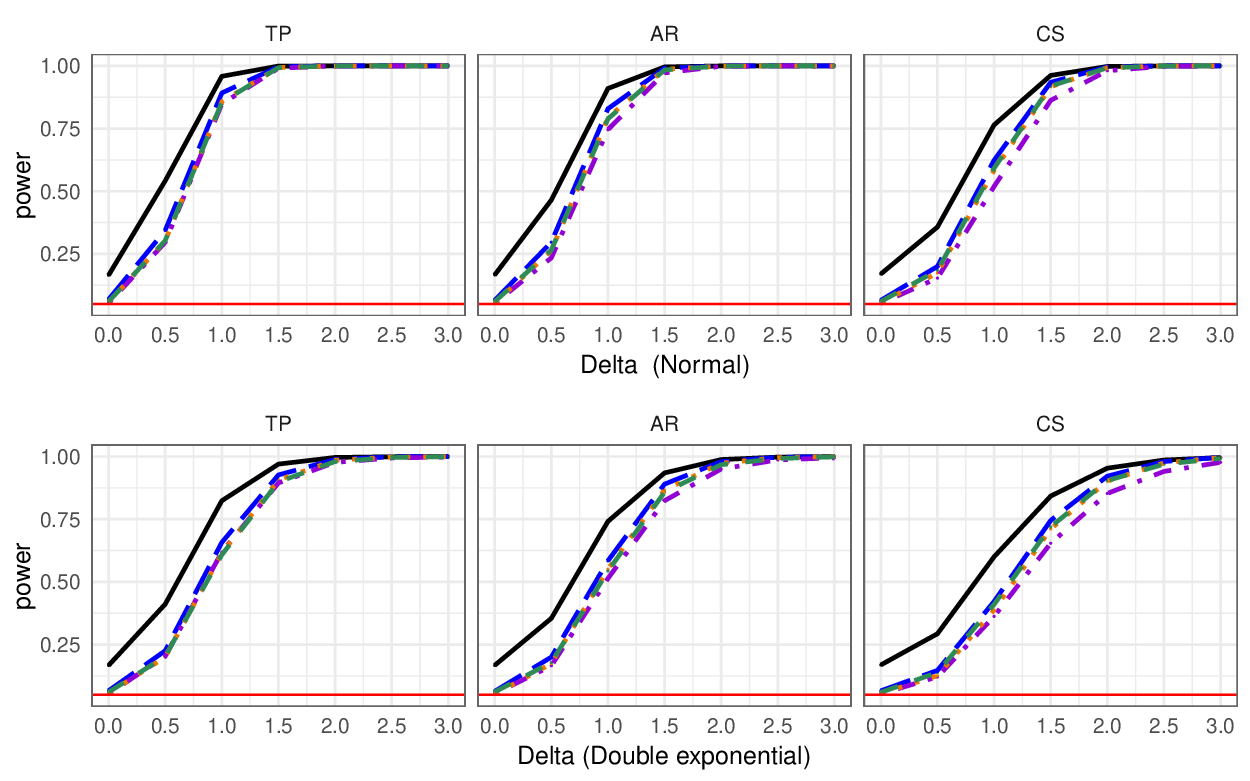}	
	\end{center}
	\caption{Power simulation results of the tests $T_W$ $({\color{black}\textbf{\textendash\textendash\textendash}})$, $T_A$ $({\color{blue} \textbf{ \textendash\textendash \quad \textendash\textendash}})$ , 	$T_W^*$ $({\color{violet}\textbf{--}\boldsymbol{\cdot}\textbf{--}})$, $T_{A}^*$ $({\color{brown}\boldsymbol{\cdots}})$, and $T_{M}^*$ $({\color{ao(english)} \textbf{-- - --} })$ under different covariance structures with sample size $n=15$  and $d=4$ under alternative $2$, for MCAR data with  missing rate $r=30\%$ with observations generated from a Normal (upper row) and a Double exponential (bottom row) distribution, respectively.}
	\label{fig:PowerNDMiss30HTalt2}	
\end{figure}

\begin{figure}[h!]
	\begin{center}
		\includegraphics[scale=0.7]{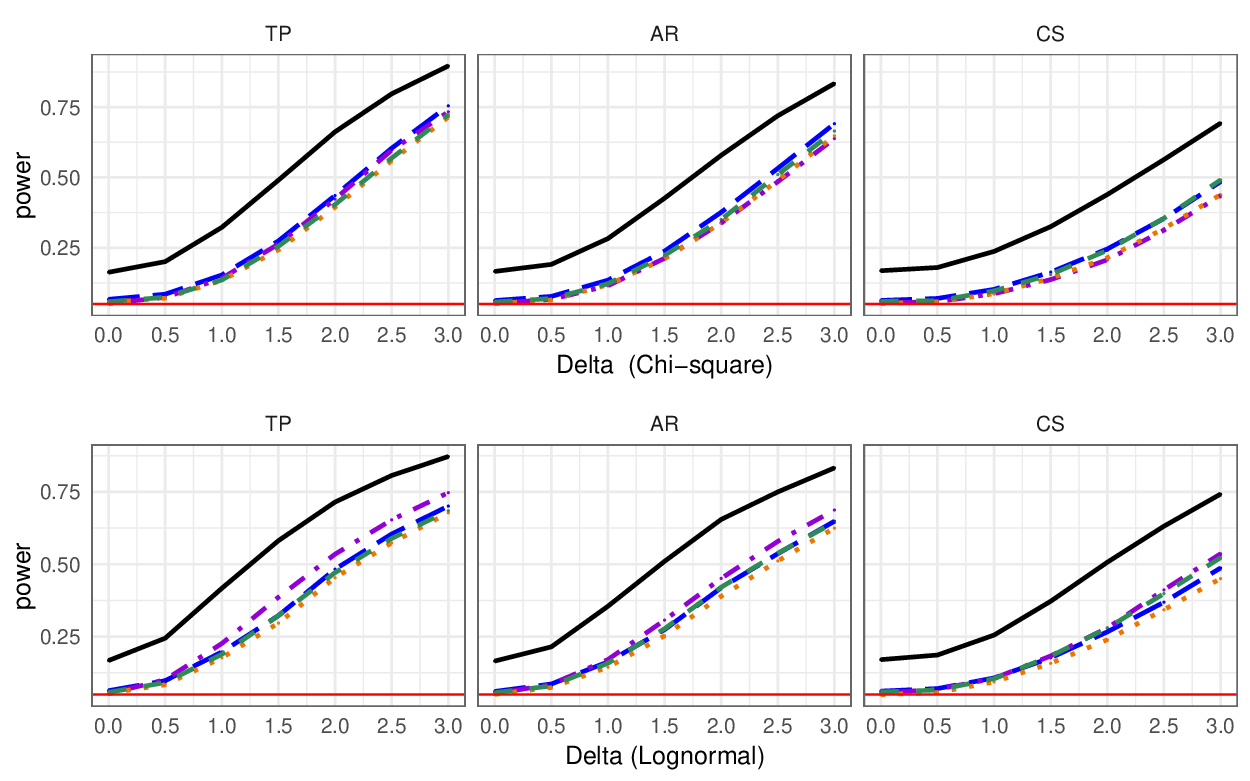}	
	\end{center}
	\caption{Power simulation results of the tests $T_W$ $({\color{black}\textbf{\textendash\textendash\textendash}})$, $T_A$ $({\color{blue} \textbf{ \textendash\textendash \quad \textendash\textendash}})$ , 	$T_W^*$ $({\color{violet}\textbf{--}\boldsymbol{\cdot}\textbf{--}})$, $T_{A}^*$ $({\color{brown}\boldsymbol{\cdots}})$, and $T_{M}^*$ $({\color{ao(english)} \textbf{-- - --} })$ under different covariance structures with sample size $n=15$  and $d=4$ under alternative $2$, for MCAR data with missing rate $r=30\%$ with observations generated from a $\chi^2_{15}$ (upper row) and a Lognormal (bottom row) distribution, respectively.}
	\label{fig:PowerCLMiss30HTalt2}	
\end{figure}

\begin{figure}[h!]
	\begin{center}
		\includegraphics[scale=0.7]{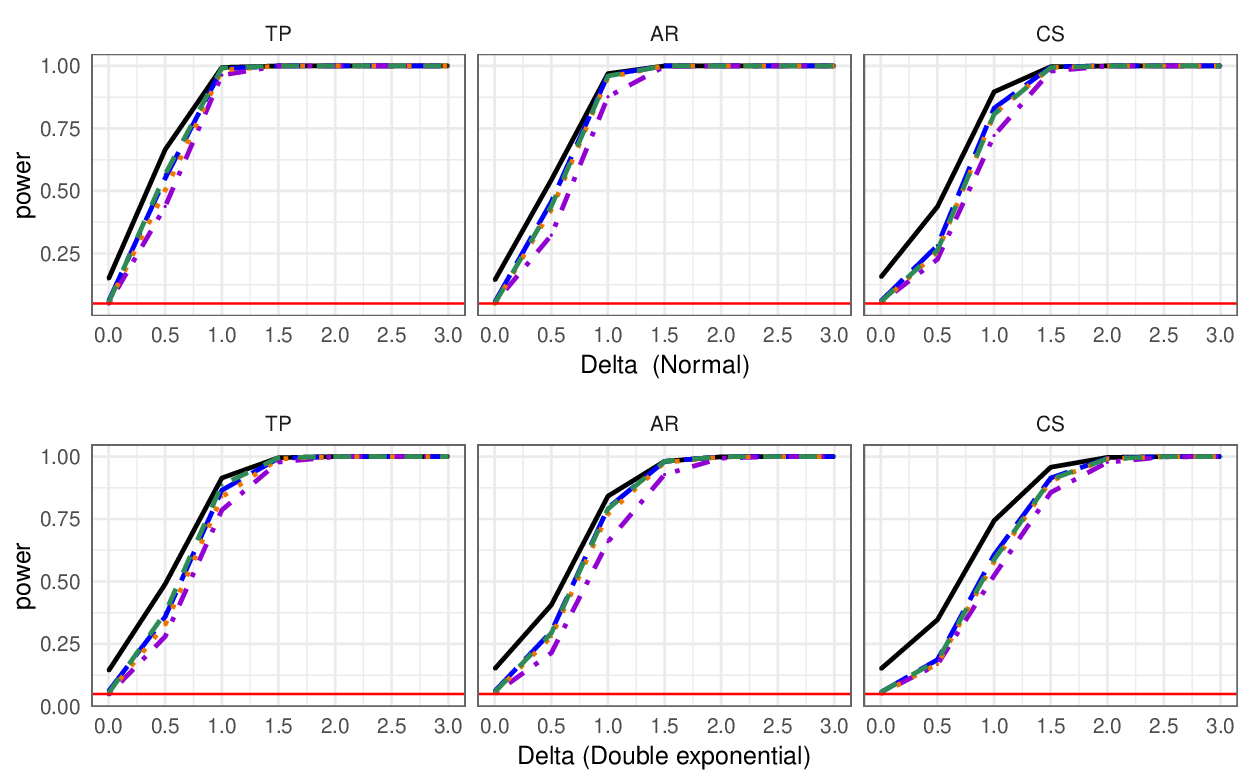}	
	\end{center}
	\caption{Power simulation results of the tests $T_W$ $({\color{black}\textbf{\textendash\textendash\textendash}})$, $T_A$ $({\color{blue} \textbf{ \textendash\textendash \quad \textendash\textendash}})$ , 	$T_W^*$ $({\color{violet}\textbf{--}\boldsymbol{\cdot}\textbf{--}})$, $T_{A}^*$ $({\color{brown}\boldsymbol{\cdots}})$, and $T_{M}^*$ $({\color{ao(english)} \textbf{-- - --} })$  under different covariance structures with sample size $n=15$ and $d=4$ under alternative $1$, under MAR1  with observations generated from a Normal (upper row) and a Double exponential (bottom row) distribution, respectively.}
	\label{fig:PowerNDMAR6HT}	
\end{figure}

\begin{figure}[h!]
	\begin{center}
		\includegraphics[scale=0.7]{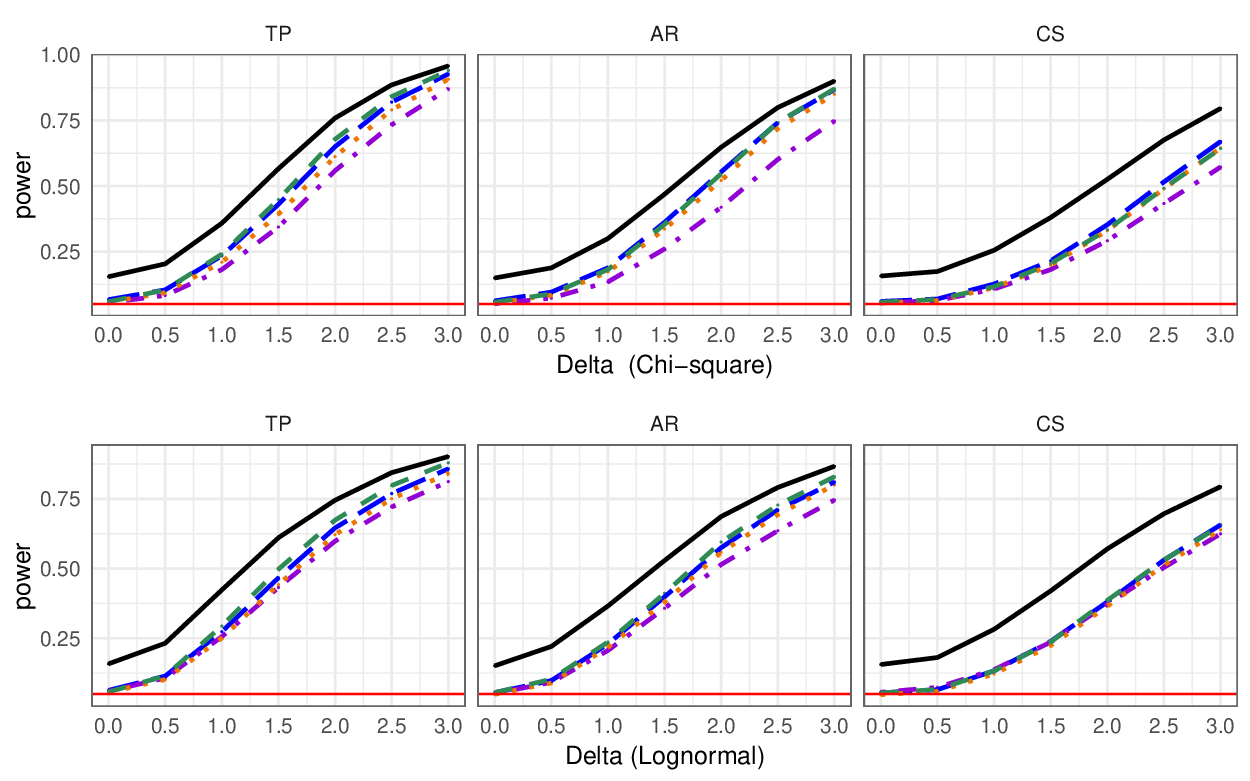}	
	\end{center}
	\caption{Power simulation results of the tests $T_W$ $({\color{black}\textbf{\textendash\textendash\textendash}})$, $T_A$ $({\color{blue} \textbf{ \textendash\textendash \quad \textendash\textendash}})$ , 	$T_W^*$ $({\color{violet}\textbf{--}\boldsymbol{\cdot}\textbf{--}})$, $T_{A}^*$ $({\color{brown}\boldsymbol{\cdots}})$, and $T_{M}^*$ $({\color{ao(english)} \textbf{-- - --} })$ under different covariance structures with sample size $n=15$ and $d=4$ under alternative $1$, under MAR1  with observations generated from a $\chi^2_{15}$ (upper row) and a Lognormal (bottom row) distribution, respectively.}
	\label{fig:PowerCLMAR6HT}	
\end{figure}

\begin{figure}[h!]
	\begin{center}
		\includegraphics[scale=0.7]{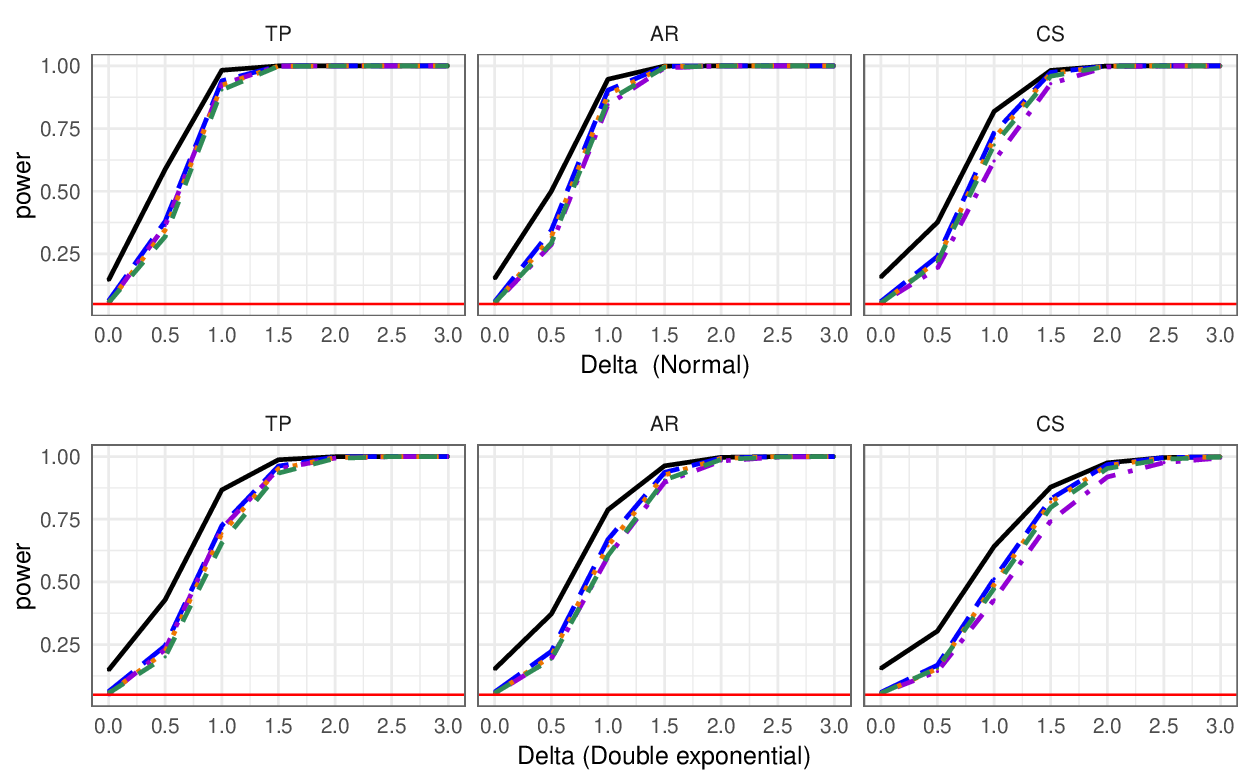}	
	\end{center}
	\caption{Power simulation results of the tests $T_W$ $({\color{black}\textbf{\textendash\textendash\textendash}})$, $T_A$ $({\color{blue} \textbf{ \textendash\textendash \quad \textendash\textendash}})$ , 	$T_W^*$ $({\color{violet}\textbf{--}\boldsymbol{\cdot}\textbf{--}})$, $T_{A}^*$ $({\color{brown}\boldsymbol{\cdots}})$, and $T_{M}^*$ $({\color{ao(english)} \textbf{-- - --} })$ under different covariance structures with sample size $n=15$ and $d=4$ under alternative $2$, under MAR1 with observations generated from a Normal (upper row) and a Double exponential (bottom row) distribution, respectively.}
	\label{fig:PowerNDMAR6HTalt2}	
\end{figure}

\begin{figure}[h!]
	\begin{center}
		\includegraphics[scale=0.7]{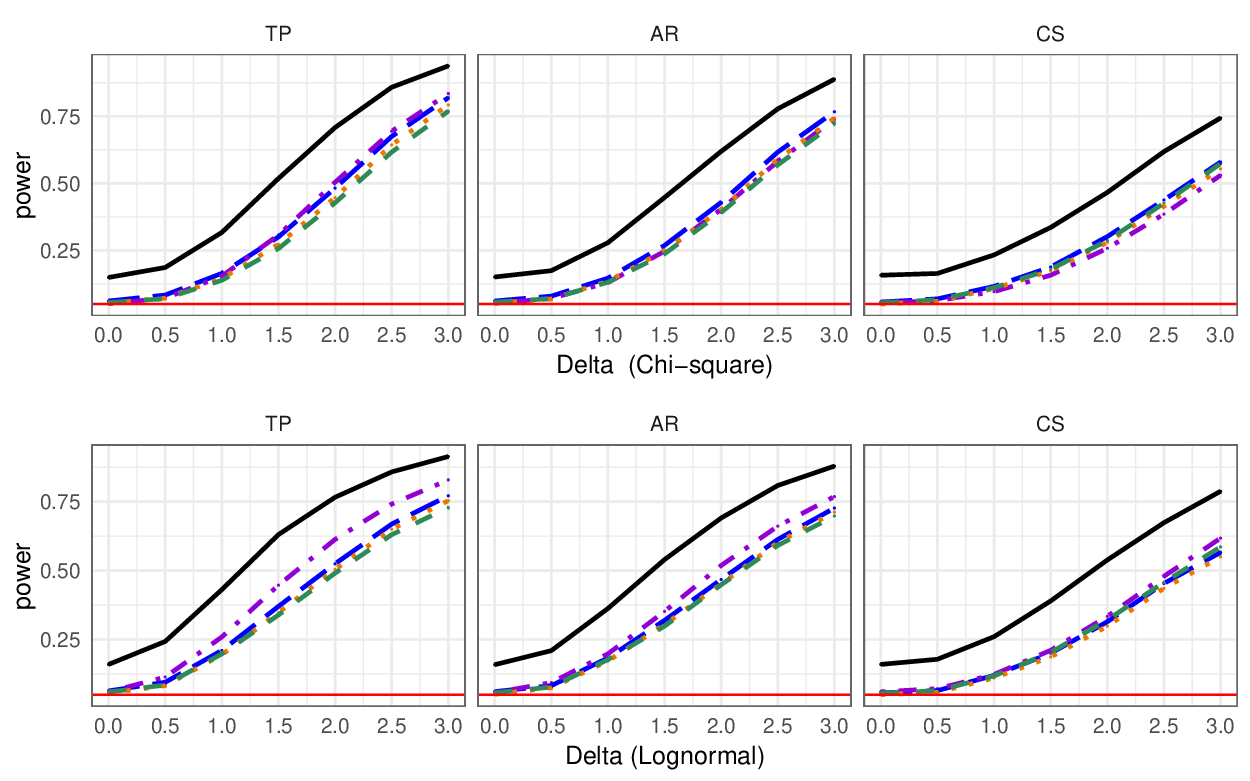}	
	\end{center}
	\caption{Power simulation results of the tests $T_W$ $({\color{black}\textbf{\textendash\textendash\textendash}})$, $T_A$ $({\color{blue} \textbf{ \textendash\textendash \quad \textendash\textendash}})$ , 	$T_W^*$ $({\color{violet}\textbf{--}\boldsymbol{\cdot}\textbf{--}})$, $T_{A}^*$ $({\color{brown}\boldsymbol{\cdots}})$, and $T_{M}^*$ $({\color{ao(english)} \textbf{-- - --} })$  under different covariance structures with sample size $n=15$ and $d=4$ under alternative $2$, under MAR1 with observations generated from a $\chi^2_{15}$ (upper row) and a Lognormal (bottom row) distribution, respectively.}
	\label{fig:PowerCLMAR6HTalt2}	
\end{figure}

\begin{figure}[h!]
	\begin{center}
		\includegraphics[scale=0.7]{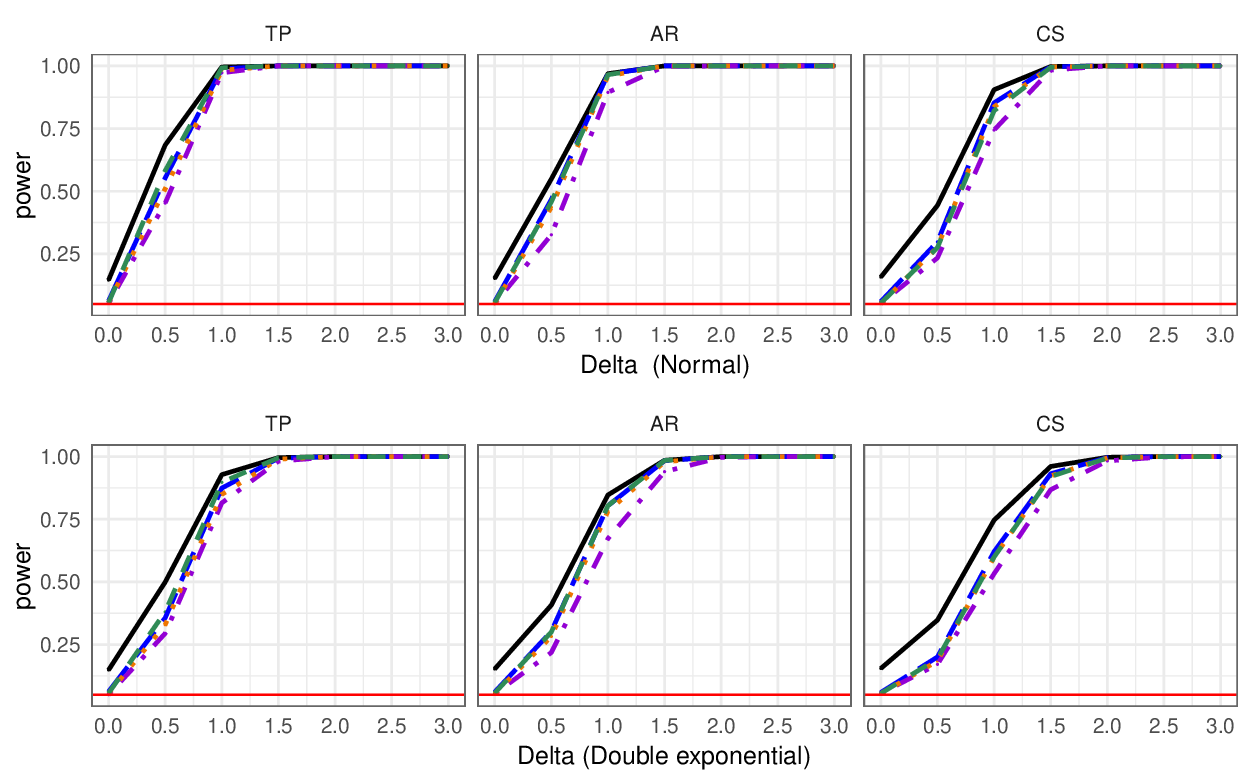}	
	\end{center}
	\caption{Power simulation results of the tests $T_W$ $({\color{black}\textbf{\textendash\textendash\textendash}})$, $T_A$ $({\color{blue} \textbf{ \textendash\textendash \quad \textendash\textendash}})$ , 	$T_W^*$ $({\color{violet}\textbf{--}\boldsymbol{\cdot}\textbf{--}})$, $T_{A}^*$ $({\color{brown}\boldsymbol{\cdots}})$, and $T_{M}^*$ $({\color{ao(english)} \textbf{-- - --} })$ under different covariance structures with sample size $n=15$ and $d=4$ under alternative $1$, under MAR2 with observations generated from a Normal (upper row) and a Double exponential (bottom row) distribution, respectively.}
	\label{fig:PowerNDMAR7HT}	
\end{figure}

\begin{figure}[h!]
	\begin{center}
		\includegraphics[scale=0.7]{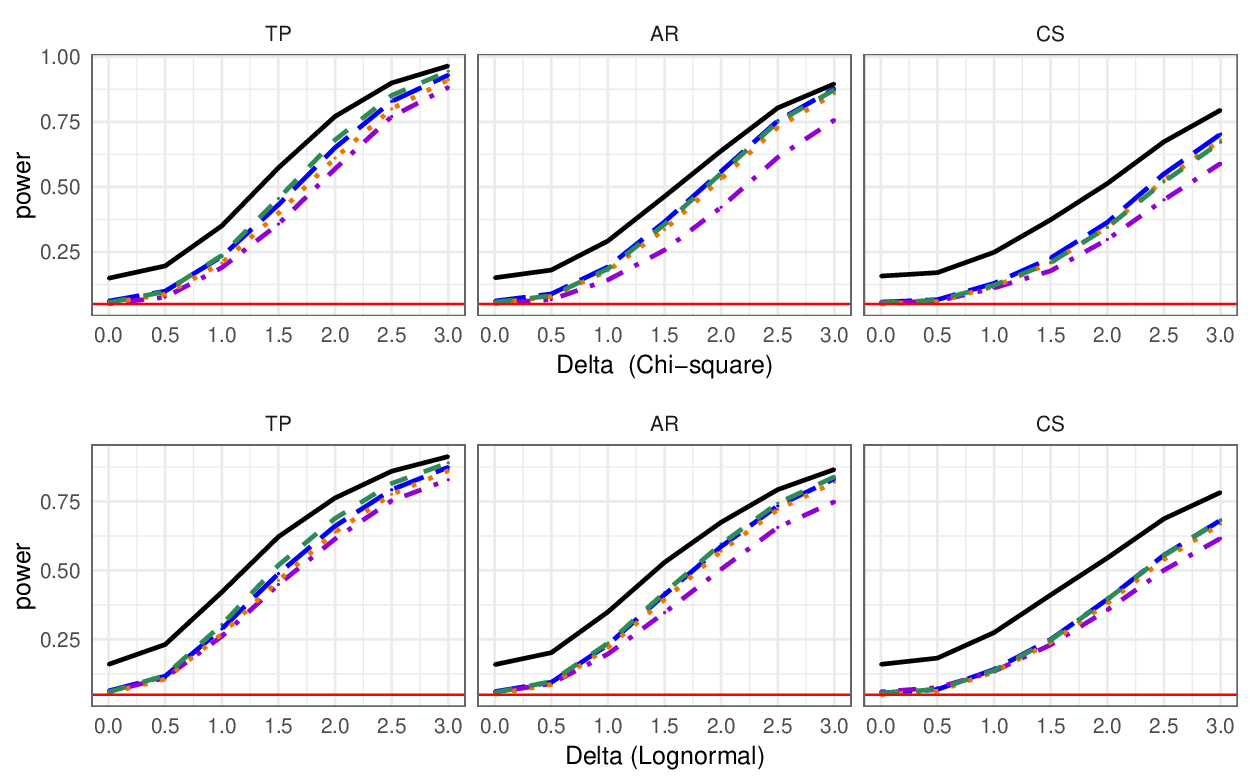}	
	\end{center}
	\caption{Power simulation results of the tests $T_W$ $({\color{black}\textbf{\textendash\textendash\textendash}})$, $T_A$ $({\color{blue} \textbf{ \textendash\textendash \quad \textendash\textendash}})$ , 	$T_W^*$ $({\color{violet}\textbf{--}\boldsymbol{\cdot}\textbf{--}})$, $T_{A}^*$ $({\color{brown}\boldsymbol{\cdots}})$, and $T_{M}^*$ $({\color{ao(english)} \textbf{-- - --} })$ under different covariance structures with sample size $n=15$ and $d=4$ under alternative $1$, under MAR2 with observations generated from a $\chi^2_{15}$ (upper row) and a Lognormal (bottom row) distribution, respectively.}
	\label{fig:PowerCLMAR7HT}	
\end{figure}

\begin{figure}[h!]
	\begin{center}
		\includegraphics[scale=0.7]{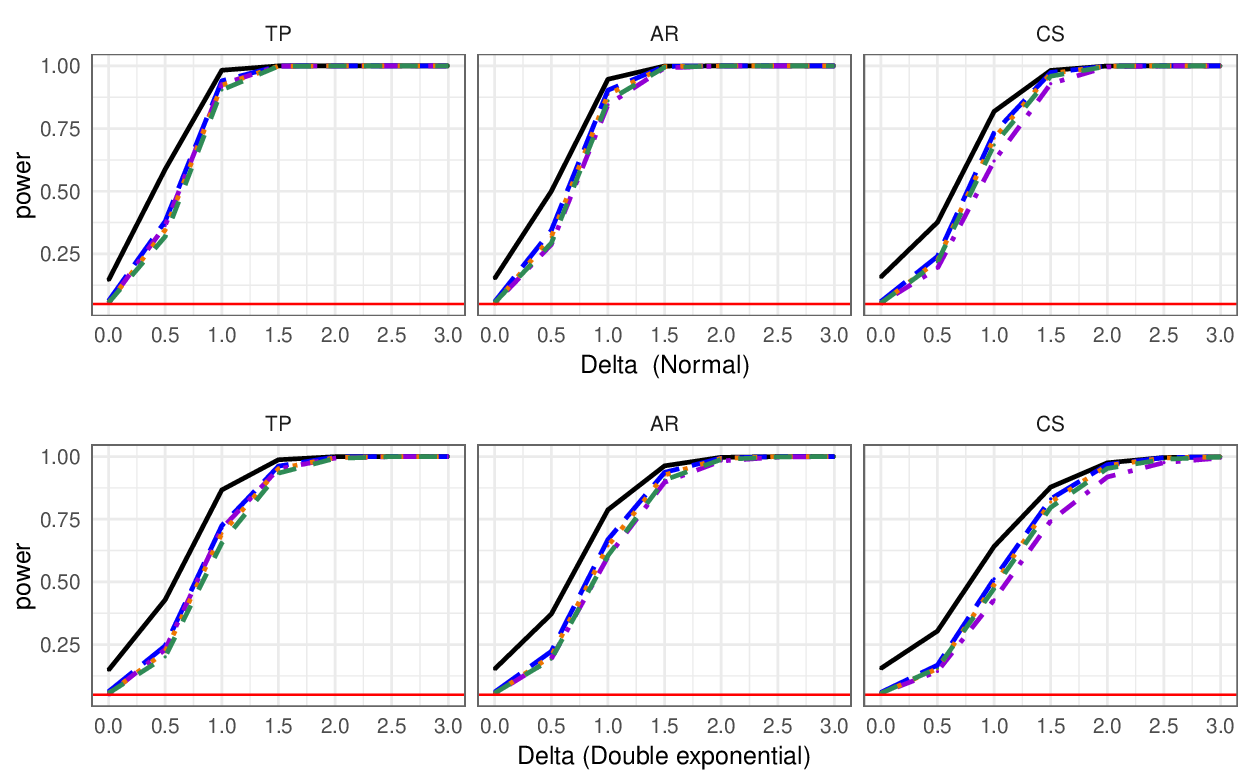}	
	\end{center}
	\caption{Power simulation results of the tests $T_W$ $({\color{black}\textbf{\textendash\textendash\textendash}})$, $T_A$ $({\color{blue} \textbf{ \textendash\textendash \quad \textendash\textendash}})$ , 	$T_W^*$ $({\color{violet}\textbf{--}\boldsymbol{\cdot}\textbf{--}})$, $T_{A}^*$ $({\color{brown}\boldsymbol{\cdots}})$, and $T_{M}^*$ $({\color{ao(english)} \textbf{-- - --} })$  under different covariance structures with sample size $n=15$ and $d=4$ under alternative $2$, under MAR2 with observations generated from a Normal (upper row) and a Double exponential (bottom row) distribution, respectively.}
	\label{fig:PowerNDMAR7HTalt2}	
\end{figure}

\begin{figure}[h!]
	\begin{center}
		\includegraphics[scale=0.7]{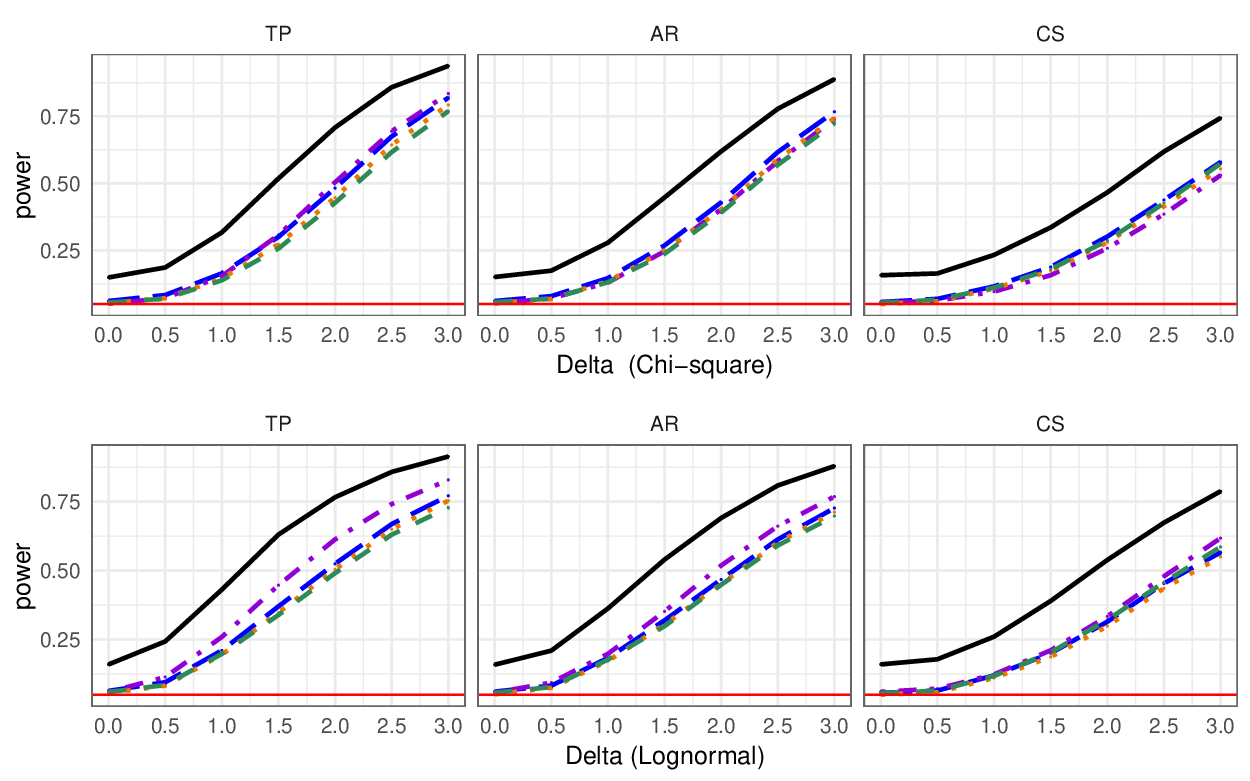}	
	\end{center}
	\caption{Power simulation results of the tests $T_W$ $({\color{black}\textbf{\textendash\textendash\textendash}})$, $T_A$ $({\color{blue} \textbf{ \textendash\textendash \quad \textendash\textendash}})$ , 	$T_W^*$ $({\color{violet}\textbf{--}\boldsymbol{\cdot}\textbf{--}})$, $T_{A}^*$ $({\color{brown}\boldsymbol{\cdots}})$, and $T_{M}^*$ $({\color{ao(english)} \textbf{-- - --} })$   under different covariance structures with sample size $n=15$ and $d=4$ under alternative $2$, under MAR2 with observations generated from a $\chi^2_{15}$ (upper row) and a Lognormal (bottom row) distribution, respectively.}
	\label{fig:PowerCLMAR7HTalt2}	
\end{figure}

 Power analysis results of the considered methods under the MCAR framework for several distributions, involving various covariance settings for detecting Alternative $1$ (as stated in the paper) are summarized in Figures \ref{fig:PowerNDMiss10HT} - \ref{fig:PowerCLMiss10HT} (missing rate $r=10\%$). The power results for investigating Alternative $2$ are displayed  in Figures \ref{figPowerNDMiss10HTalt2} - \ref{figPowerCLMiss10HTalt2}  (missing rate $r=10\%$) and Figures \ref{fig:PowerNDMiss30HTalt2} - \ref{fig:PowerCLMiss30HTalt2} ($r=30\%$).  The power analysis results under the MAR framework for detecting Alternative $1$ and $2$ are summarized in Figures \ref{fig:PowerNDMAR6HT} - \ref{fig:PowerCLMAR6HTalt2} (MAR1 scenario)  and Figures \ref{fig:PowerNDMAR7HT} - \ref{fig:PowerCLMAR7HTalt2} (MAR2 scenario). \\

\begin{table}[ht]
	\centering
	\caption{Simulation results for type-I error level ($\alpha=0.05$) for normal distribution, $d=4$ and different percentages $r$ of MCAR data. 
	}
\begin{tabular*}{0.9\linewidth}{c c  c  cc| ccc c  cc| ccc}
		\cmidrule(lr){1-14}
		& & &  \multicolumn{5}{c}{$r=10$}  && \multicolumn{5}{c}{$r=30$} \\ 
		\cmidrule(lr){4-8} \cmidrule(lr){10-14} 
		Hyp&	Cov &n  &\multicolumn{2}{c}{Classic}  & 
		\multicolumn{3}{c}{WBtstrp}  && \multicolumn{2}{c}{Classic}  & \multicolumn{3}{c}{WBtstrp} \\ \cmidrule(lr){4-5} \cmidrule(lr){6-8} 
		\cmidrule(lr){10-11} \cmidrule(lr){12-14}  
		&	&& $T_W$&$T_A$&$T_W^W$&$T_A^W$&$T_M^W$&&$T_W$&$T_A$&$T_W^W$&$T_A^W$&$T_M^W$\\[2pt]
		\cmidrule(lr){1-14}
$H_0^T$&AR&(5,5)&25.5&7&5.2&5&6.1&&33&8.6&5.9&5.2&6.2\\
&&(10,10)&12.2&5.3&4.9&4.8&4.9&&14.7&6.3&5.2&5.1&5.9\\
&&(10,20)&12.2&5.8&5.9&5.3&5.8&&14.6&6.4&6.8&5.7&6.8\\
&&(20,20)&7.6&5.1&4.8&4.9&5.1&&8.4&5.3&5&5.1&5.2\\
\cmidrule(lr){1-14}
&CS&(5,5)&25.9&6.9&5.3&4.9&6&&33.2&8.2&5.8&4.9&6.4\\
&&(10,10)&12.3&5.4&4.8&4.7&5.1&&14.7&5.8&5.3&4.9&5.7\\
&&(10,20)&12.5&5.8&5.9&5.2&6&&14.3&6.3&7&5.5&6.7\\
&&(20,20)&7.8&5.1&4.8&4.9&4.9&&8.4&5.3&5.2&5&5.3\\
\cmidrule(lr){1-14}
&TP&(5,5)&25.2&7.8&5&5.4&5.8&&33.1&9.1&5.7&5.4&6.3\\
&&(10,10)&11.8&5.7&4.7&5&5.1&&14.4&6.9&5.5&5.7&6.1\\
&&(10,20)&12.1&6&5.8&5.5&5.6&&14.7&6.6&6.8&6&6.9\\
&&(20,20)&7.6&5.1&4.7&5&5.1&&8.4&5.5&5&5.1&5.2\\

\cmidrule(lr){1-14}
$H_0^{GT}$&AR&(5,5)&26.1&6.9&5.4&4.9&6&&34.6&8.8&6.5&5.2&6.8\\
&&(10,10)&11.8&5.6&5&5.1&5.2&&15.7&6.1&5.6&5.1&5.9\\
&&(10,20)&11.9&5.8&5.7&5.4&5.8&&14.3&6.5&6.6&5.7&6.8\\
&&(20,20)&8.2&5.4&5&5.1&5.1&&8.6&5.6&5.2&5.3&5.4\\
\cmidrule(lr){1-14}
&CS&(5,5)&26&6.4&5&4.7&5.9&&34.5&8.6&6.4&5&6.6\\
&&(10,10)&12.2&5.1&4.8&4.7&4.9&&15&5.9&5.4&5&6\\
&&(10,20)&12&5.6&5.8&5.2&5.7&&14.5&6.2&6.5&5.5&6.6\\
&&(20,20)&8&5.3&5.1&5.1&5.3&&8.6&5.4&5.1&5.1&5.4\\
\cmidrule(lr){1-14}
&TP&(5,5)&26.1&7.3&5.2&5.2&5.9&&33.9&9.5&6.2&5.7&7.2\\
&&(10,10)&11.8&6.1&4.9&5.5&5.5&&15.5&6.5&5.4&5.3&5.9\\
&&(10,20)&11.9&6&5.7&5.5&5.9&&14.5&7&6.7&6.2&7.1\\
&&(20,20)&8.2&5.4&5&5&5.2&&8.8&5.7&5.2&5.4&5.6\\

		\cmidrule(lr){1-14}
		
	\end{tabular*}
	\label{TableTypErrorNormalt4MCAR}
\end{table}

\begin{table}[ht]
	\centering
	\caption{Simulation results for type-I error level ($\alpha=0.05$) for double exponential distribution, $d=4$ and different percentages $r$ of MCAR data.  
	}
	\begin{tabular*}{0.9\linewidth}{c c  c  cc| ccc c  cc| ccc}
		\cmidrule(lr){1-14}
		& & &  \multicolumn{5}{c}{$r=10$}  && \multicolumn{5}{c}{$r=30$} \\ 
		\cmidrule(lr){4-8} \cmidrule(lr){10-14} 
		Hyp&	Cov &n  &\multicolumn{2}{c}{Classic}  & 
		\multicolumn{3}{c}{WBtstrp}  && \multicolumn{2}{c}{Classic}  & \multicolumn{3}{c}{WBtstrp} \\ \cmidrule(lr){4-5} \cmidrule(lr){6-8} 
		\cmidrule(lr){10-11} \cmidrule(lr){12-14}  
		&	&& $T_W$&$T_A$&$T_W^W$&$T_A^W$&$T_M^W$&&$T_W$&$T_A$&$T_W^W$&$T_A^W$&$T_M^W$\\[2pt]
		\cmidrule(lr){1-14}
		
	$H_0^T$&AR&(5,5)&25.7&7.2&5.1&5.1&6&&32.6&8.4&6&5&6.2\\
	&&(10,10)&12.4&5.5&4.8&4.9&5.2&&14.5&6&5.4&5.2&6.1\\
	&&(10,20)&12&5.6&5.9&5.2&5.6&&14.3&6.4&6.5&5.6&6.6\\
	&&(20,20)&8&5.1&4.8&4.8&5&&8.6&5.4&5.1&5&5.4\\
	\cmidrule(lr){1-14}
	&CS&(5,5)&25.8&6.8&5.3&4.9&5.9&&33&8.3&6.1&5&6.1\\
	&&(10,10)&12.3&5.3&4.9&4.8&5.2&&14.5&5.8&5.2&5&5.5\\
	&&(10,20)&12.3&5.8&5.7&5.2&5.9&&14.3&6.4&6.7&5.7&6.6\\
	&&(20,20)&7.8&4.9&4.8&4.7&4.9&&8.5&5.2&5&4.9&5.2\\
	\cmidrule(lr){1-14}
	&TP&(5,5)&25.5&7.6&5.3&5.5&6.1&&32.7&8.9&6.3&5.3&6.2\\
	&&(10,10)&12&5.6&4.7&4.9&5.1&&14.4&6.5&5.4&5.5&6.1\\
	&&(10,20)&11.8&6&5.7&5.4&5.5&&14.5&6.6&6.6&5.8&6.8\\
	&&(20,20)&7.8&5.1&4.9&4.9&5&&8.7&5.6&5.2&5.3&5.2\\

	\cmidrule(lr){1-14}
	$H_0^{GT}$&AR&(5,5)&26.1&6.9&5.2&4.8&5.9&&34.2&8.8&6.2&5.3&6.9\\
	&&(10,10)&11.8&5.6&5&5.1&5.2&&15.2&6.2&5.5&5.3&5.9\\
	&&(10,20)&11.8&5.9&5.9&5.5&5.7&&14.2&6.3&6.4&5.5&6.8\\
	&&(20,20)&8.2&5.2&4.9&5.1&5.1&&8.7&5.6&5.1&5.2&5.5\\
	\cmidrule(lr){1-14}
	&CS&(5,5)&26&6.6&5&4.8&5.9&&34.4&8.6&6.4&5.1&6.4\\
	&&(10,10)&12&5.1&4.8&4.6&4.9&&14.8&5.9&5.3&4.8&5.8\\
	&&(10,20)&12.2&5.8&5.8&5.2&5.7&&14&6.1&6.5&5.4&6.4\\
	&&(20,20)&7.9&5.2&5&4.9&5.1&&8.5&5.3&5&5&5.3\\
	\cmidrule(lr){1-14}
	&TP&(5,5)&25.9&7.2&5.3&5.4&5.8&&33.6&9.2&6.2&5.7&7.1\\
	&&(10,10)&11.9&6.2&5&5.6&5.6&&15.6&6.6&5.6&5.6&6.1\\
	&&(10,20)&11.9&6.2&5.9&5.6&6.1&&14.3&6.7&6.6&6&6.9\\
	&&(20,20)&8&5.2&4.9&5&4.9&&8.7&5.7&5.2&5.3&5.5\\

		\cmidrule(lr){1-14}
		
	\end{tabular*}
	\label{TableTypErrorDexpt4MCAR}
\end{table}

\begin{table}[ht]
	\centering
	\caption{Simulation results for type-I error level ($\alpha=0.05$) for chi-square distribution, $d=4$ and different percentages $r$ of MCAR data.  
	}
	\begin{tabular*}{0.9\linewidth}{c c  c  cc| ccc c  cc| ccc}
		\cmidrule(lr){1-14}
		& & &  \multicolumn{5}{c}{$r=10$}  && \multicolumn{5}{c}{$r=30$} \\ 
		\cmidrule(lr){4-8} \cmidrule(lr){10-14} 
		Hyp&	Cov &n  &\multicolumn{2}{c}{Classic}  & 
		\multicolumn{3}{c}{WBtstrp}  && \multicolumn{2}{c}{Classic}  & \multicolumn{3}{c}{WBtstrp} \\ \cmidrule(lr){4-5} \cmidrule(lr){6-8} 
		\cmidrule(lr){10-11} \cmidrule(lr){12-14}  
		&	&& $T_W$&$T_A$&$T_W^W$&$T_A^W$&$T_M^W$&&$T_W$&$T_A$&$T_W^W$&$T_A^W$&$T_M^W$\\[2pt]
		\cmidrule(lr){1-14}

$H_0^T$&AR&(5,5)&25.5&7.1&5.1&5.1&5.9&&32.7&8.4&5.7&5.1&6\\
&&(10,10)&12&5.4&4.9&4.8&5&&14.4&6.2&5.3&5.2&5.9\\
&&(10,20)&12&5.8&5.6&5.3&5.8&&14.3&6.3&6.7&5.6&6.8\\
&&(20,20)&7.6&4.9&4.7&4.8&5&&8.7&5.4&4.9&5&5.3\\
\cmidrule(lr){1-14}
&CS&(5,5)&25.9&6.7&5.2&4.7&5.8&&33.1&8.4&5.8&5.1&6.4\\
&&(10,10)&11.9&5.2&4.8&4.7&5.1&&14.7&5.7&5.1&5&5.5\\
&&(10,20)&12.3&5.7&6&5.2&6&&14.3&6.4&7&5.6&6.8\\
&&(20,20)&7.9&5.1&5&5&5&&8.5&5.2&5.2&4.9&5.2\\
\cmidrule(lr){1-14}
&TP&(5,5)&25.2&7.6&5.1&5.3&5.8&&32.9&9&5.9&5.6&6.2\\
&&(10,10)&12&5.7&4.8&5.1&5.1&&14.4&6.4&5.4&5.4&6\\
&&(10,20)&11.8&6.1&5.5&5.5&5.7&&14.3&6.8&6.8&6&6.7\\
&&(20,20)&7.4&4.9&4.9&4.8&4.9&&8.6&5.7&4.9&5.2&5.2\\

\cmidrule(lr){1-14}
$H_0^{GT}$&AR&(5,5)&26.1&6.8&5&4.8&5.8&&34.3&8.9&6.1&5.3&6.7\\
&&(10,10)&11.7&5.7&4.9&5.1&5.2&&15&5.9&5.2&5&5.9\\
&&(10,20)&12.1&5.7&5.7&5.3&5.7&&14.2&6.7&6.7&5.9&6.9\\
&&(20,20)&8.3&5.2&5.1&5.1&5.1&&8.9&5.8&5.3&5.4&5.5\\
\cmidrule(lr){1-14}
&CS&(5,5)&25.9&6.4&5&4.8&5.8&&34.5&8.3&6&5&6.4\\
&&(10,10)&11.8&5.1&4.7&4.7&5&&14.8&5.9&5.4&4.9&5.7\\
&&(10,20)&12&5.5&5.7&5&5.5&&14.1&6.2&6.4&5.3&6.6\\
&&(20,20)&8&5.2&5.1&5&5.1&&8.8&5.3&5&5.1&5.3\\
\cmidrule(lr){1-14}
&TP&(5,5)&26&7.3&5.4&5.3&5.9&&33.8&9.3&6.4&5.7&6.9\\
&&(10,10)&12.1&6.2&5&5.4&5.6&&15.2&6.5&5.3&5.4&5.9\\
&&(10,20)&11.8&5.9&5.6&5.3&5.9&&14.3&7.1&6.5&6.1&7.1\\
&&(20,20)&8.2&5.2&5.1&4.9&5&&8.8&5.9&5.3&5.6&5.6\\
		\cmidrule(lr){1-14}
		
	\end{tabular*}
	\label{TableTypErrorChisqt4MCAR}
\end{table}

\begin{table}[ht]
	\centering
	\caption{Simulation results for type-I error level ($\alpha=0.05$) for lognormal distribution, $d=4$ and different percentages $r$ of MCAR data .  
	}
	\begin{tabular*}{0.9\linewidth}{c c  c  cc| ccc c  cc| ccc}
		\cmidrule(lr){1-14}
		& & &  \multicolumn{5}{c}{$r=10$}  && \multicolumn{5}{c}{$r=30$} \\ 
		\cmidrule(lr){4-8} \cmidrule(lr){10-14} 
		Hyp&	Cov &n  &\multicolumn{2}{c}{Classic}  & 
		\multicolumn{3}{c}{WBtstrp}  && \multicolumn{2}{c}{Classic}  & \multicolumn{3}{c}{WBtstrp} \\ \cmidrule(lr){4-5} \cmidrule(lr){6-8} 
		\cmidrule(lr){10-11} \cmidrule(lr){12-14}  
		&	&& $T_W$&$T_A$&$T_W^W$&$T_A^W$&$T_M^W$&&$T_W$&$T_A$&$T_W^W$&$T_A^W$&$T_M^W$\\[2pt]
		\cmidrule(lr){1-14}
$H_0^T$&AR&(5,5)&25.9&6.7&4.9&4.9&5.6&&32.3&8&5.6&4.8&5.9\\
&&(10,10)&12.3&5.1&4.9&4.7&5&&14.6&6&5.4&5&5.7\\
&&(10,20)&11.9&5.7&5.8&5.1&5.6&&14.1&6.6&6.6&5.9&6.7\\
&&(20,20)&8&4.9&5.3&4.8&5&&8.7&5.4&5.3&5.2&5.4\\
\cmidrule(lr){1-14}
&CS&(5,5)&26.2&6.3&5.2&4.4&5.4&&32.5&8&5.9&5.1&5.8\\
&&(10,10)&12.5&5&5.3&4.4&5.1&&14.3&5.6&5.3&4.8&5.2\\
&&(10,20)&12.5&5.3&6.2&5.1&5.7&&14.2&6.3&6.7&5.7&6.7\\
&&(20,20)&8.1&4.8&5.3&4.7&4.8&&8.9&5.2&5.3&4.9&5.2\\
\cmidrule(lr){1-14}
&TP&(5,5)&26.1&7.2&5.2&5&5.7&&32.8&8.5&5.9&5.2&6.1\\
&&(10,10)&12.8&5.4&5&4.9&5.2&&14.7&6.2&5.7&5.3&6\\
&&(10,20)&12.2&5.9&6.1&5.3&5.8&&14.5&6.6&6.8&5.8&6.9\\
&&(20,20)&8.4&5&5.4&4.8&5&&9&5.6&5.4&5.3&5.3\\

\cmidrule(lr){1-14}
$H_0^{GT}$&AR&(5,5)&25.6&6.6&4.4&4.9&5.8&&33.2&8.5&5.4&5.4&6.5\\
&&(10,10)&12&5.6&4.6&5.1&5.3&&14.7&5.7&5.4&4.9&5.7\\
&&(10,20)&11.5&5.4&5.6&4.9&5.3&&14&6.4&6.1&5.7&6.5\\
&&(20,20)&7.8&5.1&5.1&5&5&&8.6&5.7&5.1&5.4&5.4\\
\cmidrule(lr){1-14}
&CS&(5,5)&25.8&6.4&4.4&4.7&5.7&&33.1&8.2&5.6&5.1&6.3\\
&&(10,10)&11.8&5.2&4.5&4.8&5.2&&14.4&5.8&5.1&5&5.7\\
&&(10,20)&11.5&5.2&5.6&4.9&5.6&&13.7&6&6&5.2&6.2\\
&&(20,20)&8.1&5.2&5.1&5.1&5.2&&8.5&5.2&5&5&5.2\\
\cmidrule(lr){1-14}
&TP&(5,5)&25.2&7.2&4.5&5.2&5.7&&32.6&8.8&5.8&5.5&6.5\\
&&(10,10)&12&5.9&4.7&5.2&5.4&&14.8&6&5.4&5.1&5.8\\
&&(10,20)&11.5&5.6&5.7&5.1&5.4&&13.8&6.9&6.2&6.1&6.9\\
&&(20,20)&7.8&5.3&5.1&5.1&5.1&&8.9&5.7&5.2&5.4&5.4\\

		\cmidrule(lr){1-14}
		
	\end{tabular*}
	\label{TableTypErrorLognormt4MCAR}
\end{table}

\begin{table}[ht]
	\centering
	\caption{Simulation results for type-I error level ($\alpha=0.05$) for normal distribution, $d=8$ and different percentages $r$ of MCAR data.}
	\begin{tabular*}{0.9\linewidth}{c c  c  cc| ccc c  cc| ccc}
		\cmidrule(lr){1-14}
		& & &  \multicolumn{5}{c}{$r=10$}  && \multicolumn{5}{c}{$r=30$} \\ 
		\cmidrule(lr){4-8} \cmidrule(lr){10-14} 
		Hyp&	Cov &n  &\multicolumn{2}{c}{Classic}  & 
		\multicolumn{3}{c}{WBtstrp}  && \multicolumn{2}{c}{Classic}  & \multicolumn{3}{c}{WBtstrp} \\ \cmidrule(lr){4-5} \cmidrule(lr){6-8} 
		\cmidrule(lr){10-11} \cmidrule(lr){12-14}  
		&	&& $T_W$&$T_A$&$T_W^W$&$T_A^W$&$T_M^W$&&$T_W$&$T_A$&$T_W^W$&$T_A^W$&$T_M^W$\\[2pt]
		\cmidrule(lr){1-14}
		
	$H_0^T$&AR&(5,5)&46&5&6&4.9&5.8&&35.8&4.9&5.2&4.1&5.7\\
	&&(10,10)&34.2&5.3&5.3&5.6&5.9&&40.6&4.8&4.3&5&5.6\\
	&&(10,20)&29.4&4.9&6&5.2&5.5&&36&5.2&6.7&5.6&6.7\\
	&&(20,20)&15.2&5.2&5.2&5.4&5.4&&17.9&5.3&5.3&5.6&5.5\\
	\cmidrule(lr){1-14}
	&CS&(5,5)&49.9&3.4&6.4&3.8&5.7&&37&4.2&5.2&3.7&5.5\\
	&&(10,10)&34.3&4.2&5.1&4.9&5.5&&39.9&3.8&4.8&4.3&5.4\\
	&&(10,20)&29.5&4.3&6.3&5.2&5.8&&34.5&3.9&6.4&4.7&6.2\\
	&&(20,20)&15.3&4.4&5.2&5.2&5.3&&17.6&4.7&5.3&5.2&5.3\\
	\cmidrule(lr){1-14}
	&TP&(5,5)&41.3&6.8&5.5&5.8&6&&34&6.6&4.9&5.2&6.1\\
	&&(10,10)&34.7&6.4&4.9&6.1&6.1&&41.9&5.9&4.4&5.8&5.9\\
	&&(10,20)&30&5.7&6&5.4&5.5&&36.6&6&6.8&6&6.7\\
	&&(20,20)&14.8&5.4&5.1&5.4&5.4&&17.3&5.5&5.1&5.5&5.3\\
	\cmidrule(lr){1-14}
	$H_0^{GT}$&AR&(5,5)&46.3&5.1&6.3&4.8&6&&36.4&5.3&5.5&4.6&5.8\\
	&&(10,10)&33.5&4.9&4.7&5.3&5.8&&40.6&4.6&4.4&5&5.7\\
	&&(10,20)&30.2&4.9&6.3&5.1&5.8&&36.8&5.2&6.9&5.4&6.7\\
	&&(20,20)&15.3&5&5&5.2&5.2&&17.2&4.9&5.3&5.3&5.2\\
	\cmidrule(lr){1-14}
	&CS&(5,5)&49.1&3.7&6.2&3.9&5.7&&37.6&4.4&5.5&4&5.5\\
	&&(10,10)&33.5&4&5&4.9&5.3&&41.2&4&4.6&4.7&5.5\\
	&&(10,20)&30.3&4&6.3&4.8&5.7&&35.9&4.6&7.2&5.4&6.8\\
	&&(20,20)&15.4&4.4&5.2&5&5.1&&17&4.7&5&5.1&5.4\\
	\cmidrule(lr){1-14}
	&TP&(5,5)&42&6.6&5.7&5.5&6.1&&33.7&6.7&5&5.4&6\\
	&&(10,10)&34.5&5.7&4.7&5.4&5.4&&41.6&5.7&4.2&5.5&5.7\\
	&&(10,20)&30.4&5.8&6.4&5.5&5.8&&36.8&6.1&7.2&6&6.6\\
	&&(20,20)&15.1&5.4&4.8&5.3&5.4&&17.2&5.3&5.3&5.3&5.2\\

		\cmidrule(lr){1-14}
		
	\end{tabular*}
	\label{TableTypErrornormalt8MCAR}
\end{table}

\begin{table}[ht]
	\centering
	\caption{Simulation results for type-I error level ($\alpha=0.05$) for double exponential distribution, $d=8$ and different percentages $r$ of MCAR data.}
	\begin{tabular*}{0.9\linewidth}{c c  c  cc| ccc c  cc| ccc}
		\cmidrule(lr){1-14}
		& & &  \multicolumn{5}{c}{$r=10$}  && \multicolumn{5}{c}{$r=30$} \\ 
		\cmidrule(lr){4-8} \cmidrule(lr){10-14} 
		Hyp&	Cov &n  &\multicolumn{2}{c}{Classic}  & 
		\multicolumn{3}{c}{WBtstrp}  && \multicolumn{2}{c}{Classic}  & \multicolumn{3}{c}{WBtstrp} \\ \cmidrule(lr){4-5} \cmidrule(lr){6-8} 
		\cmidrule(lr){10-11} \cmidrule(lr){12-14}  
		&	&& $T_W$&$T_A$&$T_W^W$&$T_A^W$&$T_M^W$&&$T_W$&$T_A$&$T_W^W$&$T_A^W$&$T_M^W$\\[2pt]
		\cmidrule(lr){1-14}
$H_0^T$&AR&(5,5)&46.2&4.9&5.7&4.7&5.8&&35.8&4.9&5&4.3&5.4\\
&&(10,10)&34.4&5.2&5.2&5.7&5.9&&40.8&4.6&4.4&5&5.5\\
&&(10,20)&29&5&6.2&5.3&5.6&&36&5&6.8&5.5&6.7\\
&&(20,20)&15.4&5.1&5.3&5.4&5.5&&17.8&5.1&5.4&5.4&5.6\\
\cmidrule(lr){1-14}
&CS&(5,5)&50&3.4&6.4&3.8&5.3&&37.6&4.3&5.2&3.8&5.6\\
&&(10,10)&34.4&4&5.2&4.9&5.3&&39.7&3.8&4.8&4.4&5.2\\
&&(10,20)&29.4&4.2&6.4&5.2&5.7&&34.4&3.9&6.3&4.6&6.1\\
&&(20,20)&15.6&4.6&5.3&5.2&5.3&&17.6&4.6&5.5&5.1&5.3\\
\cmidrule(lr){1-14}
&TP&(5,5)&41.5&6.7&5.3&5.8&6.2&&33.8&6.3&4.6&5.1&5.8\\
&&(10,10)&34.9&6.4&5&6.2&6.1&&41.4&5.9&4.4&5.7&5.9\\
&&(10,20)&29.5&5.5&6&5.3&5.6&&36.3&6&7&6.1&6.8\\
&&(20,20)&15.1&5.3&5.1&5.3&5.5&&17.2&5.5&5.1&5.6&5.4\\

\cmidrule(lr){1-14}
$H_0^{GT}$&AR&(5,5)&46.2&4.9&6.1&4.6&5.9&&35.4&5.1&4.8&4.3&5.7\\
&&(10,10)&33.7&4.7&4.9&5.1&5.4&&40.6&4.6&4.4&4.9&5.6\\
&&(10,20)&30.2&4.4&6.2&5&5.7&&36.8&5.1&6.9&5.6&6.7\\
&&(20,20)&15.3&4.9&5.1&5.1&5.2&&17.1&5&5.1&5.3&5.2\\
\cmidrule(lr){1-14}
&CS&(5,5)&49.1&3.5&6.6&4&5.5&&37.5&4.3&5.3&3.9&5.5\\
&&(10,10)&33.6&4&4.9&4.8&5.3&&41.1&4.1&4.5&4.7&5.5\\
&&(10,20)&30.3&4&6.3&4.9&5.7&&35.9&4.7&7.1&5.4&6.7\\
&&(20,20)&15.1&4.3&5.3&5&5.1&&16.6&4.7&5&5.2&5.4\\
\cmidrule(lr){1-14}
&TP&(5,5)&41.8&6.5&5.4&5.5&6.2&&34.3&6.6&4.6&5.2&5.9\\
&&(10,10)&34.2&5.7&4.7&5.3&5.5&&41.9&5.6&4.2&5.5&5.6\\
&&(10,20)&29.8&5.5&6.3&5.4&5.7&&36.2&6&7.2&5.9&6.8\\
&&(20,20)&15.4&5.3&5&5.2&5.4&&17.1&5.3&5.5&5.3&5.6\\

		\cmidrule(lr){1-14}
	\end{tabular*}
	\label{TableTypErrordexpt8MCAR}
\end{table}

\begin{table}[ht]
	\centering
	\caption{Simulation results for type-I error level ($\alpha=0.05$) for chi-square distribution, $d=8$ and different percentages $r$ of MCAR data.}
	\begin{tabular*}{0.9\linewidth}{c c  c  cc| ccc c  cc| ccc}
		\cmidrule(lr){1-14}
		& & &  \multicolumn{5}{c}{$r=10$}  && \multicolumn{5}{c}{$r=30$} \\ 
		\cmidrule(lr){4-8} \cmidrule(lr){10-14} 
		Hyp&	Cov &n  &\multicolumn{2}{c}{Classic}  & 
		\multicolumn{3}{c}{WBtstrp}  && \multicolumn{2}{c}{Classic}  & \multicolumn{3}{c}{WBtstrp} \\ \cmidrule(lr){4-5} \cmidrule(lr){6-8} 
		\cmidrule(lr){10-11} \cmidrule(lr){12-14}  
		&	&& $T_W$&$T_A$&$T_W^W$&$T_A^W$&$T_M^W$&&$T_W$&$T_A$&$T_W^W$&$T_A^W$&$T_M^W$\\[2pt]
		\cmidrule(lr){1-14}
$H_0^T$&AR&(5,5)&45.6&4.9&5.9&4.8&5.8&&35.6&4.8&4.7&4.1&5.6\\
&&(10,10)&34.4&5.1&5.2&5.6&5.7&&41.1&4.8&4.3&5.1&5.5\\
&&(10,20)&29.4&5&5.9&5.3&5.6&&35.4&5.2&6.4&5.5&6.6\\
&&(20,20)&15&5.1&5&5.3&5.4&&17.5&5.1&5.4&5.4&5.4\\
\cmidrule(lr){1-14}
&CS&(5,5)&49.4&3.1&6.5&3.6&5.6&&37.3&4.3&5.1&4&5.6\\
&&(10,10)&34.8&4&5.3&4.9&5.5&&39.9&3.6&4.7&4.5&5.5\\
&&(10,20)&29.3&4.2&6.2&5.2&5.7&&34.6&4&6.2&4.7&6.1\\
&&(20,20)&15.3&4.5&5.2&5.1&5.1&&17.6&4.6&5.4&5.1&5.4\\
\cmidrule(lr){1-14}
&TP&(5,5)&41&6.9&6&5.7&6&&34.1&6.4&4.8&5.1&6\\
&&(10,10)&34.6&6.5&4.9&6.1&5.9&&42&5.8&4.3&5.6&5.9\\
&&(10,20)&29.4&5.6&5.9&5.4&5.5&&36.2&5.8&6.6&5.8&6.6\\
&&(20,20)&14.9&5.5&4.8&5.2&5.4&&17&5.4&5.2&5.4&5.3\\

\cmidrule(lr){1-14}

$H_0^{GT}$&AR&(5,5)&45.4&4.9&6.2&4.7&5.9&&36.1&5.2&5.2&4.4&5.8\\
&&(10,10)&33.8&4.9&4.7&5.3&5.7&&40.6&4.8&4.3&5.1&5.7\\
&&(10,20)&29.9&4.7&6.2&5.1&5.6&&36.2&5.1&6.7&5.4&6.6\\
&&(20,20)&15.2&4.9&5&5.2&5.3&&16.8&4.9&5.3&5.2&5.3\\
\cmidrule(lr){1-14}
&CS&(5,5)&49&3.5&6.2&3.7&5.6&&37.2&4.2&5&3.9&5.6\\
&&(10,10)&33.5&4.1&5&5&5.4&&41.1&4&4.8&4.6&5.6\\
&&(10,20)&30&4&6.2&4.9&5.7&&35.7&4.6&7.1&5.4&6.5\\
&&(20,20)&15.1&4.3&5&5&5&&16.8&4.7&5&5.1&5.2\\
\cmidrule(lr){1-14}
&TP&(5,5)&41.3&6.5&5.7&5.5&6.3&&33.5&6.7&4.6&5.2&6\\
&&(10,10)&34.6&5.7&4.6&5.4&5.6&&41.7&5.5&4.3&5.5&5.7\\
&&(10,20)&29.5&5.7&6.2&5.5&5.7&&36.3&5.9&6.4&5.8&6.5\\
&&(20,20)&14.8&5.4&5&5.3&5.4&&16.7&5.3&5.3&5.4&5.4\\

		\cmidrule(lr){1-14}
	\end{tabular*}
	\label{TableTypErrorchisqt8MCAR}
\end{table}

\begin{table}[!ht]
	\centering
	\caption{Simulation results for type-I error level ($\alpha=0.05$) for lognormal distribution, $d=8$ and different percentages $r$ of MCAR data.}
	\begin{tabular*}{0.9\linewidth}{c c  c  cc| ccc c  cc| ccc}
		\cmidrule(lr){1-14}
		& & &  \multicolumn{5}{c}{$r=10$}  && \multicolumn{5}{c}{$r=30$} \\ 
		\cmidrule(lr){4-8} \cmidrule(lr){10-14} 
		Hyp&	Cov &n  &\multicolumn{2}{c}{Classic}  & 
		\multicolumn{3}{c}{WBtstrp}  && \multicolumn{2}{c}{Classic}  & \multicolumn{3}{c}{WBtstrp} \\ \cmidrule(lr){4-5} \cmidrule(lr){6-8} 
		\cmidrule(lr){10-11} \cmidrule(lr){12-14}  
		&	&& $T_W$&$T_A$&$T_W^W$&$T_A^W$&$T_M^W$&&$T_W$&$T_A$&$T_W^W$&$T_A^W$&$T_M^W$\\[2pt]
		\cmidrule(lr){1-14}
		
$H_0^T$&AR&(5,5)&43.4&4.4&5.8&4.5&5.4&&35&4.4&4.9&4&5\\
&&(10,10)&34.5&4.8&5.4&5.3&5.8&&41.6&4.5&4.4&5.1&5.5\\
&&(10,20)&28.5&4.6&5.8&5.1&5.3&&34.5&4.6&5.9&5.1&6.4\\
&&(20,20)&15&4.7&5&5&5.1&&17&4.6&5.2&5&5.1\\
\cmidrule(lr){1-14}
&CS&(5,5)&46.4&3&5.8&3.4&4.9&&35.9&4&4.8&3.8&5.2\\
&&(10,10)&34.9&3.7&5.5&4.6&5.1&&40.8&3.6&4.6&4.6&5.5\\
&&(10,20)&28.1&3.9&5.8&5&5.5&&34.3&3.7&5.9&4.7&5.8\\
&&(20,20)&14.9&4.1&5.3&4.9&5&&17.1&4.3&5.4&5.1&5.3\\
\cmidrule(lr){1-14}
&TP&(5,5)&41.1&6.3&5.2&5.5&5.9&&33.7&5.9&4.2&4.9&5.3\\
&&(10,10)&36&6&5.5&5.9&5.9&&42.6&5.5&4.5&5.6&5.7\\
&&(10,20)&29.3&5.4&6.2&5.3&5.6&&35.4&5.7&6.5&5.8&6.8\\
&&(20,20)&15.8&5.2&6&5.2&5.2&&17.1&5.2&5.5&5.2&5.4\\

\cmidrule(lr){1-14}
$H_0^{GT}$&AR&(5,5)&42.3&4.5&5.6&4.4&5.6&&34.9&4.6&4.6&4.2&5.3\\
&&(10,10)&33.4&4.7&4.2&5.1&5.4&&41.3&4.2&4.2&4.8&5.4\\
&&(10,20)&28.2&4.4&5.4&5.1&5.6&&34.8&4.6&6.1&5.4&6.3\\
&&(20,20)&15&4.6&4.9&5&5.1&&16.2&4.7&4.9&5.1&5.3\\
\cmidrule(lr){1-14}
&CS&(5,5)&46.4&3.2&5.8&3.5&5.1&&36.3&3.8&4.6&3.6&5\\
&&(10,10)&33.6&3.8&4.8&4.6&5.1&&41.2&3.7&4.3&4.6&5.2\\
&&(10,20)&29.2&3.9&5.9&4.9&5.7&&35&4.1&6.3&5.1&6.1\\
&&(20,20)&14.9&4.2&4.7&4.9&5&&16.6&4.4&5&5&5.3\\
\cmidrule(lr){1-14}
&TP&(5,5)&40&6.1&5.2&5.3&6.2&&32.6&6.1&4.2&5.2&5.8\\
&&(10,10)&33.8&5.6&4.2&5.5&5.6&&41&5.1&4&5.2&5.8\\
&&(10,20)&28.3&5.7&5.3&5.6&5.9&&35.2&5.3&5.9&5.6&6.3\\
&&(20,20)&14.7&5.1&5.1&5.2&5.2&&16.2&5.3&4.8&5.5&5.3\\
		
		\cmidrule(lr){1-14}
	\end{tabular*}
	\label{TableTypErrorlognormalt8MCAR}
\end{table}

\begin{table}[ht]
	\centering
	\caption{Simulation results for type-I error level ($\alpha=0.05$) for normal distribution, testing the hypothesis $H_0^G$, $d\in\{4,8\}$ and different percentages $r$ of MCAR data.}
	\begin{tabular*}{0.9\linewidth}{c c  c  cc| ccc c  cc| ccc}
		\cmidrule(lr){1-14}
		& & &  \multicolumn{5}{c}{$r=10$}  && \multicolumn{5}{c}{$r=30$} \\ 
		\cmidrule(lr){4-8} \cmidrule(lr){10-14} 
		d&	Cov &n  &\multicolumn{2}{c}{Classic}  & 
		\multicolumn{3}{c}{WBtstrp}  && \multicolumn{2}{c}{Classic}  & \multicolumn{3}{c}{WBtstrp} \\ \cmidrule(lr){4-5} \cmidrule(lr){6-8} 
		\cmidrule(lr){10-11} \cmidrule(lr){12-14}  
		&	&& $T_W$&$T_A$&$T_W^W$&$T_A^W$&$T_M^W$&&$T_W$&$T_A$&$T_W^W$&$T_A^W$&$T_M^W$\\[2pt]
	\cmidrule(lr){1-14}
		4&AR&(5,5)&8.2&8.2&3.3&3.3&5.2&&9.8&9.8&4.7&4.7&4.6\\
		&&(10,10)&6.2&6.2&4.5&4.5&5.1&&7&7&4.9&4.9&5.2\\
		&&(10,20)&5.8&5.8&4.7&4.7&5.2&&6.6&6.6&5.1&5.1&5.4\\
		&&(20,20)&5.5&5.5&5&5&5.1&&5.6&5.6&5&5&5.1\\
		\cmidrule(lr){1-14}
		&CS&(5,5)&8.8&8.8&3.5&3.5&5.3&&9.8&9.8&4.8&4.8&4.9\\
		&&(10,10)&6&6&4.6&4.6&5.1&&6.6&6.6&4.8&4.8&5.1\\
		&&(10,20)&5.9&5.9&4.9&4.9&5.4&&6.4&6.4&5.1&5.1&5.2\\
		&&(20,20)&5.6&5.6&5&5&5.3&&5.7&5.7&5&5&5.1\\
		\cmidrule(lr){1-14}
		&TP&(5,5)&8.2&8.2&3.3&3.3&5.2&&9.9&9.9&4.6&4.6&5\\
		&&(10,10)&6.3&6.3&4.9&4.9&5.6&&6.7&6.7&4.7&4.7&5\\
		&&(10,20)&5.9&5.9&4.7&4.7&5.2&&6.4&6.4&4.9&4.9&5.1\\
		&&(20,20)&5.4&5.4&4.9&4.9&5.2&&5.6&5.6&5.1&5.1&5.2\\
		\cmidrule(lr){1-14}
		8&AR&(5,5)&9.3&9.3&3.9&3.9&6&&10.9&10.9&5.3&5.3&6\\
		&&(10,10)&6.6&6.6&4.8&4.8&5.8&&7.1&7.1&5&5&5.6\\
		&&(10,20)&6.3&6.3&5.1&5.1&5.6&&6.9&6.9&5&5&5.7\\
		&&(20,20)&6.1&6.1&5.3&5.3&5.8&&5.4&5.4&4.9&4.9&5.1\\
		\cmidrule(lr){1-14}
		&CS&(5,5)&10&10&4.4&4.4&6.2&&10.6&10.6&5.1&5.1&5.8\\
		&&(10,10)&6.5&6.5&4.7&4.7&5.5&&7.3&7.3&4.9&4.9&5.7\\
		&&(10,20)&6.4&6.4&5.1&5.1&5.7&&6.7&6.7&4.9&4.9&5.4\\
		&&(20,20)&6&6&5.4&5.4&5.8&&5.8&5.8&5&5&5.2\\
		\cmidrule(lr){1-14}
		&TP&(5,5)&9.5&9.5&3.7&3.7&6&&10.7&10.7&4.9&4.9&5.7\\
		&&(10,10)&6.3&6.3&4.8&4.8&5.7&&7.1&7.1&5&5&5.8\\
		&&(10,20)&6.3&6.3&5.1&5.1&5.8&&6.7&6.7&4.9&4.9&5.5\\
		&&(20,20)&5.9&5.9&5.4&5.4&5.8&&5.7&5.7&5.1&5.1&5.2\\

		\cmidrule(lr){1-14}
		
	\end{tabular*}
	\label{TableErrorMCARNormalHG}
\end{table}

\begin{table}[ht]
	\centering
	\caption{Simulation results for type-I error level ($\alpha=0.05$) for double exponential distribution, testing the hypothesis $H_0^G$, $d\in\{4,8\}$ and different percentages $r$ of MCAR data.}
	\begin{tabular*}{0.9\linewidth}{c c  c  cc| ccc c  cc| ccc}
		\cmidrule(lr){1-14}
		& & &  \multicolumn{5}{c}{$r=10$}  && \multicolumn{5}{c}{$r=30$} \\ 
		\cmidrule(lr){4-8} \cmidrule(lr){10-14} 
		d&	Cov &n  &\multicolumn{2}{c}{Classic}  & 
		\multicolumn{3}{c}{WBtstrp}  && \multicolumn{2}{c}{Classic}  & \multicolumn{3}{c}{WBtstrp} \\ \cmidrule(lr){4-5} \cmidrule(lr){6-8} 
		\cmidrule(lr){10-11} \cmidrule(lr){12-14}  
		&	&& $T_W$&$T_A$&$T_W^W$&$T_A^W$&$T_M^W$&&$T_W$&$T_A$&$T_W^W$&$T_A^W$&$T_M^W$\\[2pt]
\cmidrule(lr){1-14}
		4&AR&(5,5)&8.3&8.3&3.1&3.1&5.5&&9.8&9.8&4.9&4.9&5\\
		&&(10,10)&6.6&6.6&5&5&5.8&&6.9&6.9&4.8&4.8&5.3\\
		&&(10,20)&6.1&6.1&4.8&4.8&5.6&&6.4&6.4&5.1&5.1&5.2\\
		&&(20,20)&5.3&5.3&4.8&4.8&5.1&&5.8&5.8&5.2&5.2&5.3\\
		\cmidrule(lr){1-14}
		&CS&(5,5)&8.5&8.5&3.3&3.3&5.7&&10.2&10.2&4.7&4.7&5.4\\
		&&(10,10)&6.3&6.3&4.8&4.8&5.6&&6.9&6.9&5.1&5.1&5.6\\
		&&(10,20)&5.8&5.8&4.6&4.6&5.4&&6.4&6.4&4.8&4.8&5.3\\
		&&(20,20)&5.6&5.6&4.9&4.9&5.3&&5.9&5.9&5.1&5.1&5.3\\
		\cmidrule(lr){1-14}
		&TP&(5,5)&8.2&8.2&3.3&3.3&5.3&&9.5&9.5&4.7&4.7&4.9\\
		&&(10,10)&6.6&6.6&5&5&5.8&&6.7&6.7&4.7&4.7&5.3\\
		&&(10,20)&6&6&4.8&4.8&5.4&&6.4&6.4&4.8&4.8&5\\
		&&(20,20)&5.5&5.5&5.1&5.1&5.3&&5.8&5.8&5.2&5.2&5.3\\
		\cmidrule(lr){1-14}
		8&AR&(5,5)&9.4&9.4&4.1&4.1&6.6&&11&11&5&5&6.2\\
		&&(10,10)&6.4&6.4&4.7&4.7&6&&6.7&6.7&4.6&4.6&5.6\\
		&&(10,20)&6.5&6.5&5&5&6.1&&6.7&6.7&5.1&5.1&5.8\\
		&&(20,20)&6.2&6.2&5.5&5.5&6.1&&5.6&5.6&4.8&4.8&5.1\\
		\cmidrule(lr){1-14}
		&CS&(5,5)&9.6&9.6&4.5&4.5&7.1&&10.4&10.4&4.9&4.9&6.1\\
		&&(10,10)&6&6&4.4&4.4&5.6&&7.1&7.1&5&5&6\\
		&&(10,20)&6.4&6.4&4.9&4.9&6.1&&6.5&6.5&4.7&4.7&5.5\\
		&&(20,20)&6&6&5.4&5.4&6&&5.8&5.8&5.1&5.1&5.5\\
		\cmidrule(lr){1-14}
		&TP&(5,5)&9.8&9.8&4.2&4.2&6.6&&10.4&10.4&4.9&4.9&5.7\\
		&&(10,10)&6.6&6.6&4.9&4.9&5.9&&6.9&6.9&4.8&4.8&5.6\\
		&&(10,20)&6.5&6.5&5.2&5.2&6.2&&6.8&6.8&5&5&5.5\\
		&&(20,20)&6.2&6.2&5.6&5.6&6.1&&5.4&5.4&4.6&4.6&4.9\\	
		
		\cmidrule(lr){1-14}
		
	\end{tabular*}
	\label{TableErrorMCARDexpHG}
\end{table}

\begin{table}[ht]
	\centering
	\caption{Simulation results for type-I error level ($\alpha=0.05$) for chi-square distribution, testing the hypothesis $H_0^G$, $d\in\{4,8\}$ and different percentages $r$ of MCAR data.}
	\begin{tabular*}{0.9\linewidth}{c c  c  cc| ccc c  cc| ccc}
		\cmidrule(lr){1-14}
		& & &  \multicolumn{5}{c}{$r=10$}  && \multicolumn{5}{c}{$r=30$} \\ 
		\cmidrule(lr){4-8} \cmidrule(lr){10-14} 
		d&	Cov &n  &\multicolumn{2}{c}{Classic}  & 
		\multicolumn{3}{c}{WBtstrp}  && \multicolumn{2}{c}{Classic}  & \multicolumn{3}{c}{WBtstrp} \\ \cmidrule(lr){4-5} \cmidrule(lr){6-8} 
		\cmidrule(lr){10-11} \cmidrule(lr){12-14}  
		&	&& $T_W$&$T_A$&$T_W^W$&$T_A^W$&$T_M^W$&&$T_W$&$T_A$&$T_W^W$&$T_A^W$&$T_M^W$\\[2pt]
	\cmidrule(lr){1-14}
		4&AR&(5,5)&8.4&8.4&3.6&3.6&5.3&&10.1&10.1&4.8&4.8&4.9\\
		&&(10,10)&6.5&6.5&4.9&4.9&5.8&&6.7&6.7&4.8&4.8&5.2\\
		&&(10,20)&6.3&6.3&5.1&5.1&5.6&&6.8&6.8&5.1&5.1&5.6\\
		&&(20,20)&5.5&5.5&4.9&4.9&5.2&&5.6&5.6&5.1&5.1&5.2\\
		\cmidrule(lr){1-14}
		&CS&(5,5)&8.5&8.5&3.5&3.5&5.6&&9.9&9.9&4.6&4.6&5.1\\
		&&(10,10)&6.3&6.3&4.6&4.6&5.7&&6.9&6.9&4.9&4.9&5.4\\
		&&(10,20)&6.2&6.2&5&5&5.7&&6.8&6.8&5.1&5.1&5.4\\
		&&(20,20)&5.4&5.4&4.9&4.9&5.3&&5.9&5.9&5.3&5.3&5.4\\
		\cmidrule(lr){1-14}
		&TP&(5,5)&8.2&8.2&3.3&3.3&5.5&&9.8&9.8&4.8&4.8&5.1\\
		&&(10,10)&6.6&6.6&4.8&4.8&5.8&&6.3&6.3&4.7&4.7&5.1\\
		&&(10,20)&6&6&5&5&5.5&&6.4&6.4&4.9&4.9&5\\
		&&(20,20)&5.5&5.5&5.1&5.1&5.3&&5.7&5.7&5.2&5.2&5.3\\
		\cmidrule(lr){1-14}
		8&AR&(5,5)&9.2&9.2&4&4&6.5&&10.8&10.8&5.1&5.1&6.4\\
		&&(10,10)&6.4&6.4&4.7&4.7&5.7&&7.2&7.2&4.8&4.8&5.6\\
		&&(10,20)&6.6&6.6&5.2&5.2&6.2&&6.9&6.9&5.1&5.1&5.7\\
		&&(20,20)&6.2&6.2&5.6&5.6&6.1&&5.5&5.5&4.7&4.7&5\\
		\cmidrule(lr){1-14}
		&CS&(5,5)&9.9&9.9&4.5&4.5&7.1&&10.8&10.8&5&5&6.2\\
		&&(10,10)&6.5&6.5&4.6&4.6&5.8&&7.2&7.2&4.9&4.9&6\\
		&&(10,20)&6.6&6.6&5.4&5.4&6.2&&6.4&6.4&4.9&4.9&5.4\\
		&&(20,20)&5.6&5.6&5.2&5.2&5.6&&5.7&5.7&5&5&5.3\\
		\cmidrule(lr){1-14}
		&TP&(5,5)&9.4&9.4&3.8&3.8&6.4&&10.7&10.7&5&5&6\\
		&&(10,10)&6.6&6.6&4.6&4.6&5.8&&7.1&7.1&4.9&4.9&5.7\\
		&&(10,20)&6.7&6.7&5.1&5.1&6.1&&6.8&6.8&5&5&5.7\\
		&&(20,20)&5.9&5.9&5.3&5.3&5.7&&5.4&5.4&4.9&4.9&5.2\\

		\cmidrule(lr){1-14}
		
	\end{tabular*}
	\label{TableErrorMCARChisqHG}
\end{table}

\begin{table}[ht]
	\centering
	\caption{Simulation results for type-I error level ($\alpha=0.05$) for lognormal distribution, testing the hypothesis $H_0^G$, $d\in\{4,8\}$ and different percentages $r$ of MCAR data.}
	\begin{tabular*}{0.9\linewidth}{c c  c  cc| ccc c  cc| ccc}
		\cmidrule(lr){1-14}
		& & &  \multicolumn{5}{c}{$r=10$}  && \multicolumn{5}{c}{$r=30$} \\ 
		\cmidrule(lr){4-8} \cmidrule(lr){10-14} 
		d&	Cov &n  &\multicolumn{2}{c}{Classic}  & 
		\multicolumn{3}{c}{WBtstrp}  && \multicolumn{2}{c}{Classic}  & \multicolumn{3}{c}{WBtstrp} \\ \cmidrule(lr){4-5} \cmidrule(lr){6-8} 
		\cmidrule(lr){10-11} \cmidrule(lr){12-14}  
		&	&& $T_W$&$T_A$&$T_W^W$&$T_A^W$&$T_M^W$&&$T_W$&$T_A$&$T_W^W$&$T_A^W$&$T_M^W$\\[2pt]
		\cmidrule(lr){1-14}

		4&AR&(5,5)&8.3&8.3&3.3&3.3&6.3&&10.2&10.2&4.6&4.6&5.7\\
		&&(10,10)&6.8&6.8&4.9&4.9&6.6&&6.5&6.5&4.4&4.4&5.3\\
		&&(10,20)&6.5&6.5&5.4&5.4&6.4&&6.8&6.8&5.3&5.3&5.9\\
		&&(20,20)&5.5&5.5&5.1&5.1&5.6&&5.9&5.9&5.1&5.1&5.4\\
		\cmidrule(lr){1-14}
		&CS&(5,5)&8.8&8.8&3.7&3.7&6.9&&10.2&10.2&4.9&4.9&6.2\\
		&&(10,10)&6.7&6.7&4.8&4.8&6.6&&6.8&6.8&4.7&4.7&5.7\\
		&&(10,20)&7&7&5.5&5.5&6.6&&6.9&6.9&5.3&5.3&6.1\\
		&&(20,20)&5.7&5.7&4.9&4.9&5.8&&6.2&6.2&5.5&5.5&6\\
		\cmidrule(lr){1-14}
		&TP&(5,5)&8.4&8.4&3.4&3.4&6.1&&10&10&4.8&4.8&5.7\\
		&&(10,10)&6.7&6.7&4.9&4.9&6.3&&6.6&6.6&4.4&4.4&5.4\\
		&&(10,20)&6.4&6.4&5.2&5.2&6.2&&6.9&6.9&5.3&5.3&5.9\\
		&&(20,20)&5.5&5.5&5&5&5.5&&5.7&5.7&5.1&5.1&5.3\\
		\cmidrule(lr){1-14}
		8&AR&(5,5)&8.6&8.6&4.2&4.2&7.7&&10.6&10.6&4.8&4.8&7.5\\
		&&(10,10)&6.4&6.4&4.6&4.6&6.4&&6.5&6.5&4.8&4.8&6.2\\
		&&(10,20)&7.2&7.2&5.9&5.9&7.3&&6.8&6.8&5.3&5.3&6.4\\
		&&(20,20)&6.1&6.1&5.4&5.4&6.3&&5.2&5.2&4.3&4.3&5.1\\
		\cmidrule(lr){1-14}
		&CS&(5,5)&9&9&3.9&3.9&8.1&&10.1&10.1&4.6&4.6&7.7\\
		&&(10,10)&6.3&6.3&4.2&4.2&6.3&&7.1&7.1&4.9&4.9&6.8\\
		&&(10,20)&6.9&6.9&5.6&5.6&7&&6.7&6.7&5.1&5.1&6.3\\
		&&(20,20)&6&6&5.4&5.4&6&&5.3&5.3&4.5&4.5&5.4\\
		\cmidrule(lr){1-14}
		&TP&(5,5)&9.2&9.2&3.8&3.8&7.3&&10.9&10.9&5&5&6.9\\
		&&(10,10)&6.5&6.5&4.5&4.5&6.2&&6.5&6.5&4.8&4.8&6\\
		&&(10,20)&6.9&6.9&5.5&5.5&6.7&&7&7&5.2&5.2&6.2\\
		&&(20,20)&6.1&6.1&5.5&5.5&6.3&&5.1&5.1&4.5&4.5&5\\

		\cmidrule(lr){1-14}
		
	\end{tabular*}
	\label{TableErrorMCARLognormHG}
\end{table}

\begin{table}[ht]
	\centering
	\caption{Simulation results for type-I error level ($\alpha=0.05$) for normal distribution, $d=4$ and two different MAR mechanisms.}
	\begin{tabular*}{0.9\linewidth}{c c  c  cc| ccc c  cc| ccc}
		\cmidrule(lr){1-14}
		& & &  \multicolumn{5}{c}{$MAR1$}  && \multicolumn{5}{c}{$MAR2$} \\ 
		\cmidrule(lr){4-8} \cmidrule(lr){10-14} 
		Hyp&	Cov &n  &\multicolumn{2}{c}{Classic}  & 
		\multicolumn{3}{c}{WBtstrp}  && \multicolumn{2}{c}{Classic}  & \multicolumn{3}{c}{WBtstrp} \\ \cmidrule(lr){4-5} \cmidrule(lr){6-8} 
		\cmidrule(lr){10-11} \cmidrule(lr){12-14}  
		&	&& $T_W$&$T_A$&$T_W^W$&$T_A^W$&$T_M^W$&&$T_W$&$T_A$&$T_W^W$&$T_A^W$&$T_M^W$\\[2pt]
		\cmidrule(lr){1-14}
		
	$H_0^T$&AR&(5,5)&29&7.9&5.9&5.4&6.7&&25.2&6.9&5&5&5.9\\
	&&(10,10)&13.6&6&5.3&5.2&5.5&&11.5&5.4&4.5&4.8&5\\
	&&(10,20)&13.6&6.3&6.3&5.7&6.4&&12.1&6&5.6&5.4&5.9\\
	&&(20,20)&7.8&5.1&4.8&4.9&5&&8.4&5.6&5.3&5.4&5.5\\
	\cmidrule(lr){1-14}
	&CS&(5,5)&29.5&7.6&5.7&5.2&6.4&&26&6.7&5&4.8&6.2\\
	&&(10,10)&13.3&5.7&5.1&4.9&5.3&&11.7&5.1&4.7&4.5&4.9\\
	&&(10,20)&12.8&5.7&6&5.3&5.9&&12.2&5.5&5.8&5.1&5.8\\
	&&(20,20)&8.6&5.5&5.3&5.2&5.4&&8.4&5.7&5.3&5.3&5.4\\
	\cmidrule(lr){1-14}
	&TP&(5,5)&30&8.5&6.2&6.1&7&&25.4&7.2&5&5.2&5.8\\
	&&(10,10)&12.6&5.9&4.7&5.2&5.6&&10.9&5.8&4.4&5.2&5\\
	&&(10,20)&13.4&6.5&6.2&5.8&6.2&&11.8&6&5.2&5.5&5.7\\
	&&(20,20)&7.8&5.1&4.9&4.9&5&&8.1&5.4&5.2&5.2&5.3\\
	\cmidrule(lr){1-14}
	$H_0^{GT}$&AR&(5,5)&30.1&7.7&5.9&5.1&6.7&&26.7&7.1&5.2&5&6.2\\
	&&(10,10)&13.4&6.3&5.4&5.5&5.8&&12.2&5.5&4.8&5&5.1\\
	&&(10,20)&13.1&6.2&6.2&5.6&6.1&&12&5.4&5.7&4.9&5.4\\
	&&(20,20)&8.4&5.3&5.1&5.1&5.1&&8.6&5.9&5.3&5.7&5.6\\
	\cmidrule(lr){1-14}
	&CS&(5,5)&29.6&7.3&5.9&4.9&6.2&&25.7&6.1&5.2&4.3&6\\
	&&(10,10)&13.4&5.7&5.1&5&5.3&&12.3&5.2&4.8&4.7&5\\
	&&(10,20)&12.9&5.9&6.1&5.4&6&&12&5.4&5.6&5&5.6\\
	&&(20,20)&8.2&5.5&5.1&5.3&5.3&&8.2&5.4&5.2&5.2&5.3\\
	\cmidrule(lr){1-14}
	&TP&(5,5)&30.4&8.8&6.3&6.1&7.1&&27&7.7&5.5&5.5&6.2\\
	&&(10,10)&13.4&6.4&5.4&5.6&5.6&&12.4&5.9&5&5.2&5.3\\
	&&(10,20)&13.3&6.5&6.2&6&6.2&&11.8&5.7&5.4&5.2&5.5\\
	&&(20,20)&8.1&5.4&5&5&5.2&&8&5.8&5.1&5.4&5.4\\

		\cmidrule(lr){1-14}
		
	\end{tabular*}
	\label{TableErrorMARNormalt4}
\end{table}

\begin{table}[ht]
	\centering
	\caption{Simulation results for type-I error level ($\alpha=0.05$) for double exponential distribution, $d=4$ and two different MAR mechanisms.}
	\begin{tabular*}{0.9\linewidth}{c c  c  cc| ccc c  cc| ccc}
		\cmidrule(lr){1-14}
		& & &  \multicolumn{5}{c}{$MAR1$}  && \multicolumn{5}{c}{$MAR2$} \\ 
		\cmidrule(lr){4-8} \cmidrule(lr){10-14} 
		Hyp&	Cov &n  &\multicolumn{2}{c}{Classic}  & 
		\multicolumn{3}{c}{WBtstrp}  && \multicolumn{2}{c}{Classic}  & \multicolumn{3}{c}{WBtstrp} \\ \cmidrule(lr){4-5} \cmidrule(lr){6-8} 
		\cmidrule(lr){10-11} \cmidrule(lr){12-14}  
		&	&& $T_W$&$T_A$&$T_W^W$&$T_A^W$&$T_M^W$&&$T_W$&$T_A$&$T_W^W$&$T_A^W$&$T_M^W$\\[2pt]
		\cmidrule(lr){1-14}
		
$H_0^T$&AR&(5,5)&29.9&7.8&5.9&5.4&6.7&&25.8&6.6&5.1&4.8&5.6\\
&&(10,10)&12.9&5.9&4.8&5.2&5.4&&11.6&5.3&4.6&4.8&4.9\\
&&(10,20)&12.2&5.6&5.6&5.1&5.8&&11.7&5.8&5.6&5.2&5.6\\
&&(20,20)&9&5.9&5.5&5.7&5.7&&8.5&5.6&5.5&5.4&5.4\\
\cmidrule(lr){1-14}
&CS&(5,5)&29.7&7.4&5.6&5.1&6.3&&25.6&6.8&4.8&4.8&5.9\\
&&(10,10)&13.5&5.9&5.2&5.2&5.7&&11.9&5&4.4&4.6&5\\
&&(10,20)&12.9&5.7&5.8&5.2&6&&11.8&5.4&5.6&5&5.8\\
&&(20,20)&8.2&5.2&5.2&5&5.3&&8.4&5.7&5.3&5.4&5.4\\
\cmidrule(lr){1-14}
&TP&(5,5)&29.6&8.4&5.9&5.7&6.5&&25.4&7&5.1&5.1&5.6\\
&&(10,10)&12.9&6.2&4.8&5.5&5.4&&11.4&5.7&4.4&5.1&5\\
&&(10,20)&12.6&6.3&6.2&5.6&6.2&&11.6&6&5.4&5.4&5.9\\
&&(20,20)&8.1&5.5&4.8&5.3&5.2&&8.4&5.6&5.4&5.3&5.2\\
\cmidrule(lr){1-14}
$H_0^{GT}$&AR&(5,5)&30.1&7.4&5.7&4.9&6.1&&27&6.9&5.4&4.9&6.1\\
&&(10,10)&13.1&5.7&4.7&5&5.1&&11.9&5.6&5&4.9&5.1\\
&&(10,20)&13.1&6&6.1&5.6&6.2&&11.6&5.4&5.4&4.9&5.3\\
&&(20,20)&8.2&5.8&5.2&5.6&5.4&&8.4&5.8&5.3&5.6&5.4\\
\cmidrule(lr){1-14}
&CS&(5,5)&28.9&7.3&6.1&4.9&6.6&&26.1&6.1&5.3&4.4&5.9\\
&&(10,10)&14&5.9&5.6&5.3&5.9&&12.3&5.2&5.1&4.9&5.2\\
&&(10,20)&12.5&5.6&5.6&5.1&5.7&&11.5&5.3&5.7&4.9&5.5\\
&&(20,20)&8.3&5.3&5.2&5.1&5.3&&8.2&5.4&5&5.1&5.1\\
\cmidrule(lr){1-14}
&TP&(5,5)&29.9&8&6.3&5.5&6.7&&27.2&7.7&5.7&5.2&6.2\\
&&(10,10)&13&6.1&5.1&5.3&5.4&&12.2&5.9&4.9&5.1&5.2\\
&&(10,20)&13.3&6&6.3&5.5&6.2&&11.7&5.6&5.3&5.1&5.3\\
&&(20,20)&8.2&5.4&5.2&5.3&5.4&&8.1&5.9&5&5.6&5.5\\

		\cmidrule(lr){1-14}
		
	\end{tabular*}
	\label{TableErrorMARDexpt4}
\end{table}

\begin{table}[ht]
	\centering
	\caption{Simulation results for type-I error level ($\alpha=0.05$) for chi-square distribution, $d=4$ and two different MAR mechanisms.}
	\begin{tabular*}{0.9\linewidth}{c c  c  cc| ccc c  cc| ccc}
		\cmidrule(lr){1-14}
		& & &  \multicolumn{5}{c}{$MAR1$}  && \multicolumn{5}{c}{$MAR2$} \\ 
		\cmidrule(lr){4-8} \cmidrule(lr){10-14} 
		Hyp&	Cov &n  &\multicolumn{2}{c}{Classic}  & 
		\multicolumn{3}{c}{WBtstrp}  && \multicolumn{2}{c}{Classic}  & \multicolumn{3}{c}{WBtstrp} \\ \cmidrule(lr){4-5} \cmidrule(lr){6-8} 
		\cmidrule(lr){10-11} \cmidrule(lr){12-14}  
		&	&& $T_W$&$T_A$&$T_W^W$&$T_A^W$&$T_M^W$&&$T_W$&$T_A$&$T_W^W$&$T_A^W$&$T_M^W$\\[2pt]
			\cmidrule(lr){1-14}
		
	$H_0^T$&AR&(5,5)&29.5&7.8&5.6&5.4&6.5&&25&6.7&5.1&4.9&5.8\\
	&&(10,10)&13.1&5.8&5.2&5.1&5.4&&11.5&5.5&4.5&5&5\\
	&&(10,20)&12.6&6&5.7&5.4&5.8&&11.8&5.9&5.5&5.4&5.9\\
	&&(20,20)&8.2&5.3&4.8&5.2&5.3&&8.1&5.6&5.2&5.4&5.4\\
	\cmidrule(lr){1-14}
	&CS&(5,5)&28.3&7&5.7&4.6&6.2&&25.9&6.6&4.9&4.7&6.1\\
	&&(10,10)&13.2&5.9&5.4&5.3&5.7&&11.6&5&4.7&4.5&4.8\\
	&&(10,20)&12.7&5.4&5.7&5.1&5.7&&11.9&5.7&5.7&5.1&5.7\\
	&&(20,20)&8.2&5.1&4.7&4.9&5&&8.5&5.4&5.1&5.2&5.3\\
	\cmidrule(lr){1-14}
	&TP&(5,5)&30.2&8.6&6.3&5.9&6.8&&25.5&7.3&4.9&5.4&6\\
	&&(10,10)&12.5&5.9&4.9&5.1&5.3&&11.3&5.7&4.3&5&4.9\\
	&&(10,20)&12.8&6.7&6.2&6.1&6.5&&11.5&6.2&5.5&5.7&5.9\\
	&&(20,20)&7.8&5.3&4.8&5.2&5.3&&7.9&5.4&5.1&5.3&5.4\\
	\cmidrule(lr){1-14}
	$H_0^{GT}$&AR&(5,5)&28.8&7.6&6.3&5.1&6.5&&26.3&6.9&5.3&5&6.2\\
	&&(10,10)&12.8&5.7&5.1&5&5.2&&12.4&5.6&5&5.1&5.4\\
	&&(10,20)&13.1&6.1&6.2&5.5&6.2&&11.7&5.6&5.8&4.9&5.6\\
	&&(20,20)&8.6&5.4&5.2&5.2&5.2&&8.5&5.8&5.4&5.8&5.7\\
	\cmidrule(lr){1-14}
	&CS&(5,5)&29&7.2&5.7&4.7&6.4&&26.1&6.1&5.2&4.4&6.1\\
	&&(10,10)&13.1&5.8&5.1&5.1&5.3&&12.7&5.4&5&4.8&5.1\\
	&&(10,20)&12.7&5.7&5.9&5.4&5.8&&12&5.5&5.7&5.2&5.5\\
	&&(20,20)&8.8&5.3&5.3&5.1&5.4&&8.1&5.3&5&5&5.1\\
	\cmidrule(lr){1-14}
	&TP&(5,5)&29.5&8.6&6.2&5.9&6.5&&26.8&7.7&5.1&5.3&6.4\\
	&&(10,10)&12.4&6&4.8&5.2&5.5&&11.8&5.7&4.7&5.1&5.3\\
	&&(10,20)&13.2&6.3&6.3&5.6&6.2&&11.9&5.6&5.4&4.9&5.6\\
	&&(20,20)&8.5&5.6&5.4&5.3&5.4&&8.4&5.9&5.4&5.6&5.6\\

		\cmidrule(lr){1-14}
		
	\end{tabular*}
	\label{TableErrorMARChisqt4}
\end{table}

\begin{table}[ht]
	\centering
	\caption{Simulation results for type-I error level ($\alpha=0.05$) for lognormal distribution, $d=4$ and two different MAR mechanisms.}
	\begin{tabular*}{0.9\linewidth}{c c  c  cc| ccc c  cc| ccc}
		\cmidrule(lr){1-14}
		& & &  \multicolumn{5}{c}{$MAR1$}  && \multicolumn{5}{c}{$MAR2$} \\ 
		\cmidrule(lr){4-8} \cmidrule(lr){10-14} 
		Hyp&	Cov &n  &\multicolumn{2}{c}{Classic}  & 
		\multicolumn{3}{c}{WBtstrp}  && \multicolumn{2}{c}{Classic}  & \multicolumn{3}{c}{WBtstrp} \\ \cmidrule(lr){4-5} \cmidrule(lr){6-8} 
		\cmidrule(lr){10-11} \cmidrule(lr){12-14}  
		&	&& $T_W$&$T_A$&$T_W^W$&$T_A^W$&$T_M^W$&&$T_W$&$T_A$&$T_W^W$&$T_A^W$&$T_M^W$\\[2pt]
		\cmidrule(lr){1-14}
		
	$H_0^T$&AR&(5,5)&29.3&7.2&5.4&5.2&6&&26.2&6.7&5.1&4.6&5.8\\
	&&(10,10)&13.9&5.8&5.4&5.2&5.4&&11.9&5.3&4.9&4.7&4.9\\
	&&(10,20)&12.8&5.9&6.1&5.4&5.9&&12.1&5.6&5.7&5.1&5.7\\
	&&(20,20)&8.6&5.5&5.3&5.2&5.3&&8.1&5.3&5.5&5.3&5.5\\
	\cmidrule(lr){1-14}
	&CS&(5,5)&30.1&6.9&5.7&4.7&5.8&&26.4&5.9&5.1&4.4&5.2\\
	&&(10,10)&13.2&5.4&5.5&4.9&5.3&&11.9&4.7&4.9&4.2&4.7\\
	&&(10,20)&12.6&5.5&6&5.1&5.9&&12.1&5.3&5.8&4.9&5.5\\
	&&(20,20)&8.6&5.5&5.4&5.3&5.3&&8.3&5.1&5.4&5&5.1\\
	\cmidrule(lr){1-14}
	&TP&(5,5)&30.4&8&6&5.3&6.3&&26.7&7.1&5.3&5.2&5.9\\
	&&(10,10)&13.7&6.1&5.8&5.4&5.7&&11.9&5.5&4.7&5&5.2\\
	&&(10,20)&12.9&6.5&6.4&5.9&6.4&&12.2&5.7&5.8&5.4&5.6\\
	&&(20,20)&8.5&5.5&5.5&5.2&5.2&&8.7&5.5&5.7&5.3&5.4\\
	\cmidrule(lr){1-14}
	$H_0^{GT}$&AR&(5,5)&29&7.3&5.6&4.9&6.3&&26.2&6.6&4.9&4.6&5.9\\
	&&(10,10)&13&5.8&4.7&5.2&5.6&&12.3&5.4&4.8&4.8&5.4\\
	&&(10,20)&12.4&6.1&5.9&5.6&6&&11.8&5.6&5.6&5.1&5.7\\
	&&(20,20)&8.4&5.5&5.4&5.3&5.5&&8.4&5.7&5.4&5.5&5.4\\
	\cmidrule(lr){1-14}
	&CS&(5,5)&29.2&7.2&5.3&4.9&6.4&&25.9&6.2&4.9&4.5&6.2\\
	&&(10,10)&13.2&5.6&5&5.1&5.4&&12.3&5.1&4.9&4.8&5.1\\
	&&(10,20)&12.7&5.5&6&5.2&5.9&&11.9&5.6&5.4&5.1&5.6\\
	&&(20,20)&8.1&5.1&5&4.9&5.2&&8.3&5.4&5.2&5.3&5.3\\
	\cmidrule(lr){1-14}
	&TP&(5,5)&28.9&7.4&5.4&5.1&6&&26.6&7.4&5&5.2&6\\
	&&(10,10)&13&6.3&5.2&5.7&5.8&&11.9&5.5&4.9&5&5.2\\
	&&(10,20)&12.1&6.2&5.5&5.6&5.9&&11.6&5.5&5.4&5.1&5.5\\
	&&(20,20)&8.8&5.7&5.4&5.6&5.8&&8&5.8&5&5.5&5.4\\

		\cmidrule(lr){1-14}
		
	\end{tabular*}
	\label{TableErrorMARLognormalt4}
\end{table}

\begin{table}[ht]
	\centering
	\caption{Simulation results for type-I error level ($\alpha=0.05$) for normal distribution, $d=8$ and two different MAR mechanisms.}
	\begin{tabular*}{0.9\linewidth}{c c  c  cc| ccc c  cc| ccc}
		\cmidrule(lr){1-14}
		& & &  \multicolumn{5}{c}{$MAR1$}  && \multicolumn{5}{c}{$MAR2$} \\ 
		\cmidrule(lr){4-8} \cmidrule(lr){10-14} 
		Hyp&	Cov &n  &\multicolumn{2}{c}{Classic}  & 
		\multicolumn{3}{c}{WBtstrp}  && \multicolumn{2}{c}{Classic}  & \multicolumn{3}{c}{WBtstrp} \\ \cmidrule(lr){4-5} \cmidrule(lr){6-8} 
		\cmidrule(lr){10-11} \cmidrule(lr){12-14}  
		&	&& $T_W$&$T_A$&$T_W^W$&$T_A^W$&$T_M^W$&&$T_W$&$T_A$&$T_W^W$&$T_A^W$&$T_M^W$\\[2pt]
			\cmidrule(lr){1-14}
		
	$H_0^T$&AR&(5,5)&39.8&5.5&5.3&5.3&6.4&&44.9&5.2&6.2&4.9&6.4\\
	&&(10,10)&37&4.9&5.2&5.3&5.5&&34.3&5.1&5&5.4&5.7\\
	&&(10,20)&32.2&5.7&6.5&6&6.4&&29.8&5&6.6&5.1&5.7\\
	&&(20,20)&15.4&5&5.2&5.3&5.4&&15.8&5.1&4.9&5.3&5.3\\
	\cmidrule(lr){1-14}
	&CS&(5,5)&41.8&4&5.1&4&6.2&&48.4&3.5&6.7&3.8&5.6\\
	&&(10,10)&35.2&4.1&4.8&4.8&5.5&&33.8&4.1&5.2&5&5.3\\
	&&(10,20)&30.9&4.3&5.9&5&5.8&&30&4.2&6.4&5.1&5.9\\
	&&(20,20)&16&4.2&5.2&4.8&5&&15.7&4.4&4.9&5.1&5.2\\
	\cmidrule(lr){1-14}
	&TP&(5,5)&37.5&8&5&6.4&7.1&&41.9&6.9&5.9&5.8&6.7\\
	&&(10,10)&37.2&5.9&4.6&5.6&5.6&&34.2&6&5&5.7&5.7\\
	&&(10,20)&32.3&6.3&6&6&6.3&&28.9&5.8&5.6&5.7&5.8\\
	&&(20,20)&15.7&5.5&5&5.5&5.3&&15.8&5.3&5.1&5.2&5.2\\
	\cmidrule(lr){1-14}
	$H_0^{GT}$&AR&(5,5)&40.6&5.2&5.4&4.7&6.3&&44.3&4.9&6.3&4.7&6.3\\
	&&(10,10)&36&5.1&4.8&5.3&5.5&&34.1&5.2&4.8&5.4&5.6\\
	&&(10,20)&31.7&5.3&6.7&5.6&6.3&&30.6&5.3&6.6&5.6&6.2\\
	&&(20,20)&15.9&5&5.3&5.2&5.2&&15.4&5&5.2&5.3&5.6\\
	\cmidrule(lr){1-14}
	&CS&(5,5)&41.6&4&5.4&4.2&6.2&&48.6&3.6&6.8&4.1&6\\
	&&(10,10)&36.1&4.3&4.8&5.2&5.4&&34.4&4.1&4.9&5&5.5\\
	&&(10,20)&31.5&4&6.1&4.9&5.5&&30.7&4.6&6.6&5.6&6.2\\
	&&(20,20)&16&4.5&5.1&5.2&5.1&&15&4.6&5.1&5.2&5.5\\
	\cmidrule(lr){1-14}
	&TP&(5,5)&38.5&7.5&5.3&6.2&6.9&&40.6&6.6&5.7&5.6&6\\
	&&(10,10)&37.2&5.8&4.9&5.6&5.6&&35.4&5.9&4.9&5.6&5.5\\
	&&(10,20)&32.7&5.9&6.8&5.7&6.4&&31.4&6&6.5&5.7&6.2\\
	&&(20,20)&15.9&5.7&5.2&5.6&5.6&&15.6&5.7&5&5.7&5.5\\

		\cmidrule(lr){1-14}
		
	\end{tabular*}
	\label{TableErrorMARNormalt8}
\end{table}

\begin{table}[ht]
	\centering
	\caption{Simulation results for type-I error level ($\alpha=0.05$) for double exponential distribution, $d=8$ and two different MAR mechanisms.}
	\begin{tabular*}{0.9\linewidth}{c c  c  cc| ccc c  cc| ccc}
		\cmidrule(lr){1-14}
		& & &  \multicolumn{5}{c}{$MAR1$}  && \multicolumn{5}{c}{$MAR2$} \\ 
		\cmidrule(lr){4-8} \cmidrule(lr){10-14} 
		Hyp&	Cov &n  &\multicolumn{2}{c}{Classic}  & 
		\multicolumn{3}{c}{WBtstrp}  && \multicolumn{2}{c}{Classic}  & \multicolumn{3}{c}{WBtstrp} \\ \cmidrule(lr){4-5} \cmidrule(lr){6-8} 
		\cmidrule(lr){10-11} \cmidrule(lr){12-14}  
		&	&& $T_W$&$T_A$&$T_W^W$&$T_A^W$&$T_M^W$&&$T_W$&$T_A$&$T_W^W$&$T_A^W$&$T_M^W$\\[2pt]
	\cmidrule(lr){1-14}
	$H_0^T$&AR&(5,5)&40&5.3&5.5&4.9&6.1&&44.4&5.2&6.4&5&6.1\\
	&&(10,10)&36.5&5.4&5&5.6&5.7&&33.9&5&5.2&5.3&5.6\\
	&&(10,20)&31.1&4.6&6.1&4.9&5.6&&29.9&4.9&6.4&5.2&5.7\\
	&&(20,20)&16.3&5.1&5.2&5.3&5.2&&15.7&5.1&5.3&5.4&5.4\\
	\cmidrule(lr){1-14}
	&CS&(5,5)&41.5&4&5.6&4.1&5.9&&48.7&3.4&6.8&3.9&5.4\\
	&&(10,10)&35.6&4.2&4.8&4.9&5.6&&34.3&4.2&5.2&5.1&5.4\\
	&&(10,20)&31.6&4.3&6.4&5.1&5.8&&30.3&4.2&6.4&5.1&5.6\\
	&&(20,20)&16.1&4.9&5.6&5.4&5.5&&15.5&4.3&4.8&5&5.1\\
	\cmidrule(lr){1-14}
	&TP&(5,5)&37.9&7&5.2&5.8&6.7&&41.2&6.8&6&5.8&6.5\\
	&&(10,10)&36.2&6&4.3&5.6&5.5&&34.1&6&4.6&5.6&5.5\\
	&&(10,20)&31.6&6.1&6.2&5.9&6.2&&28.8&5.9&5.6&5.7&5.8\\
	&&(20,20)&15.3&5.8&4.9&5.6&5.5&&15.9&5.6&5.2&5.3&5.4\\
	\cmidrule(lr){1-14}
	$H_0^{GT}$&AR&(5,5)&40.5&5.2&5.6&4.8&6.2&&44&5.1&6&4.6&6.6\\
	&&(10,10)&36.3&5.3&5.1&5.5&5.9&&34.1&5.1&5&5.4&5.4\\
	&&(10,20)&31.1&5.1&5.9&5.4&5.9&&30.2&5.1&6.6&5.4&5.9\\
	&&(20,20)&15.5&4.9&5.1&5&5.1&&15.4&5.3&5.3&5.5&5.6\\
	\cmidrule(lr){1-14}
	&CS&(5,5)&42&4.3&5.5&4.3&6&&49.1&3.5&7&3.9&5.8\\
	&&(10,10)&35&4&4.6&4.8&5.3&&34.4&3.9&5&4.9&5.2\\
	&&(10,20)&30.6&3.9&6.2&4.7&5.5&&30.3&4.5&6.3&5.5&6.4\\
	&&(20,20)&16.2&4.8&5.2&5.3&5.3&&14.8&4.1&5&4.8&5\\
	\cmidrule(lr){1-14}
	&TP&(5,5)&37.5&6.9&5.3&5.6&6.6&&40.7&6.7&5.6&5.5&6\\
	&&(10,10)&37.7&6.3&4.8&5.9&6.1&&35.1&5.9&4.6&5.6&5.7\\
	&&(10,20)&32.8&5.9&6.6&5.9&6.2&&30.6&5.9&6.7&5.8&6\\
	&&(20,20)&15.8&5.6&5.2&5.5&5.6&&15.3&5.6&4.9&5.5&5.5\\

		\cmidrule(lr){1-14}
		
	\end{tabular*}
	\label{TableErrorMARDExpt8}
\end{table}

\begin{table}[ht]
	\centering
	\caption{Simulation results for type-I error level ($\alpha=0.05$) for chi-square distribution, $d=8$ and two different MAR mechanisms.}
	\begin{tabular*}{0.9\linewidth}{c c  c  cc| ccc c  cc| ccc}
		\cmidrule(lr){1-14}
		& & &  \multicolumn{5}{c}{$MAR1$}  && \multicolumn{5}{c}{$MAR2$} \\ 
		\cmidrule(lr){4-8} \cmidrule(lr){10-14} 
		Hyp&	Cov &n  &\multicolumn{2}{c}{Classic}  & 
		\multicolumn{3}{c}{WBtstrp}  && \multicolumn{2}{c}{Classic}  & \multicolumn{3}{c}{WBtstrp} \\ \cmidrule(lr){4-5} \cmidrule(lr){6-8} 
		\cmidrule(lr){10-11} \cmidrule(lr){12-14}  
		&	&& $T_W$&$T_A$&$T_W^W$&$T_A^W$&$T_M^W$&&$T_W$&$T_A$&$T_W^W$&$T_A^W$&$T_M^W$\\[2pt]
		\cmidrule(lr){1-14}
		
	$H_0^T$&AR&(5,5)&39.6&5.5&5.3&5.3&6.5&&43.7&5.1&5.8&4.9&6.4\\
	&&(10,10)&36&5&4.8&5.3&5.5&&34.7&5.2&4.9&5.5&5.5\\
	&&(10,20)&30.7&5&6.1&5.3&6&&29.9&5&6.3&5.4&5.7\\
	&&(20,20)&15.7&4.9&5&5.1&5.2&&15.4&5.1&5.2&5.4&5.4\\
	\cmidrule(lr){1-14}
	&CS&(5,5)&42.3&3.9&6&4&5.9&&48.5&3.4&6.4&3.8&5.8\\
	&&(10,10)&35.6&4.2&4.8&4.9&5.3&&34.5&4.2&5.3&5.1&5.4\\
	&&(10,20)&31.5&4.2&6.2&5&5.7&&29.6&4.2&6.2&5&5.9\\
	&&(20,20)&16.3&4.7&5.6&5.3&5.4&&15.6&4.4&4.9&5.1&5.2\\
	\cmidrule(lr){1-14}
	&TP&(5,5)&37.8&7.2&5.3&5.9&6.7&&41.2&6.9&6&6&6.6\\
	&&(10,10)&36.9&5.9&4.2&5.6&5.8&&34.7&6.2&4.4&5.9&5.9\\
	&&(10,20)&31.2&5.9&6.3&5.9&6.2&&28.1&5.9&5.8&5.6&5.9\\
	&&(20,20)&15.4&5.7&5.1&5.8&5.7&&15.5&5.4&5.1&5.4&5.4\\
	\cmidrule(lr){1-14}
	$H_0^{GT}$&AR&(5,5)&39.5&4.9&5.2&4.4&6.3&&43.8&5&6.1&4.6&6.1\\
	&&(10,10)&35.9&4.7&4.3&5.1&5.2&&33.6&4.9&4.8&5.3&5.5\\
	&&(10,20)&31.5&5.1&6.5&5.4&6.3&&30.2&5.3&6.3&5.7&6.1\\
	&&(20,20)&16.2&5.1&5.1&5.4&5.3&&14.9&5.3&4.9&5.4&5.6\\
	\cmidrule(lr){1-14}
	&CS&(5,5)&42.1&4&5.7&4.2&5.9&&48.6&3.5&6.5&3.9&5.9\\
	&&(10,10)&36.4&4&5&4.9&5.3&&34&4&4.7&4.9&5.2\\
	&&(10,20)&31.2&4&6.6&5&5.9&&30.9&4.4&6.5&5.3&5.9\\
	&&(20,20)&15.6&4.5&5.3&5&5.2&&15.1&4.4&5&5.1&5.3\\
	\cmidrule(lr){1-14}
	&TP&(5,5)&37.4&6.3&4.8&5.1&6.2&&39.8&6.4&5.5&5.5&6\\
	&&(10,10)&36.3&6&4.7&5.9&5.6&&35.3&5.8&4.6&5.4&5.4\\
	&&(10,20)&31.4&6&6.2&5.8&6.3&&30.2&5.9&6.1&5.8&6.1\\
	&&(20,20)&15.2&5.6&5&5.6&5.4&&15.2&5.7&5&5.7&5.7\\

		\cmidrule(lr){1-14}
		
	\end{tabular*}
	\label{TableErrorMARChisqt8}
\end{table}

\begin{table}[ht]
	\centering
	\caption{Simulation results for type-I error level ($\alpha=0.05$) for lognormal distribution, $d=8$ and two different MAR mechanisms.}
	\begin{tabular*}{0.9\linewidth}{c c  c  cc| ccc c  cc| ccc}
		\cmidrule(lr){1-14}
		& & &  \multicolumn{5}{c}{$MAR1$}  && \multicolumn{5}{c}{$MAR2$} \\ 
		\cmidrule(lr){4-8} \cmidrule(lr){10-14} 
		Hyp&	Cov &n  &\multicolumn{2}{c}{Classic}  & 
		\multicolumn{3}{c}{WBtstrp}  && \multicolumn{2}{c}{Classic}  & \multicolumn{3}{c}{WBtstrp} \\ \cmidrule(lr){4-5} \cmidrule(lr){6-8} 
		\cmidrule(lr){10-11} \cmidrule(lr){12-14}  
		&	&& $T_W$&$T_A$&$T_W^W$&$T_A^W$&$T_M^W$&&$T_W$&$T_A$&$T_W^W$&$T_A^W$&$T_M^W$\\[2pt]
		\cmidrule(lr){1-14}
		
	$H_0^T$&AR&(5,5)&39.6&5&5.1&4.7&5.9&&42.3&4.6&5.7&4.6&5.8\\
	&&(10,10)&36.6&4.3&4.9&4.9&5.1&&35&5&4.9&5.6&5.9\\
	&&(10,20)&30.2&4.6&5.7&5.2&5.7&&29.4&4.8&5.9&5.4&5.7\\
	&&(20,20)&15.8&4.8&5.6&5.1&5.4&&15.3&5.1&5.4&5.4&5.5\\
	\cmidrule(lr){1-14}
	&CS&(5,5)&40.5&3&4.9&3.4&5.1&&46.1&3.2&6.1&3.5&5.2\\
	&&(10,10)&35.7&3.9&4.9&5&5.4&&34.8&4&5&5.1&5.2\\
	&&(10,20)&29.8&3.8&6.1&4.7&5.5&&29.4&4&6&5&5.7\\
	&&(20,20)&15.6&4.4&5.2&5.1&5.1&&15.8&4.1&5.2&4.8&4.9\\
	\cmidrule(lr){1-14}
	&TP&(5,5)&37.3&6.4&4.9&5.4&6.2&&39.1&6.6&5.5&5.8&6.1\\
	&&(10,10)&38.2&5.6&5&5.5&5.7&&35.6&6.1&4.9&5.9&5.9\\
	&&(10,20)&31.9&5.9&6.4&5.9&6.2&&28.5&5.5&6.1&5.5&5.7\\
	&&(20,20)&17.3&5.6&6.1&5.7&5.6&&15.9&5.3&6.1&5.4&5.3\\
	\cmidrule(lr){1-14}
	$H_0^{GT}$&AR&(5,5)&38.5&4.7&5.2&4.5&6&&41.2&4.6&5.8&4.4&5.9\\
	&&(10,10)&36.5&4.9&4.2&5.4&5.9&&34.1&4.7&4.2&5.2&5.4\\
	&&(10,20)&29.6&4.7&5.9&5.1&5.5&&29&4.8&5.8&5.4&5.9\\
	&&(20,20)&14.9&5.1&5.1&5.6&5.2&&14.9&5.2&4.8&5.6&5.8\\
	\cmidrule(lr){1-14}
	&CS&(5,5)&39.3&3.8&5.4&4.2&5.6&&45.7&3.1&6&3.6&5.4\\
	&&(10,10)&36.1&3.7&4.5&4.7&5.5&&33.7&3.6&4.4&4.7&4.8\\
	&&(10,20)&29.9&4.2&6.1&5.2&5.8&&29&4&5.8&5.1&5.8\\
	&&(20,20)&15.7&4.7&5&5.4&5.5&&14.7&4.1&5.2&5&5\\
	\cmidrule(lr){1-14}
	&TP&(5,5)&36.7&6.3&5&5.2&6.2&&38.4&6.3&5.1&5.4&6\\
	&&(10,10)&37.8&5.7&4.6&5.6&5.7&&34&5.6&3.8&5.4&5.5\\
	&&(10,20)&30.6&5.7&5.5&5.6&6&&29.2&5.8&5.9&5.9&6.2\\
	&&(20,20)&15.4&5.4&4.9&5.2&5.3&&14.9&5.8&4.8&5.8&5.8\\

		\cmidrule(lr){1-14}
		
	\end{tabular*}
	\label{TableErrorMARlognormt8}
\end{table}

\begin{table}[ht]
	\centering
	\caption{Simulation results for type-I error level ($\alpha=0.05$) for normal distribution, testing the hypothesis $H_0^G$, $d\in\{4,8\}$ and two different MAR mechanisms.}
	\begin{tabular*}{0.9\linewidth}{c c  c  cc| ccc c  cc| ccc}
		\cmidrule(lr){1-14}
		& & &  \multicolumn{5}{c}{$MAR1$}  && \multicolumn{5}{c}{$MAR2$} \\ 
		\cmidrule(lr){4-8} \cmidrule(lr){10-14} 
	d&	Cov &n  &\multicolumn{2}{c}{Classic}  & 
		\multicolumn{3}{c}{WBtstrp}  && \multicolumn{2}{c}{Classic}  & \multicolumn{3}{c}{WBtstrp} \\ \cmidrule(lr){4-5} \cmidrule(lr){6-8} 
		\cmidrule(lr){10-11} \cmidrule(lr){12-14}  
		&	&& $T_W$&$T_A$&$T_W^W$&$T_A^W$&$T_M^W$&&$T_W$&$T_A$&$T_W^W$&$T_A^W$&$T_M^W$\\[2pt]
		\cmidrule(lr){1-14}
4&AR&(5,5)&6&6&4.4&4.4&5.1&&5.8&5.8&4.5&4.5&4.9\\
&&(10,10)&5.6&5.6&4.4&4.4&4.9&&5.7&5.7&4.7&4.7&4.9\\
&&(10,20)&5.5&5.5&5.2&5.2&5.2&&5.1&5.1&4.9&4.9&4.9\\
&&(20,20)&8.8&8.8&3.9&3.9&5.5&&8.4&8.4&3.3&3.3&5\\
\cmidrule(lr){1-14}
&CS&(5,5)&6.2&6.2&4.7&4.7&5.5&&6&6&4.4&4.4&5.1\\
&&(10,10)&5.5&5.5&4.3&4.3&4.5&&5.5&5.5&4.4&4.4&4.9\\
&&(10,20)&5.3&5.3&4.9&4.9&4.9&&5.1&5.1&4.8&4.8&4.9\\
&&(20,20)&9.1&9.1&4.3&4.3&5.8&&8&8&3.1&3.1&5.1\\
\cmidrule(lr){1-14}
&TP&(5,5)&6.6&6.6&4.8&4.8&5.4&&5.9&5.9&4.5&4.5&5.1\\
&&(10,10)&6&6&4.7&4.7&5.1&&5.5&5.5&4.4&4.4&4.9\\
&&(10,20)&5&5&4.7&4.7&4.7&&5&5&4.9&4.9&4.9\\
&&(20,20)&8.7&8.7&3.7&3.7&4.9&&8.2&8.2&3.3&3.3&5.1\\

\cmidrule(lr){1-14}
8&AR&(5,5)&6.4&6.4&4.7&4.7&5.3&&6.3&6.3&4.7&4.7&5.4\\
&&(10,10)&6.4&6.4&4.8&4.8&5.6&&6.3&6.3&5.1&5.1&5.7\\
&&(10,20)&6.2&6.2&5.8&5.8&6&&5.5&5.5&5&5&5.3\\
&&(20,20)&10.8&10.8&5&5&6.4&&9&9&3.6&3.6&5.2\\
\cmidrule(lr){1-14}
&CS&(5,5)&7.1&7.1&5.1&5.1&5.9&&6.5&6.5&4.8&4.8&5.7\\
&&(10,10)&6.1&6.1&4.6&4.6&5.2&&6.2&6.2&4.8&4.8&5.6\\
&&(10,20)&5.8&5.8&5.2&5.2&5.5&&5.4&5.4&4.9&4.9&5.2\\
&&(20,20)&10.3&10.3&4.9&4.9&6.5&&9.3&9.3&4&4&6\\
\cmidrule(lr){1-14}
&TP&(5,5)&6.6&6.6&4.7&4.7&5.3&&6.5&6.5&4.8&4.8&5.6\\
&&(10,10)&6.2&6.2&4.7&4.7&5.3&&6&6&4.8&4.8&5.4\\
&&(10,20)&5.8&5.8&5.2&5.2&5.4&&5.4&5.4&5&5&5.3\\
&&(20,20)&10&10&4.6&4.6&5.8&&8.8&8.8&3.4&3.4&5.2\\

		\cmidrule(lr){1-14}
		
	\end{tabular*}
	\label{TableErrorMARNormalHG}
\end{table}

\begin{table}[ht]
	\centering
	\caption{Simulation results for type-I error level ($\alpha=0.05$) for double exponential distribution,testing the hypothesis $H_0^G$, $d\in\{4,8\}$ and two different MAR mechanisms.}
	\begin{tabular*}{0.9\linewidth}{c c  c  cc| ccc c  cc| ccc}
		\cmidrule(lr){1-14}
		& & &  \multicolumn{5}{c}{$MAR1$}  && \multicolumn{5}{c}{$MAR2$} \\ 
		\cmidrule(lr){4-8} \cmidrule(lr){10-14} 
		d&	Cov &n  &\multicolumn{2}{c}{Classic}  & 
		\multicolumn{3}{c}{WBtstrp}  && \multicolumn{2}{c}{Classic}  & \multicolumn{3}{c}{WBtstrp} \\ \cmidrule(lr){4-5} \cmidrule(lr){6-8} 
		\cmidrule(lr){10-11} \cmidrule(lr){12-14}  
		&	&& $T_W$&$T_A$&$T_W^W$&$T_A^W$&$T_M^W$&&$T_W$&$T_A$&$T_W^W$&$T_A^W$&$T_M^W$\\[2pt]
		\cmidrule(lr){1-14}
4&AR&(5,5)&6.3&6.3&4.4&4.4&5.4&&5.9&5.9&4.4&4.4&5.2\\
&&(10,10)&5.9&5.9&4.7&4.7&5.2&&5.4&5.4&4.3&4.3&4.8\\
&&(10,20)&5.4&5.4&4.9&4.9&5.2&&5.4&5.4&5&5&5.1\\
&&(20,20)&9.4&9.4&4.3&4.3&6&&8.1&8.1&3.3&3.3&5.2\\
\cmidrule(lr){1-14}
&CS&(5,5)&6.5&6.5&4.8&4.8&5.7&&5.9&5.9&4.6&4.6&5.4\\
&&(10,10)&6.2&6.2&4.9&4.9&5.6&&5.6&5.6&4.4&4.4&5\\
&&(10,20)&5.3&5.3&4.9&4.9&5.1&&5.2&5.2&4.9&4.9&5.2\\
&&(20,20)&8.9&8.9&4.1&4.1&5.8&&8.2&8.2&3.5&3.5&5.7\\
\cmidrule(lr){1-14}
&TP&(5,5)&5.6&5.6&4.2&4.2&4.8&&6.1&6.1&4.8&4.8&5.5\\
&&(10,10)&6&6&4.8&4.8&5.2&&5.5&5.5&4.4&4.4&4.8\\
&&(10,20)&4.9&4.9&4.6&4.6&4.7&&5.6&5.6&5.2&5.2&5.4\\
&&(20,20)&8.8&8.8&4&4&5.4&&8.2&8.2&3.4&3.4&5.1\\

\cmidrule(lr){1-14}
8&AR&(5,5)&6.7&6.7&4.8&4.8&6&&6.2&6.2&4.7&4.7&5.7\\
&&(10,10)&6.7&6.7&5.2&5.2&6.1&&6.3&6.3&5&5&6\\
&&(10,20)&6.1&6.1&5.3&5.3&5.8&&5.5&5.5&5&5&5.5\\
&&(20,20)&10.4&10.4&4.8&4.8&6.8&&9.2&9.2&4.1&4.1&6.4\\
\cmidrule(lr){1-14}
&CS&(5,5)&6.7&6.7&4.9&4.9&6.1&&6.7&6.7&4.9&4.9&6.1\\
&&(10,10)&6.5&6.5&5&5&6&&6.2&6.2&5&5&5.8\\
&&(10,20)&5.5&5.5&4.8&4.8&5.4&&5.4&5.4&4.8&4.8&5.4\\
&&(20,20)&10.4&10.4&5&5&7.3&&9.7&9.7&4.4&4.4&7.2\\
\cmidrule(lr){1-14}
&TP&(5,5)&6.8&6.8&4.9&4.9&5.6&&6.4&6.4&4.7&4.7&5.7\\
&&(10,10)&6.3&6.3&4.7&4.7&5.4&&6.2&6.2&4.9&4.9&5.6\\
&&(10,20)&5.7&5.7&5.2&5.2&5.5&&5.4&5.4&4.9&4.9&5.3\\
&&(20,20)&9.9&9.9&4.2&4.2&5.8&&8.5&8.5&3.3&3.3&5.7\\

		\cmidrule(lr){1-14}
		
	\end{tabular*}
	\label{TableErrorMARDexpHG}
\end{table}

\begin{table}[ht]
	\centering
	\caption{Simulation results for type-I error level ($\alpha=0.05$) for chi-square distribution,testing the hypothesis $H_0^G$, $d\in\{4,8\}$ and two different MAR mechanisms.}
	\begin{tabular*}{0.9\linewidth}{c c  c  cc| ccc c  cc| ccc}
		\cmidrule(lr){1-14}
		& & &  \multicolumn{5}{c}{$MAR1$}  && \multicolumn{5}{c}{$MAR2$} \\ 
		\cmidrule(lr){4-8} \cmidrule(lr){10-14} 
		d&	Cov &n  &\multicolumn{2}{c}{Classic}  & 
		\multicolumn{3}{c}{WBtstrp}  && \multicolumn{2}{c}{Classic}  & \multicolumn{3}{c}{WBtstrp} \\ \cmidrule(lr){4-5} \cmidrule(lr){6-8} 
		\cmidrule(lr){10-11} \cmidrule(lr){12-14}  
		&	&& $T_W$&$T_A$&$T_W^W$&$T_A^W$&$T_M^W$&&$T_W$&$T_A$&$T_W^W$&$T_A^W$&$T_M^W$\\[2pt]
		\cmidrule(lr){1-14}
		
	4&AR&(5,5)&6.1&6.1&4.6&4.6&5.2&&6.3&6.3&4.8&4.8&5.5\\
	&&(10,10)&6&6&4.8&4.8&5.3&&6.3&6.3&5.3&5.3&5.8\\
	&&(10,20)&5.2&5.2&4.7&4.7&4.8&&5.5&5.5&5.2&5.2&5.3\\
	&&(20,20)&9&9&4&4&5.7&&8.3&8.3&3.5&3.5&5.5\\
	\cmidrule(lr){1-14}
	&CS&(5,5)&6.3&6.3&4.6&4.6&5.4&&6.2&6.2&4.7&4.7&5.4\\
	&&(10,10)&6.5&6.5&5.2&5.2&5.8&&5.5&5.5&4.3&4.3&4.9\\
	&&(10,20)&5.4&5.4&4.8&4.8&5.1&&5.7&5.7&5.3&5.3&5.7\\
	&&(20,20)&9.2&9.2&4.1&4.1&6&&8.9&8.9&3.6&3.6&5.9\\
	\cmidrule(lr){1-14}
	&TP&(5,5)&6&6&4.3&4.3&5&&6&6&4.6&4.6&5.3\\
	&&(10,10)&5.9&5.9&4.7&4.7&5.1&&5.9&5.9&4.9&4.9&5.3\\
	&&(10,20)&5.5&5.5&5&5&5.2&&5.3&5.3&5&5&5.2\\
	&&(20,20)&9&9&4.2&4.2&5.4&&8.4&8.4&3.8&3.8&5.6\\

	\cmidrule(lr){1-14}
	8&AR&(5,5)&6.9&6.9&4.8&4.8&5.9&&6.5&6.5&4.7&4.7&5.8\\
	&&(10,10)&6.5&6.5&4.9&4.9&5.5&&6.3&6.3&5&5&5.9\\
	&&(10,20)&6&6&5.4&5.4&5.8&&5.4&5.4&4.9&4.9&5.3\\
	&&(20,20)&10.7&10.7&4.8&4.8&6.7&&9.5&9.5&4&4&6.2\\
	\cmidrule(lr){1-14}
	&CS&(5,5)&6.6&6.6&4.6&4.6&5.8&&6.7&6.7&4.8&4.8&6.1\\
	&&(10,10)&6.4&6.4&5.1&5.1&5.7&&6.3&6.3&5.1&5.1&5.9\\
	&&(10,20)&5.7&5.7&5&5&5.5&&5.9&5.9&5.3&5.3&5.7\\
	&&(20,20)&10.2&10.2&5&5&7.2&&9.9&9.9&4.1&4.1&7\\
	\cmidrule(lr){1-14}
	&TP&(5,5)&6.4&6.4&4.5&4.5&5.2&&6.4&6.4&4.5&4.5&5.7\\
	&&(10,10)&6.4&6.4&5&5&5.6&&6.6&6.6&5.2&5.2&6\\
	&&(10,20)&5.4&5.4&4.8&4.8&5.1&&5.6&5.6&5&5&5.5\\
	&&(20,20)&10.8&10.8&4.7&4.7&5.9&&8.9&8.9&3.6&3.6&5.9\\

		\cmidrule(lr){1-14}
		
	\end{tabular*}
	\label{TableErrorMARChisqHG}
\end{table}

\begin{table}[ht]
	\centering
	\caption{Simulation results for type-I error level ($\alpha=0.05$) for lognormal distribution,testing the hypothesis $H_0^G$, $d\in\{4,8\}$ and two different MAR mechanisms.}
	\begin{tabular*}{0.9\linewidth}{c c  c  cc| ccc c  cc| ccc}
		\cmidrule(lr){1-14}
		& & &  \multicolumn{5}{c}{$MAR1$}  && \multicolumn{5}{c}{$MAR2$} \\ 
		\cmidrule(lr){4-8} \cmidrule(lr){10-14} 
		d&	Cov &n  &\multicolumn{2}{c}{Classic}  & 
		\multicolumn{3}{c}{WBtstrp}  && \multicolumn{2}{c}{Classic}  & \multicolumn{3}{c}{WBtstrp} \\ \cmidrule(lr){4-5} \cmidrule(lr){6-8} 
		\cmidrule(lr){10-11} \cmidrule(lr){12-14}  
		&	&& $T_W$&$T_A$&$T_W^W$&$T_A^W$&$T_M^W$&&$T_W$&$T_A$&$T_W^W$&$T_A^W$&$T_M^W$\\[2pt]
		\cmidrule(lr){1-14}
		
4&AR&(5,5)&6.2&6.2&4.2&4.2&5.6&&6.7&6.7&4.8&4.8&6.2\\
&&(10,10)&6.8&6.8&5.4&5.4&6.2&&6.5&6.5&5.2&5.2&6.2\\
&&(10,20)&5.7&5.7&5.1&5.1&5.7&&5.7&5.7&5&5&5.7\\
&&(20,20)&9.2&9.2&4.1&4.1&6.5&&8.9&8.9&4&4&6.6\\
\cmidrule(lr){1-14}
&CS&(5,5)&6.7&6.7&4.7&4.7&6.4&&6.7&6.7&4.8&4.8&6.4\\
&&(10,10)&6.9&6.9&5.5&5.5&6.4&&6.7&6.7&5.2&5.2&6.3\\
&&(10,20)&5.8&5.8&5.1&5.1&5.7&&5.7&5.7&4.8&4.8&5.4\\
&&(20,20)&10&10&4.5&4.5&7.4&&9.4&9.4&4.1&4.1&7.4\\
\cmidrule(lr){1-14}
&TP&(5,5)&6.1&6.1&4.2&4.2&5.5&&6.2&6.2&4.4&4.4&5.9\\
&&(10,10)&6.1&6.1&4.9&4.9&5.7&&6.5&6.5&5.3&5.3&6.1\\
&&(10,20)&5.3&5.3&4.7&4.7&5.3&&5.5&5.5&4.9&4.9&5.6\\
&&(20,20)&9.2&9.2&4.3&4.3&6.5&&9.4&9.4&3.8&3.8&6.6\\

\cmidrule(lr){1-14}
8&AR&(5,5)&6.9&6.9&4.8&4.8&6.7&&6.3&6.3&4.7&4.7&6.4\\
&&(10,10)&6.7&6.7&5.3&5.3&6.5&&6.7&6.7&5.4&5.4&6.6\\
&&(10,20)&5.7&5.7&4.9&4.9&5.8&&5.8&5.8&5.1&5.1&5.8\\
&&(20,20)&10.3&10.3&4.9&4.9&8.4&&9.5&9.5&4.2&4.2&8.1\\
\cmidrule(lr){1-14}
&CS&(5,5)&7.2&7.2&5.1&5.1&7.3&&6.4&6.4&4.5&4.5&6.5\\
&&(10,10)&7&7&5.6&5.6&6.9&&6.9&6.9&5.3&5.3&6.8\\
&&(10,20)&5.8&5.8&5&5&6&&5.7&5.7&4.8&4.8&5.8\\
&&(20,20)&9.6&9.6&4.3&4.3&8.2&&9.2&9.2&4.3&4.3&8.2\\
\cmidrule(lr){1-14}
&TP&(5,5)&6.7&6.7&4.7&4.7&6.1&&6.6&6.6&4.8&4.8&6.4\\
&&(10,10)&6.7&6.7&5.2&5.2&6.3&&6.8&6.8&5.6&5.6&6.6\\
&&(10,20)&5.4&5.4&4.9&4.9&5.5&&5.6&5.6&5&5&5.7\\
&&(20,20)&10&10&4.7&4.7&6.9&&9.3&9.3&4.2&4.2&7.2\\

		\cmidrule(lr){1-14}
		
	\end{tabular*}
	\label{TableErrorMARLognormHG}
\end{table}

\begin{table}[ht]
	\centering
	\caption{Simulation results for type-I error level ($\alpha=0.05$) of the tests for different distributions under varying covariance structures with sample sizes ($n=299, d=3$) and missingness inspired from the fluvoxamine study data. }
	\begin{tabular*}{0.6\linewidth}{c   c  cc| ccc}
	\cmidrule(lr){1-7} 
		Dist&	Cov &\multicolumn{2}{c}{Classic}  & 
		\multicolumn{3}{c}{WBtstrp}  \\ \cmidrule(lr){3-4} \cmidrule(lr){5-7} 
		&	& $T_W$&$T_A$&$T_W^W$&$T_A^W$&$T_M^W$\\[2pt]
		\cmidrule{1-7}	
		Ordinal&--&5.2&4.9&5.1&5.2&5.1\\	\cmidrule(lr){1-7}
		Normal&AR&5&4.8&4.9&4.9&4.9\\
		Normal&CS&5.3&5.2&5.1&5.3&5.2\\
		Normal&TP&4.7&4.6&4.6&4.8&4.7\\	\cmidrule(lr){1-7}
		Double exponential&AR&5&4.8&4.9&4.8&4.9\\
		Double exponential&CS&5.2&5.3&5&5.2&5\\
			Double exponential&TP&4.9&4.6&4.7&4.7&4.7\\ 	\cmidrule(lr){1-7}
		Chi-square&AR&5.1&4.9&4.9&5&4.9\\
		Chi-square&CS&5.1&5.1&5&5.1&5.2\\
		Chi-square&TP&4.9&4.8&4.8&4.9&5\\ 	\cmidrule(lr){1-7}
		Lognormal&AR&6.5&5.8&6.3&5.7&5.9\\
		Lognormal&CS&5.3&5&5.2&5.1&5.1\\
		Lognormal&TP&8.9&7&8.6&6.9&7.4\\
		\cmidrule(lr){1-7}	
	\end{tabular*}
	\label{Tablea1Flov}
\end{table}

\begin{table}[!ht]
	\centering
	\caption{Simulation results for type-I error level ($\alpha=0.05$) of the tests for different distributions under varying covariance structures with sample sizes ($n_1=88,  n_2=84, d=3$) and MCAR data with missing rate ($r=30\%$) inspired from the skin disorder study data. }
	\begin{tabular*}{0.7\textwidth}{c l  c  cc| ccc}
		\cmidrule(lr){1-8} 
	Hyp	&Dist&	Cov &\multicolumn{2}{c}{Classic}  & 
		\multicolumn{3}{c}{WBtstrp}  \\ \cmidrule(lr){4-5} \cmidrule(lr){6-8} 
		&&	& $T_W$&$T_A$&$T_W^W$&$T_A^W$&$T_M^W$\\[2pt]
		\cmidrule{1-8}	
		$H_0^{G}$&Ordinal&--&5.4&5.4&5.4&5.4&5.5\\\cmidrule(lr){2-8} 
&Normal&AR&4.9&4.9&4.9&4.9&4.9\\
&Normal&CS&4.8&4.8&4.8&4.8&4.8\\
&Normal&TP&4.8&4.8&4.8&4.8&4.8\\\cmidrule(lr){2-8}  
&Double exponential&AR&5&5&5&5&5\\
		&Double exponential&CS&4.9&4.9&4.9&4.9&4.8\\
\end{tabular*}
	\label{TableToenail}
\end{table}

\clearpage
\bibliographystyle{apa}
\bibliography{References}